\documentclass[aps,pre,longbibliography,12pt,preprint,floatfix]{revtex4-1}
 \usepackage{bm}
\usepackage{natbib}
\usepackage{url}
\usepackage[intlimits]{amsmath}
\usepackage{graphicx}
\usepackage{fancyhdr}
\usepackage{amsfonts}
\usepackage{amssymb}
\usepackage{float}
\usepackage[caption=false]{subfig}
\usepackage{microtype}

\usepackage[colorlinks=false, pdfborder={0 0 0}]{hyperref}
\usepackage{cleveref}

\begin{document}
\title{A new resonance mechanism in the Swift--Hohenberg equation with time-periodic forcing}
\author{Punit Gandhi}
 \email{punit\_gandhi@berkeley.edu}
\author{Edgar Knobloch}
\affiliation{Department of Physics, University of California, Berkeley CA 94720, USA}
\author{C\'edric Beaume}
\affiliation{Department of Aeronautics, Imperial College London, South Kensington, London SW7 2AZ, UK}
\date{\today}

\date{\today}

\begin{abstract}
The generalized Swift--Hohenberg equation with a quadratic-cubic nonlinearity is used to study the persistence and decay of localized patterns in the presence of time-periodic parametric forcing. A novel resonance phenomenon between the forcing period and the time required to nucleate one wavelength of the pattern outside the pinning region is identified. The resonance generates distinct regions in parameter space characterized by the net number of wavelengths gained or lost in one forcing cycle. These regions are well described by an asymptotic theory based on the wavelength nucleation/annihilation time near the boundaries of the pinning region. The resulting theory leads to predictions that are qualitatively correct and, in some cases,  provide quantitative agreement with numerical simulations.
\end{abstract}
\maketitle
%








\section{Introduction}

Spatially localized structures in physical, chemical and biological systems \cite{champneys1998homoclinic,knobloch2008spatially} often consist of a time-independent spatial pattern embedded in a homogeneous background. The theory behind the origin and properties of such structures is well understood, at least in one spatial dimension. In this theory the localized structures are described in terms of heteroclinic cycles connecting the homogeneous state to the patterned state and back again \cite{balmforth1995solitary,woods1999heteroclinic}. Such cycles may be structurally stable, and so persist over an interval of parameter values located within the region of bistability between the homogeneous and spatially periodic states. This so-called snaking region \cite{pomeau1986front,hunt2000cellular} typically contains two or four families of homoclinic solutions connecting the homogeneous state to itself, organized within a snakes-and-ladders structure \cite{burke2006,burke2007snakes}. These correspond to spatially localized states of ever greater length and accumulate on the heteroclinic cycle as their length increases. Examples of this behavior have been identified in both gradient and nongradient systems, including buckling of slender structures \cite{hunt1989structural,hunt2000cellular}, shear flows \cite{Schneider2010snakes}, doubly diffusive convection \cite{Mercader11,Beaume2013convectons,Beaume2013nonsnaking}, porous media convection~\cite{Lojacono13} and rotating convection \cite{Beaume2013localized}, among others.

Localized states are also encountered in systems with a fluctuating or noisy background \cite{schapers2000interaction,Prigent02,Barkley05,clerc2005additive} as well in periodically driven systems \cite{Umbanhowar95,Vilar1997spatiotemporal,Lioubashevski99,binder2000observation,Rucklidge2009design,timonen13}. Temporal focing has, in general, a number of consequences. In extended systems it may destabilize existing patterns or lead to resonant excitation of new patterns and a variety of phase-locking phenomena \cite{yochelis2005frequency,marts2006resonant}. In addition, new structures may be generated by rapid switching between two coexisting attractors \cite{buceta2001stationary,belykh2013multistable}. Localized structures may be impacted in two different ways. First, the temporal forcing may render existing localized structures time-dependent, and second, it may generate bistability between a homogeneous state and an extended parametrically driven spatially periodic pattern.  The latter case creates a parameter regime where spatially localized time-dependent patterns may be found \cite{burke2008classification}. 

In the present paper we focus on time-independent systems supporting spatially localized states and study the effect of time-periodic forcing on these states. Our interest in this type of problem is motivated in part by recent studies of the growth of vegetation patterns near the transition to desertification \cite{sherratt2005analysis,tlidi2008vegetation,sherratt2010pattern,kletter2012ostwald,Meron12}. Simple models of this process predict the presence of patchy patterns \cite{Meron12} and the properties of such patterns are yet to be examined when seasonal variation in growth conditions is included.

Most of the systems mentioned above can be modeled using equations of Swift--Hohenberg-type despite the fact that this equation is of gradient type. This is because the snakes-and-ladders structure of the snaking region in gradient and nongradient systems is identical (although the stability properties of the solutions may differ \cite{BurkeDawes}). Consequently, we adopt here a model of this type and investigate the effect of time-dependent forcing on the existence and stability of localized states within this model. We identify a number of new structures in this system, including time-dependent breathing states and structures that grow or shrink in an episodic manner. In particular, we identify a novel resonance phenomenon between the forcing period and the time required to nucleate a new wavelength, triggered whenever the forcing parameter falls outside the pinning region. This resonance leads to a complex partitioning of the parameter space whose structure can be understood qualitatively, and in some cases quantitatively, using appropriate asymptotics.

This paper is organized as follows. In the next section we summarize pertinent results concerning the autonomous Swift--Hohenberg equation with competing quadratic and cubic nonlinearities. In section III we consider the effect of high frequency temporal forcing of this equation, and then in section IV we focus on the different breathing states present for intermediate frequencies. This section forms the bulk of the paper. Low frequency forcing is considered in section V, followed by brief conclusions in section VI.

\section{The Swift--Hohenberg equation}

The quadratic-cubic Swift--Hohenberg equation (SHE) serves as a model for pattern formation in a broad range of physical systems. This equation which, in one dimension, takes the form
\begin{equation}
u_t= r u-\left(1+\partial_{x}^2\right)^2u+bu^2-u^3 \label{eq:SH},
\end{equation}
describes the dynamics of a real field $u(x,t)$ in time. The parameter $r$ specifies the strength of the forcing while the parameter $b>\sqrt{27/38}$ determines the extent of the bistability region between the homogeneous state $u_h\equiv 0$ and the patterned state $u_p(x)$, $u_p(x)=u_p(x+2\pi)$ for all $x$. The equation can be written in terms of a Lyapunov functional $\mathcal{F}[u]$, referred to as the {\it free energy},
\begin{equation}
u_t=-\frac{\delta\mathcal{F}[u]}{\delta u}, \qquad  \mathcal{F}[u]=\frac{1}{\Gamma}\int_{-\Gamma/2}^{\Gamma/2}-\frac{1}{2}r u^2+\tfrac{1}{2} \left[(1+\partial_x^2)u\right]^2 - b\frac{u^3}{3} + \frac{u^4}{4} \;d x.
\end{equation}
Thus, on a domain of finite spatial period $\Gamma$ all initial conditions approach a steady state corresponding to a local minimum of the free energy.
 
In order to study the effects of time-dependence, we write $r = r_0+\rho \sin\omega t$, where $r_0$, $\rho$, and $T=2\pi/\omega$ define the offset, amplitude, and period of the oscillation. Note that this type of parametric forcing leaves the homogeneous state $u_h=0$ unchanged. In the following we take $b=1.8$ \cite{burke09,kao14} and use periodic boundary conditions on a domain of $\Gamma=80\pi$ (i.e. $40$ characteristic wavelengths), unless otherwise noted. In addition we impose the symmetry $x \rightarrow -x$ of \cref{eq:SH} on all solutions of the system thereby focusing on {\it even} solutions.  This procedure allows us to perform computations on the half domain. We integrate the equation forward in time using a fourth order exponential time differencing scheme~\cite{cox2002} on an equidistributed mesh. Our calculations are performed in Fourier space and fully dealiased. In cases where a larger domain was necessary, the spatial density of grid points was kept constant. Steady state solutions of the constant forcing case were computed using the numerical continuation software AUTO~\cite{doedel1981auto}.

\subsection{Stationary localized states}

\begin{figure}
  \centering
\mbox{
      \subfloat[]{\label{fig:shesnaking:a}\includegraphics[width=75mm]{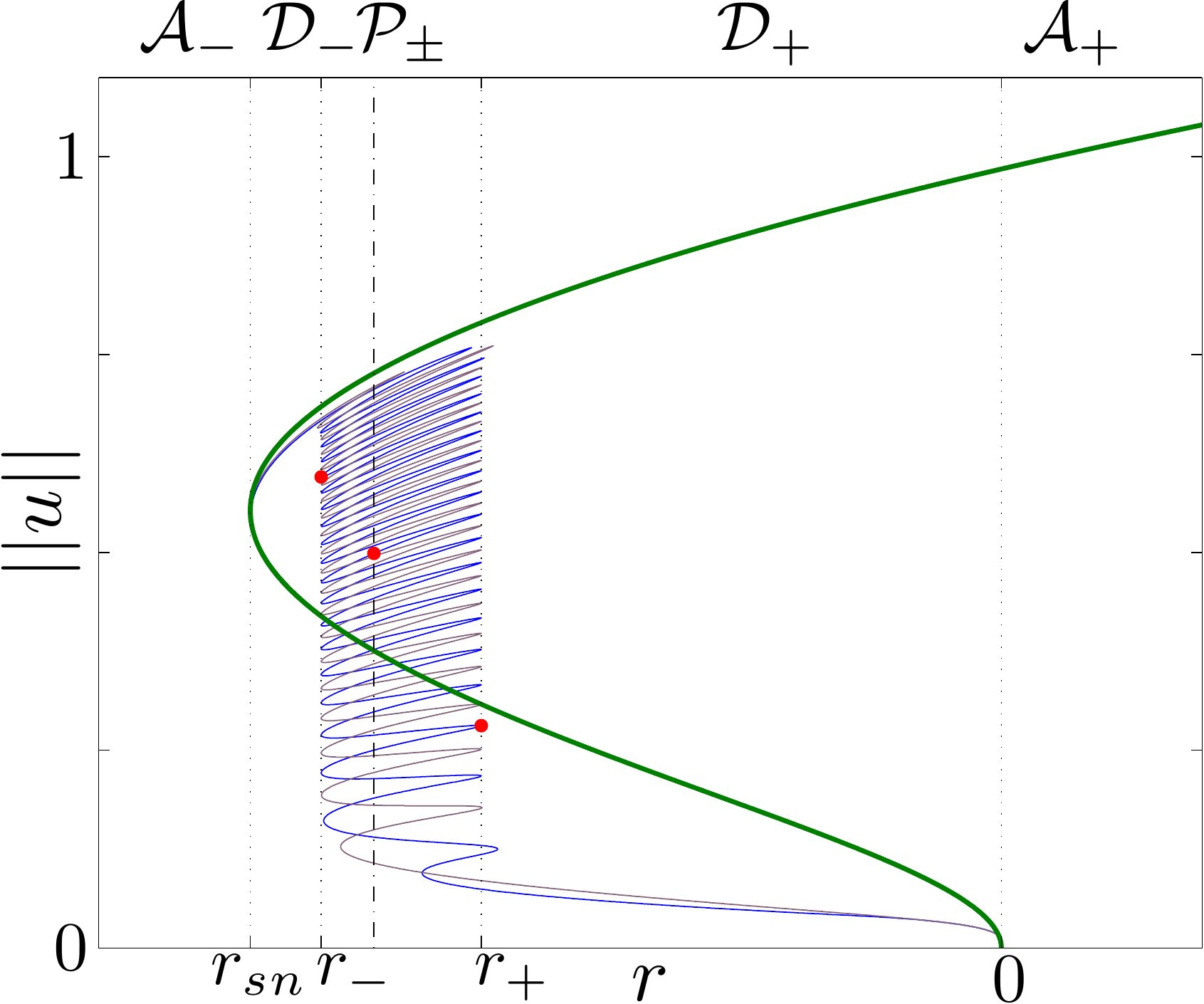}} \quad
      \subfloat[]{\label{fig:shesnaking:b}\includegraphics[width=75mm]{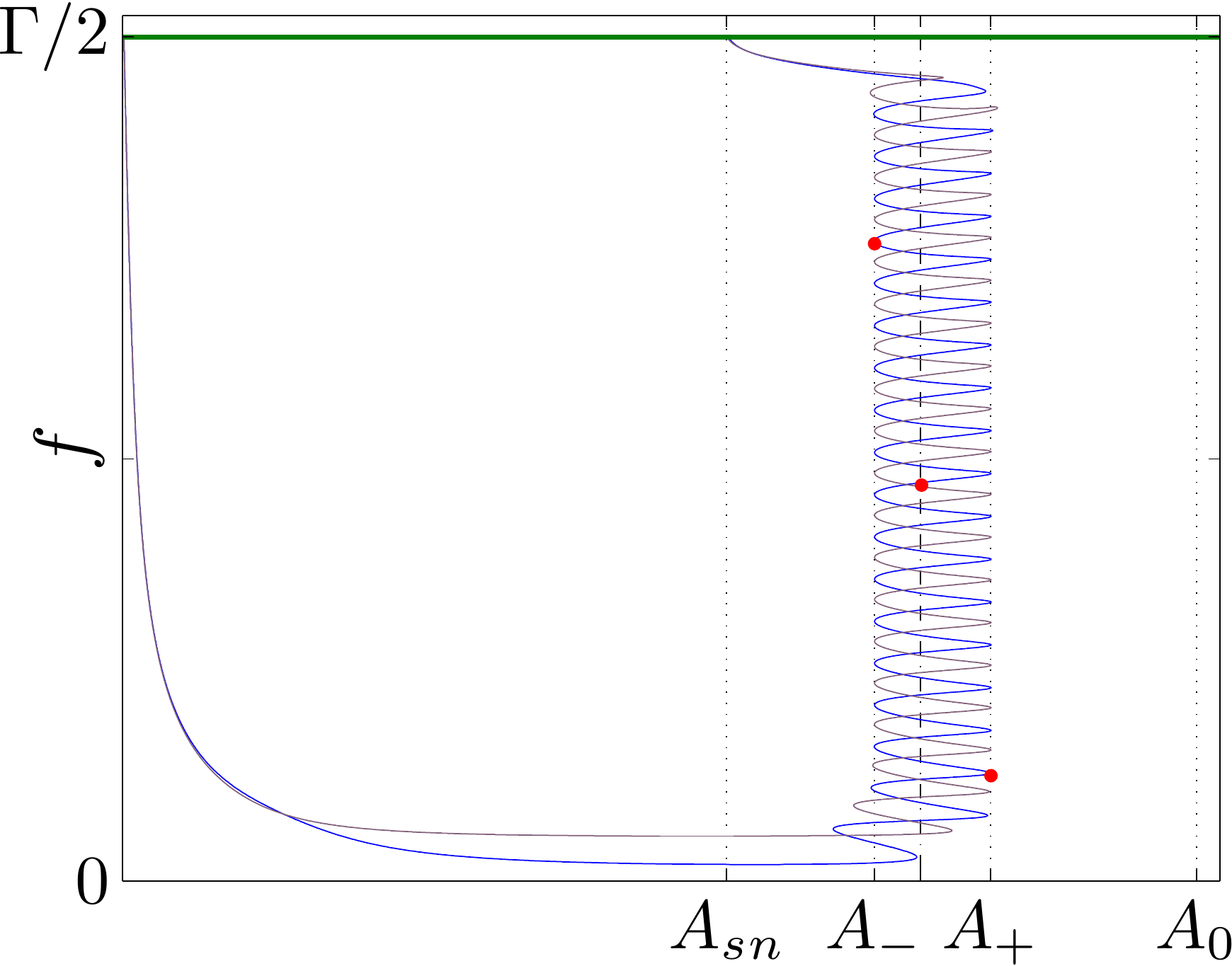}}
} \\
    \subfloat[]{\label{fig:shesnaking:c}\includegraphics[width=150mm]{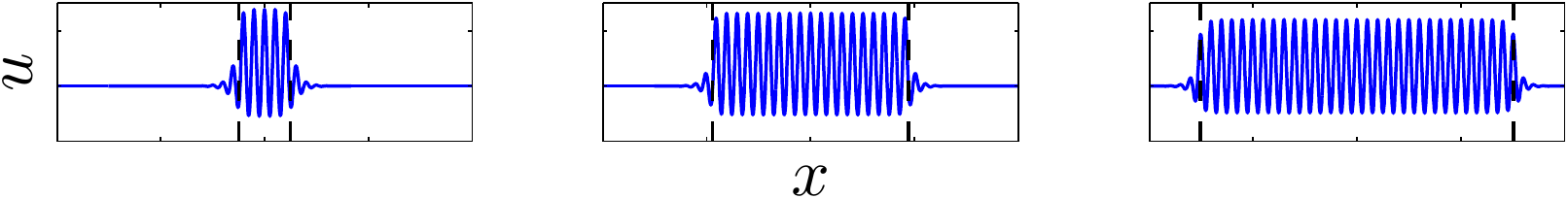}}

    \caption{(a) Bifurcation diagram showing the normalized $L_2$-norm $||u||=\sqrt{\tfrac{1}{\Gamma}\int_{-\Gamma/2}^{\Gamma/2}u^2\;dx}$ of time-independent solutions of \cref{eq:SH} as a function of the forcing parameter $r$. Vertical dashed lines delimit the amplitude regime $\mathcal{A}_-$, the depinning regimes $\mathcal{D}_{\pm}$ and the pinning region $\mathcal{P}_{\pm}$. The characteristics of each regime are described in the text. (b) The same as (a) but projected on the amplitude $A = \max_x(u)$ and the position $f>0$ of the right front, as defined in the text. (c) Solutions $u(x)$ corresponding to the red circles in (a) and (b), with black dashed lines indicating the locations $x=\pm f$ of the fronts.}
    \label{fig:shesnaking}
\end{figure}

For $b=1.8$ and $r \equiv r_0$, a spatially periodic solution $u_p$ bifurcates subcritically from $u_h$ at $r=0$. The periodic state passes through a saddle-node at $r_{sn} \approx -0.3744$ gaining stability and creating a region of bistability with $u_h$ in $-0.3744<r<0$. The Maxwell point is located at $r_M \approx -0.3126$ within the bistability region and corresponds to the point where $\mathcal{F}[u_p]=\mathcal{F}[u_h]=0$. The pinning or snaking region $r_-<r<r_+$ straddles this point ($r_- \approx -0.3390$, $r_+ \approx -0.2593$), and contains a pair of intertwined branches (\cref{fig:shesnaking:a}) of even parity spatially localized states with maxima (hereafter $L_0$) or minima (hereafter $L_{\pi}$) at $x=0$ as described in \cite{burke2006}. In a finite domain snaking continues until the domain is (almost) filled with pattern; thereafter the solution branches exit the pinning region and terminate on branches of periodic states near their saddle-node. 
Throughout this study we present our results in terms of the amplitude of the pattern, $A = \max_x(u)$, and the location $x=f$ of the front connecting the pattern to the homogeneous state relative to the axis of symmetry $x=0$ of the pattern,
\begin{equation}
f=2\frac{ \int_{0}^{\Gamma/2}  x u^2 \;d x}{ \int_{0}^{\Gamma/2}  u^2 \;d x}.
\end{equation}
As the amplitude $A$ of snaking localized solutions is comparable to that of the periodic state at the same parameter values, larger values of $f$ indicate broader localized structures. However, between the pinning region and $r=0$ the solutions broaden to fill the available domain as their amplitude decreases to zero. Thus $f$ increases without bound as $A\to0$ (\cref{fig:shesnaking:b}).

\subsection{Temporal dynamics of localized initial conditions}

Spatially localized initial conditions of \cref{eq:SH} eventually settle on a steady state, but the type of steady state and the transient leading to it depend on $r$ and the initial condition. The relevant regimes organized around the presence of steady spatially localized states can be identified in \cref{fig:shesnaking} and are summarized below:

\begin{figure}
\centering
\subfloat[$\mathcal{A}_-$, $r=-0.40$]{\includegraphics[width=60mm]{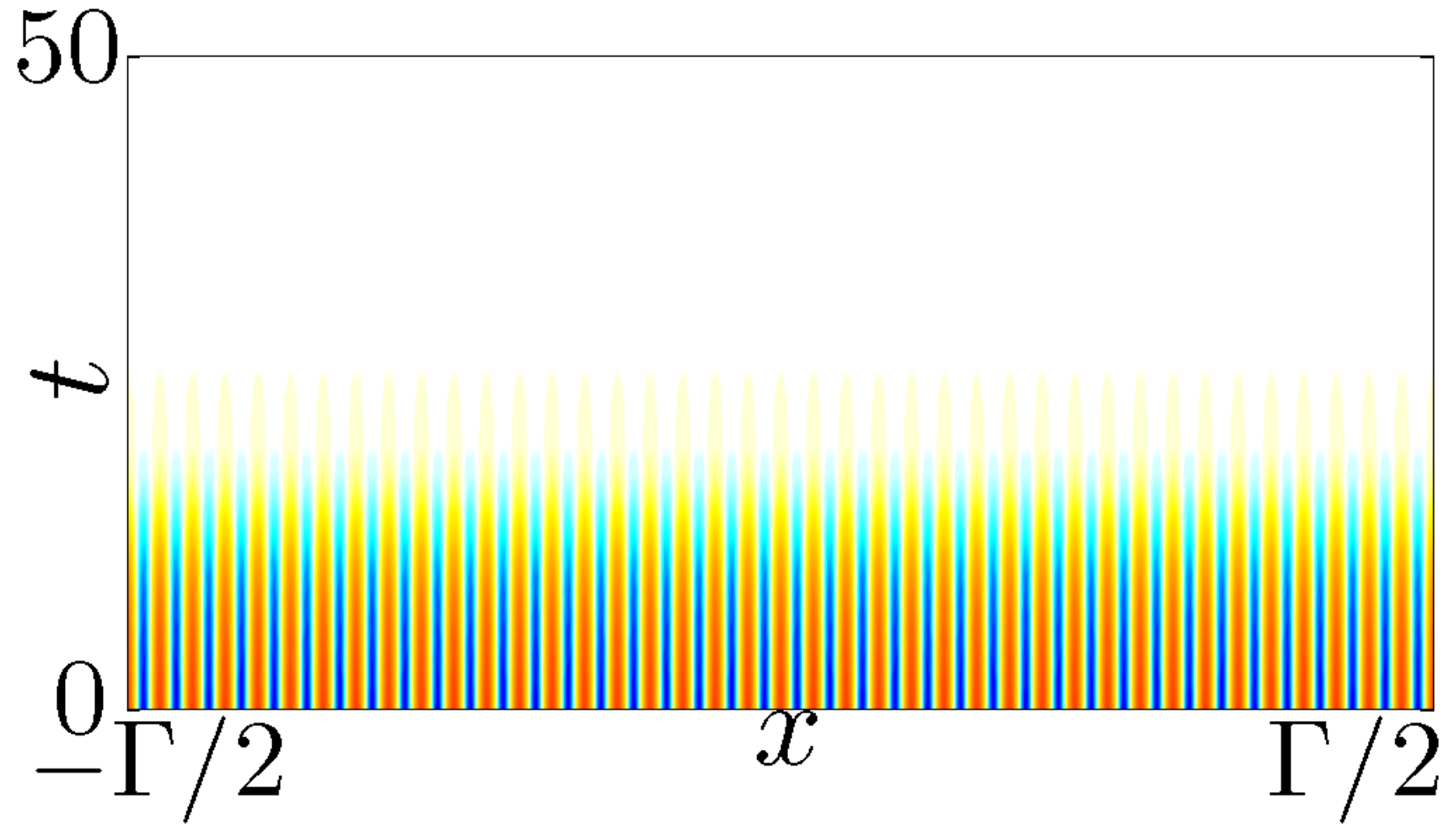}
						\includegraphics[width=60mm ]{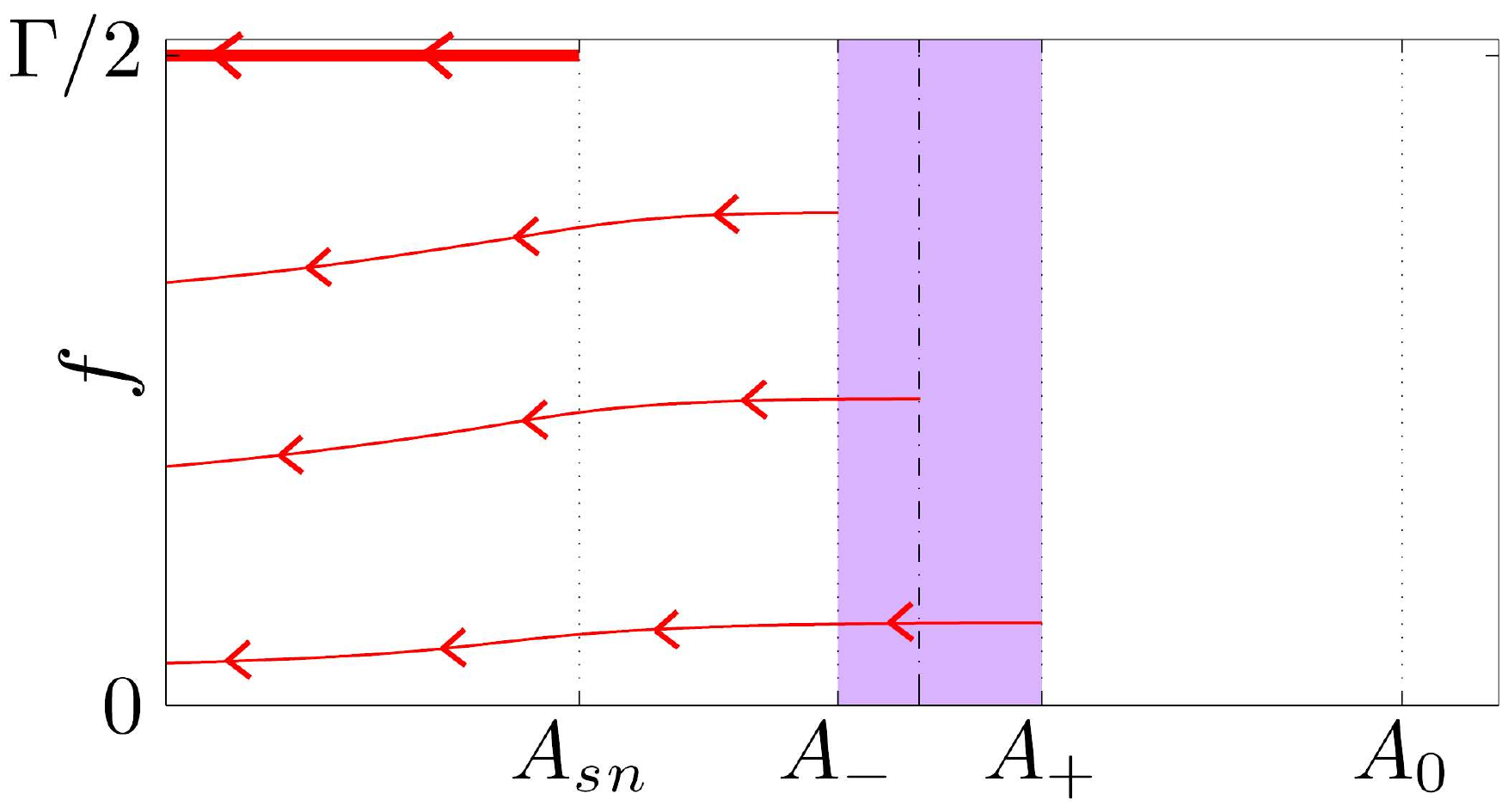}
		}
\\
\subfloat[$\mathcal{D}_-$, $r=-0.36$]{\includegraphics[width=60mm]{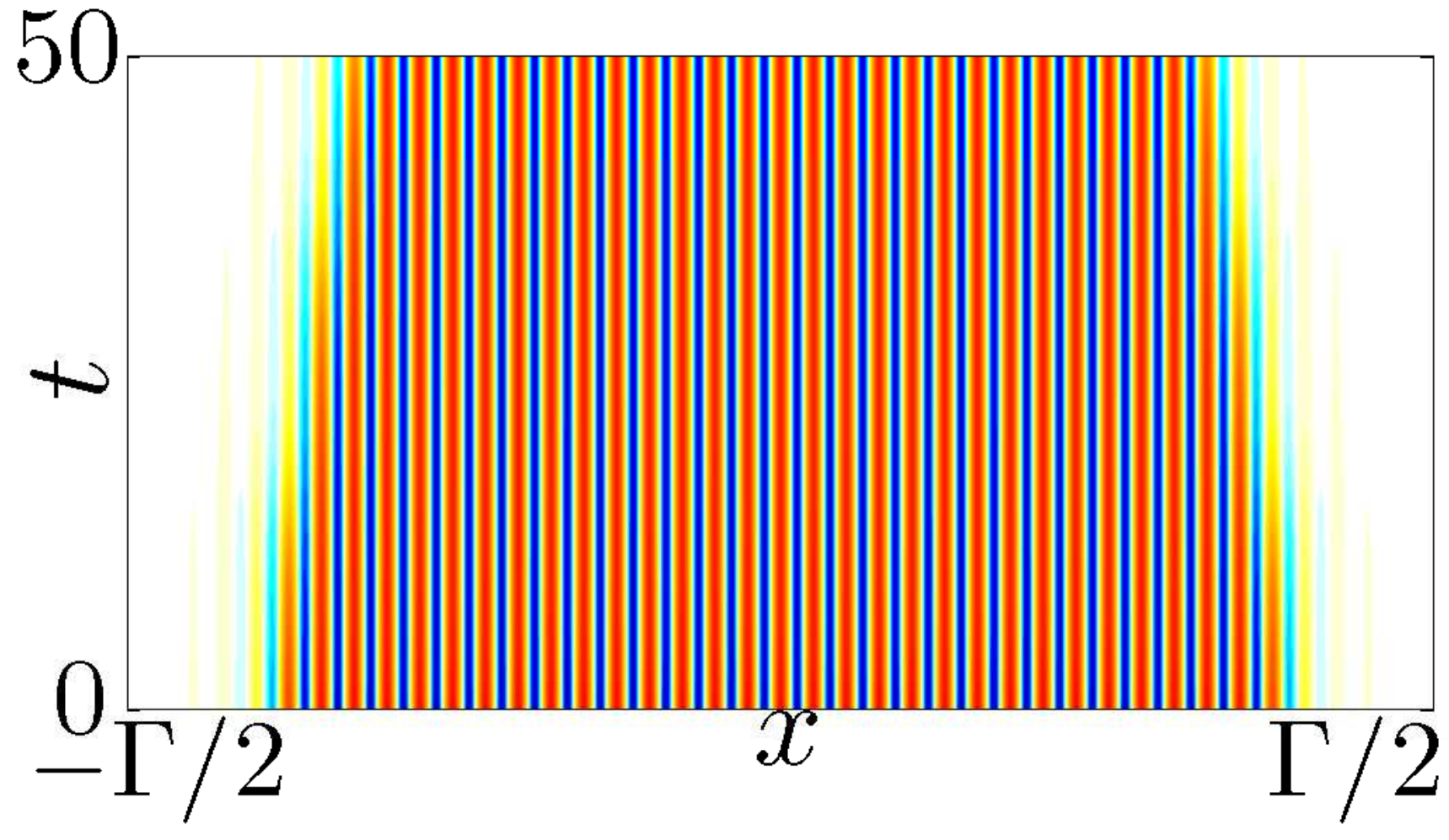}
						\includegraphics[width=60mm]{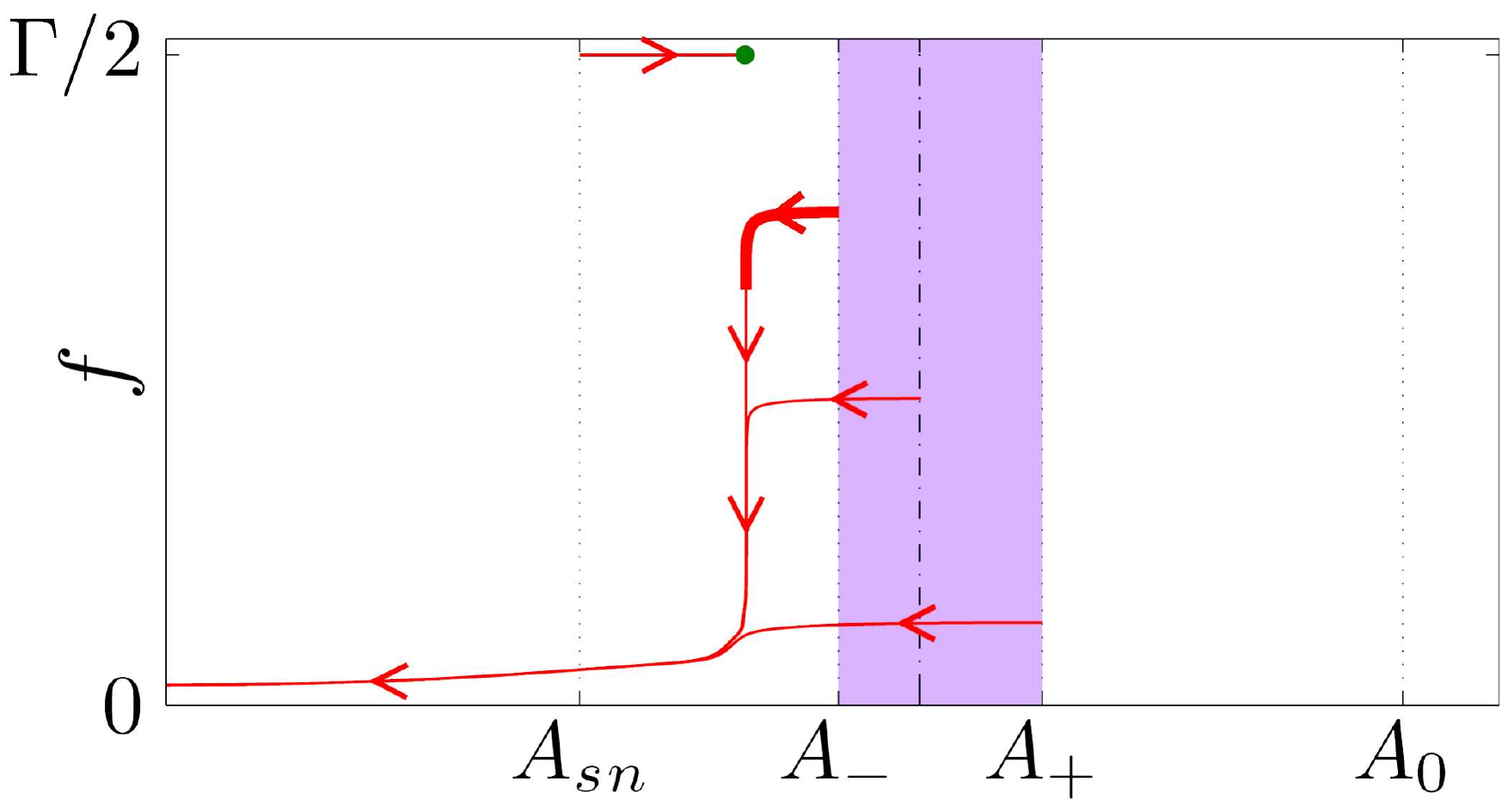}
		}
\\
\subfloat[$\mathcal{P}_+$, $r=-0.28$]{\includegraphics[width=60mm]{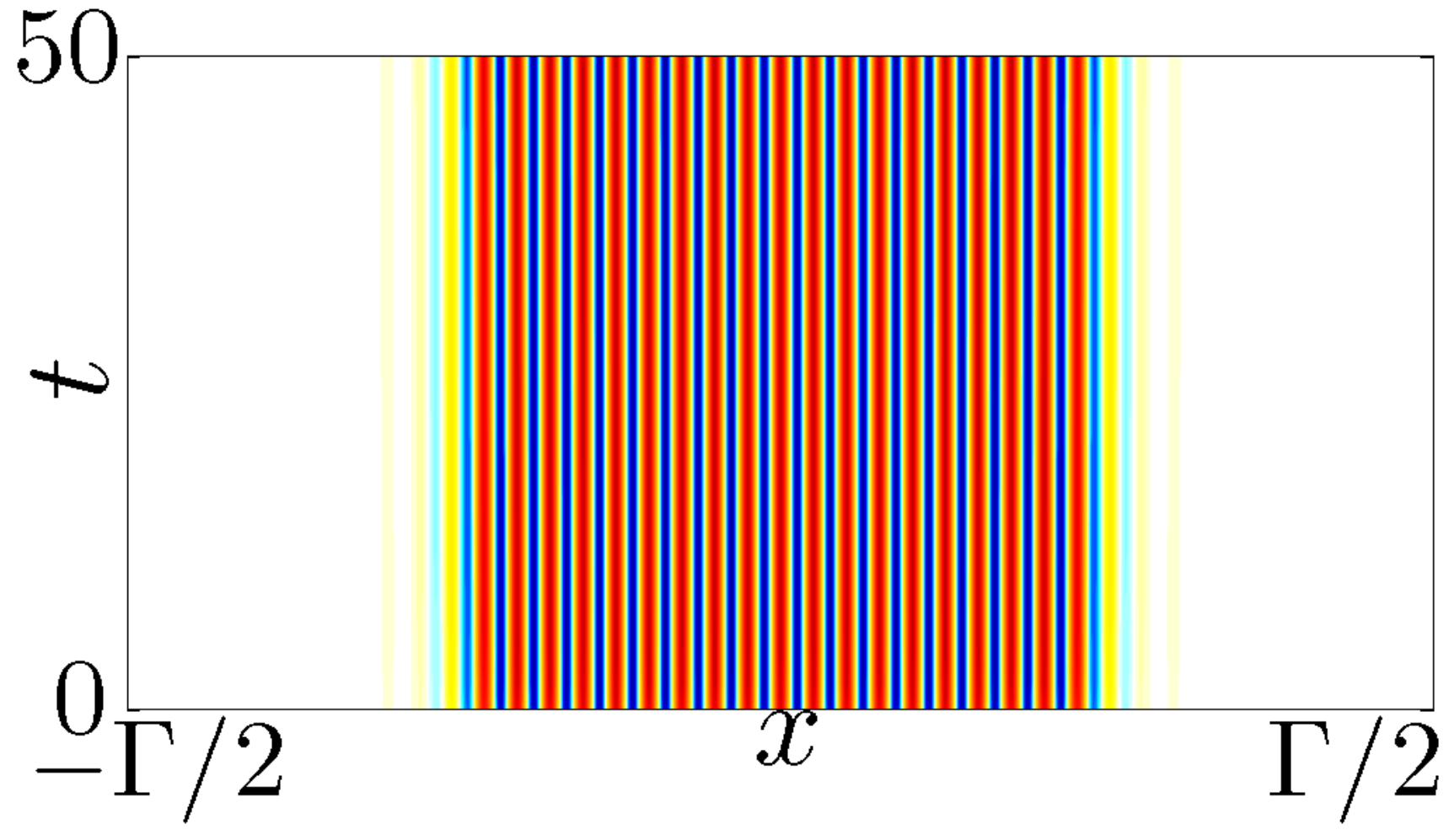}
						\includegraphics[width=60mm]{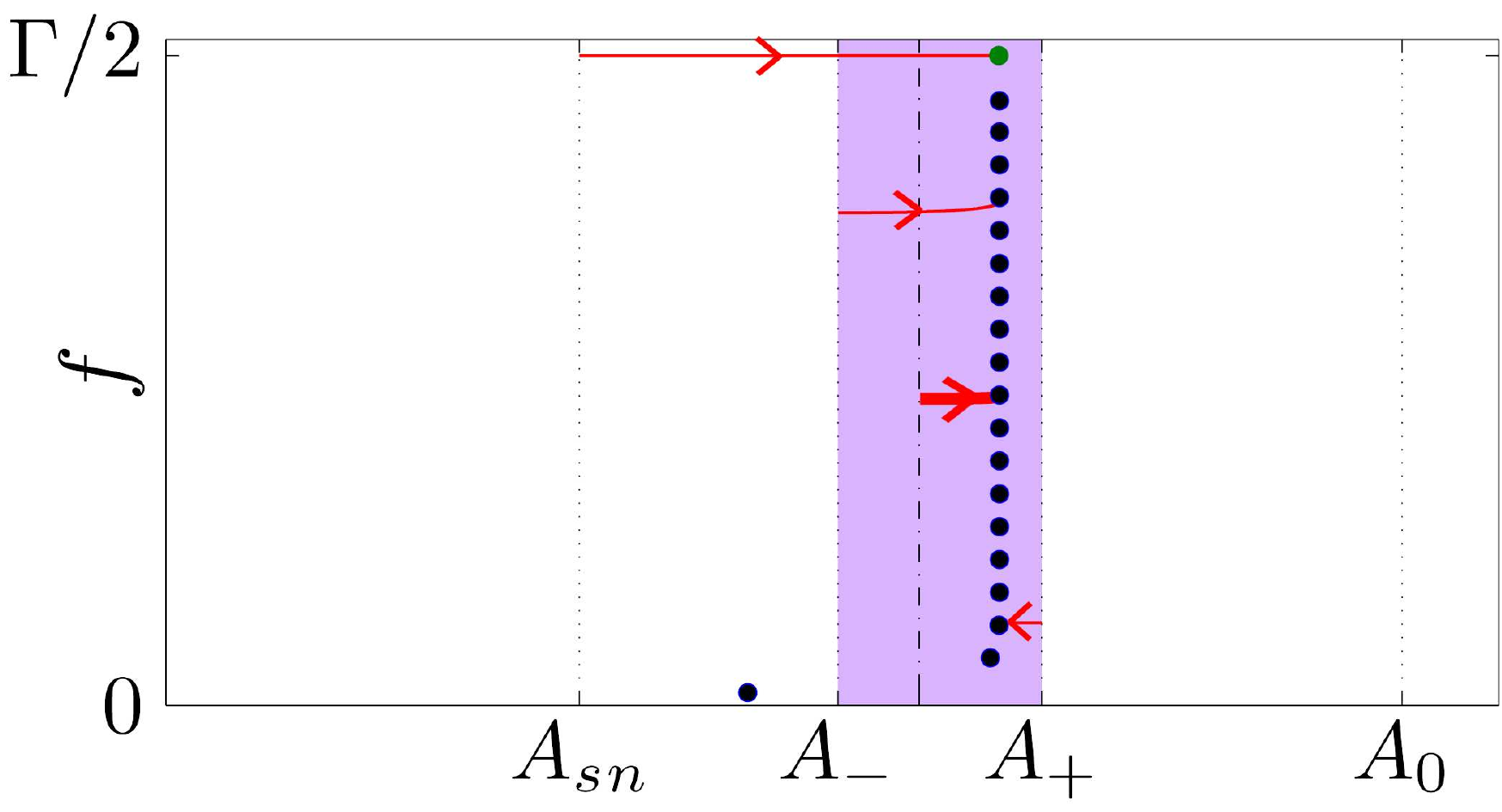}
		}
\\
\subfloat[$\mathcal{D}_+$, $r=-0.20$]{\includegraphics[width=60mm]{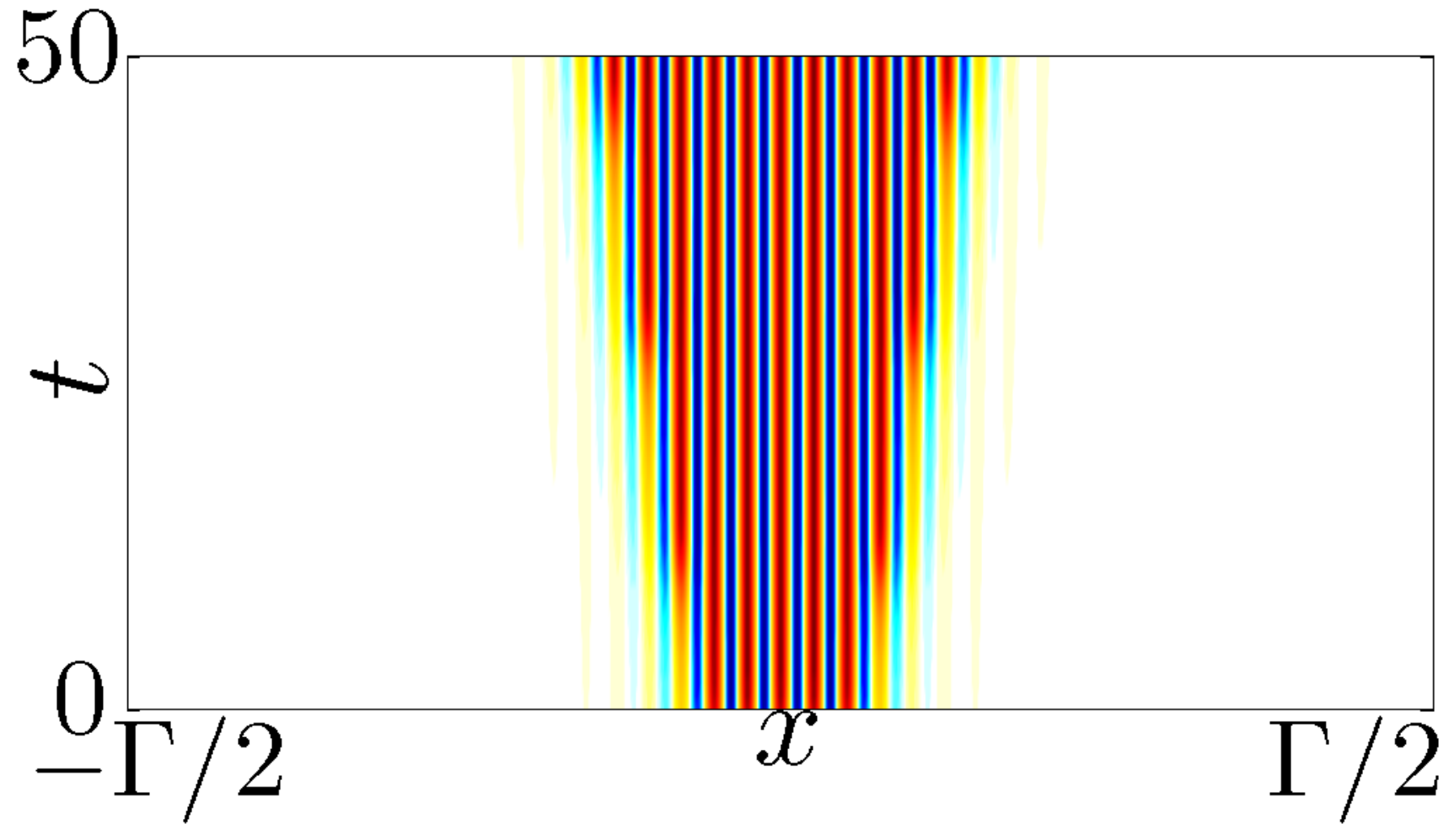}
						\includegraphics[width=60mm]{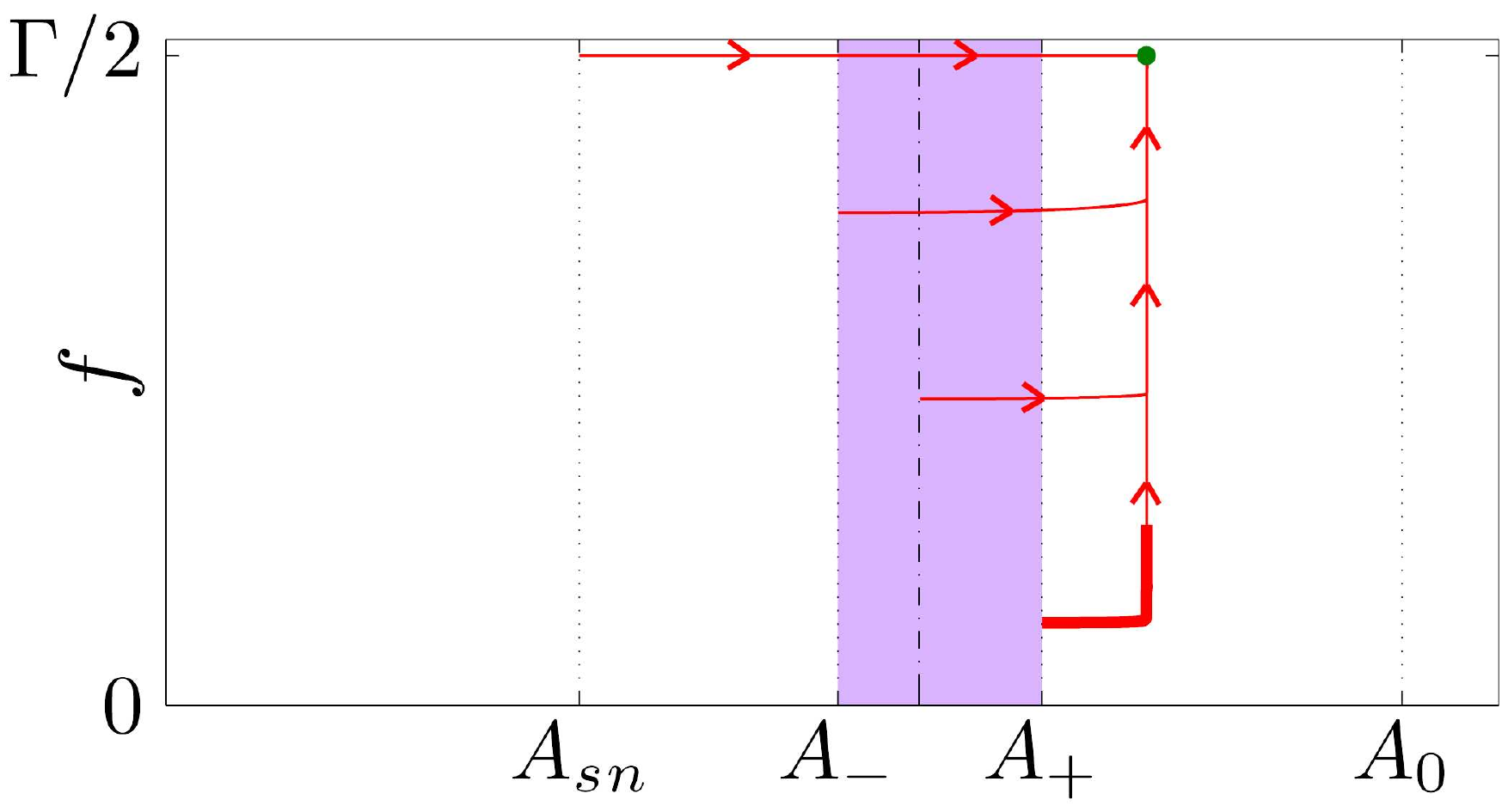}
		}
\caption{Space-time plots (left panels) and sample phase space trajectories (right panels) illustrating the dynamics of localized solutions of $L_{0}$ type in the different parameter regimes in \cref{fig:shesnaking}, initialized with different values of $r$.  Green dots indicate stable periodic states for the given forcing while blue dots indicate stable localized states. The purple region shows the pinning region.  }
\label{fig:portraits}
\end{figure}

\begin{itemize}

\item Regime $\mathcal{A}_-$: $r<r_{sn}$. Only the trivial state is stable. The dynamics is dominated by an overall amplitude (or body) mode and the amplitude of any localized initial condition decays homogeneously to zero.
\item Regime $\mathcal{D}_-$: $r_{sn} \le r \le r_-$. Two stable states are present: $u_h$ and $u_p$, with $\mathcal{F}[u_h]<\mathcal{F}[u_p]$. Spatially localized initial conditions evolve via a depinning (or edge) mode responsible for the progressive loss of spatial periods while keeping their amplitude constant. The solution collapses to the trivial state only when its extent becomes comparable to one wavelength.
\item Regime $\mathcal{P}_{\pm}$: $r_- \le r \le r_+$. There is a large number of coexisting stable and unstable states: trivial, spatially periodic and spatially localized with different numbers of periods. The long-time behavior of the system is determined by the basins of attraction of the stable states and hence by the initial conditions provided.
\item Regime $\mathcal{D}_+$: $r_+ \le r \le 0$. The situation is similar to that in $\mathcal{D}_-$ but this time $\mathcal{F}[u_p]<\mathcal{F}[u_h]$. Spatially localized initial conditions nucleate additional wavelengths under the influence of the depinning mode, and in periodic domains evolve into the spatially periodic state.
\item Regime $\mathcal{A}_+$: $r>0$. The only stable state is the spatially periodic state.
\end{itemize}
These regimes are depicted in the phase portraits in \cref{fig:portraits}. The computations use $L_{0}$ localized solutions from the snaking region, hereafter $u_0(x)$, as initial conditions. These evolve first in $A$ to the appropriate amplitude, followed by depinning if $r$ falls outside the pinning region. 

If $r=r_{\pm}+\delta$, where $|\delta| \ll 1$, the resulting front propagates at an overall constant speed determined by the nucleation time $T^{\text{dpn}}\propto |\delta|^{-1/2}$ computed in reference~\cite{burke2006}.  In this calculation, the solution takes the form $u(x,t)=u_0(x)+\sqrt{|\delta|} a(t) v_{\pm}(x)+\mathcal{O}(|\delta|)$, where $a$ is the time-dependent amplitude associated with the eigenmode $v_{\pm}$ that is responsible for triggering a nucleation ($+$) or annihilation ($-$) event.  The equation that governs the dynamics of $a$ is
\begin{equation}
\alpha_1\dot{a}= \sqrt{|\delta|}(\alpha_2\,\mathrm{sgn}(\delta)+\alpha_3 a^2),
\label{eq:depinningacf}
\end{equation}
where $\mathrm{sgn}(\delta)$ is the sign of $\delta$ and the coefficients $\alpha_j$ for each of the two cases ($r_{\pm}$) are computed numerically from the following integrals:
\begin{equation}
\alpha_1=\int_0^{\Gamma/2} v_{\pm}(x)^2 \; dx, \quad
\alpha_2=\int_0^{\Gamma/2} v_{\pm}(x) u_0(x)\; dx,\quad
\alpha_3=\int_0^{\Gamma/2} v_{\pm}(x)^3(b- 3u_0(x))\; dx.\label{eq:alpha}
\end{equation}

\begin{figure}
\centering
\mbox{
\subfloat[$\delta<0$]{\begin{tabular}[c]{c}
		\includegraphics[width=40mm]{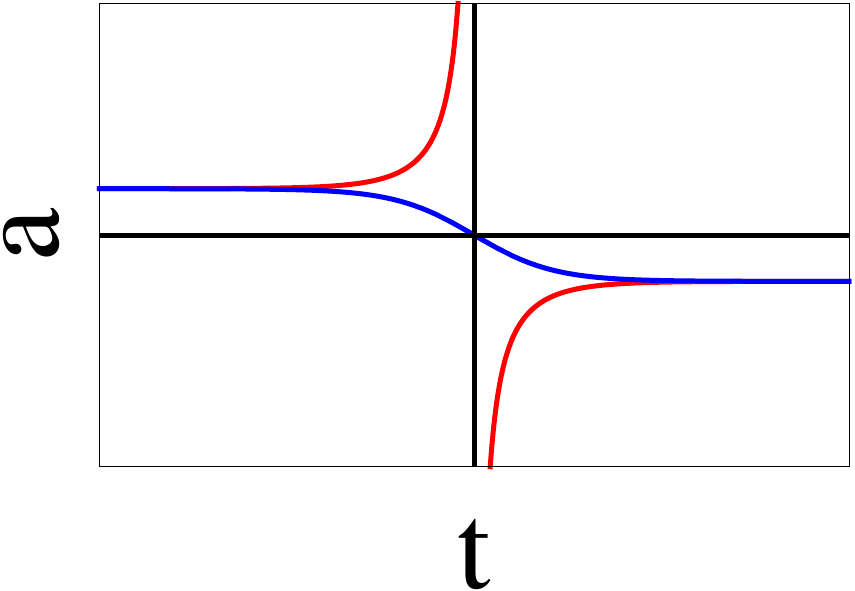}\\
		\includegraphics[width=40mm]{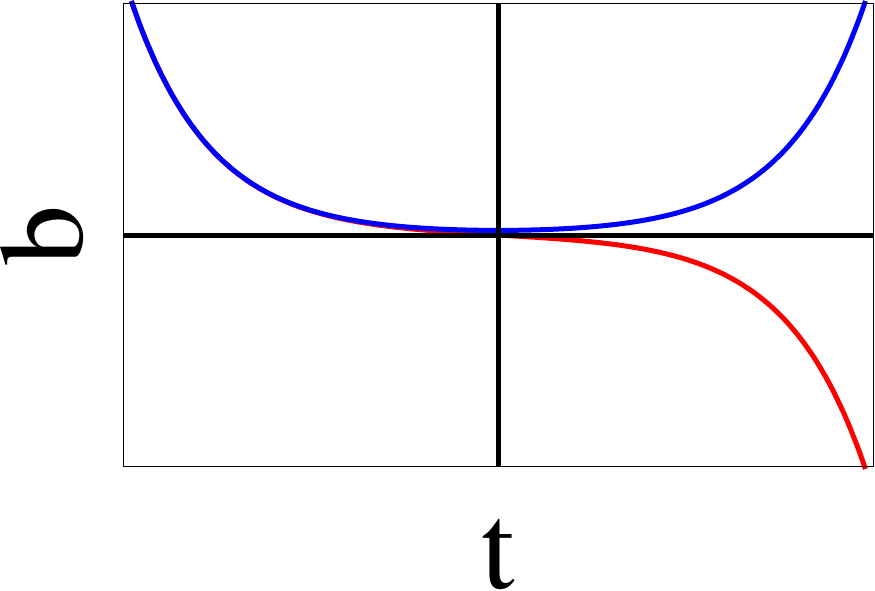}
		\end{tabular}}\quad
\subfloat[Stability diagram]{\begin{tabular}[c]{c}
		\includegraphics[width=55mm]{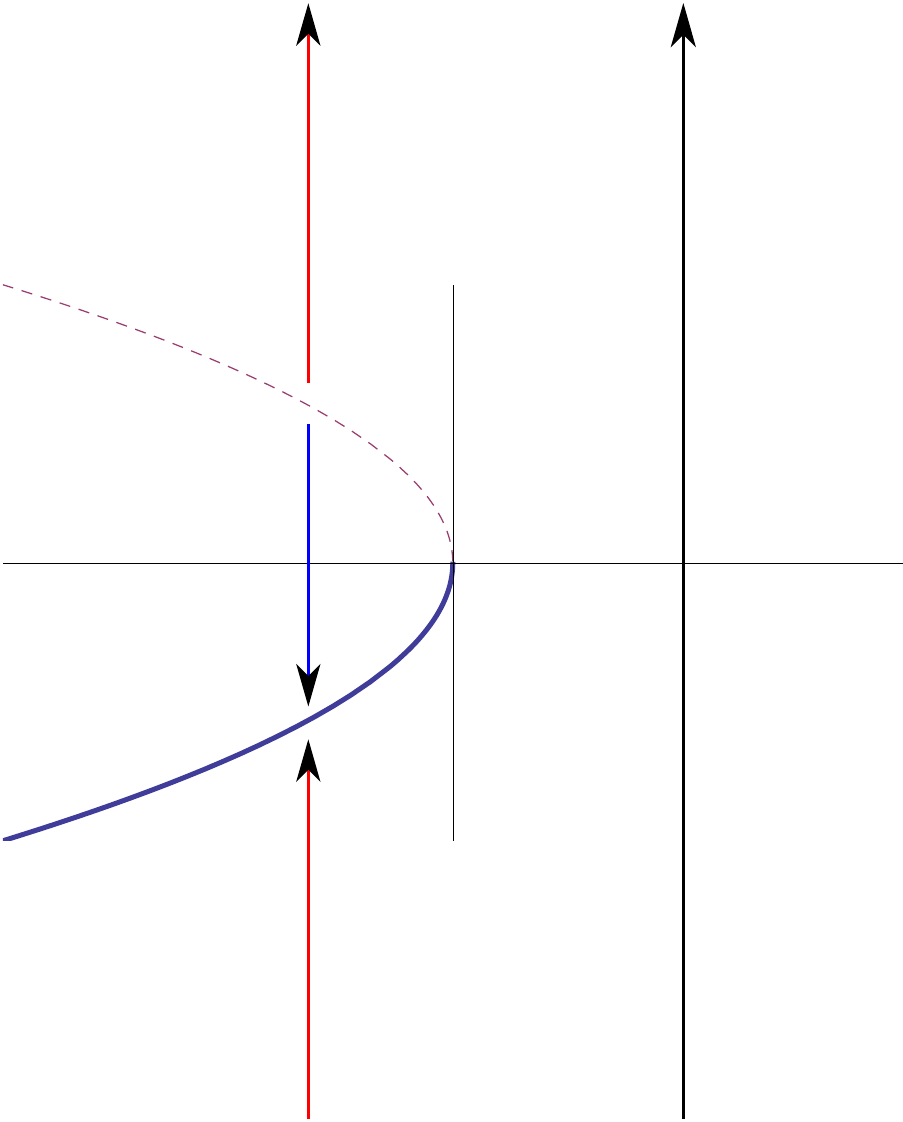}
		\end{tabular}}\quad
\subfloat[$\delta>0$]{\begin{tabular}[c]{c}
		\includegraphics[width=40mm]{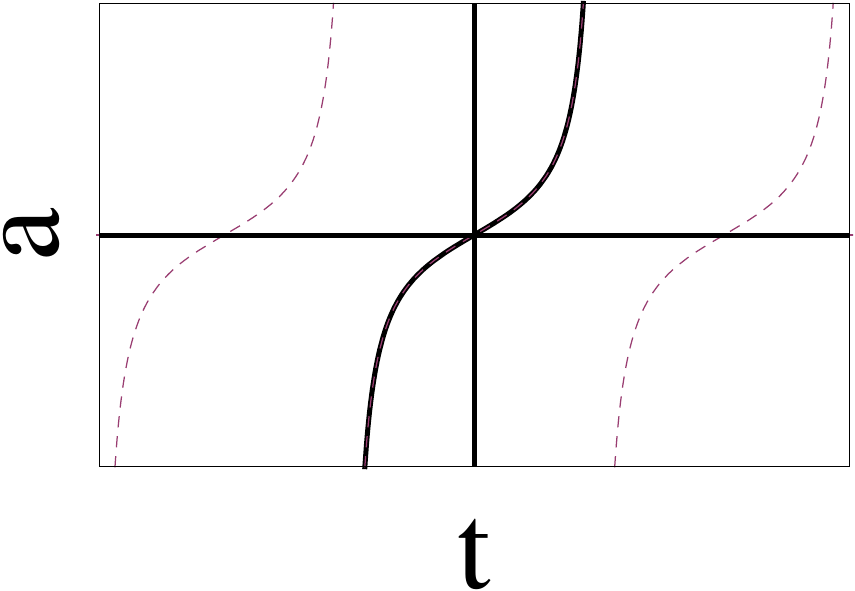}\\
		\includegraphics[width=40mm]{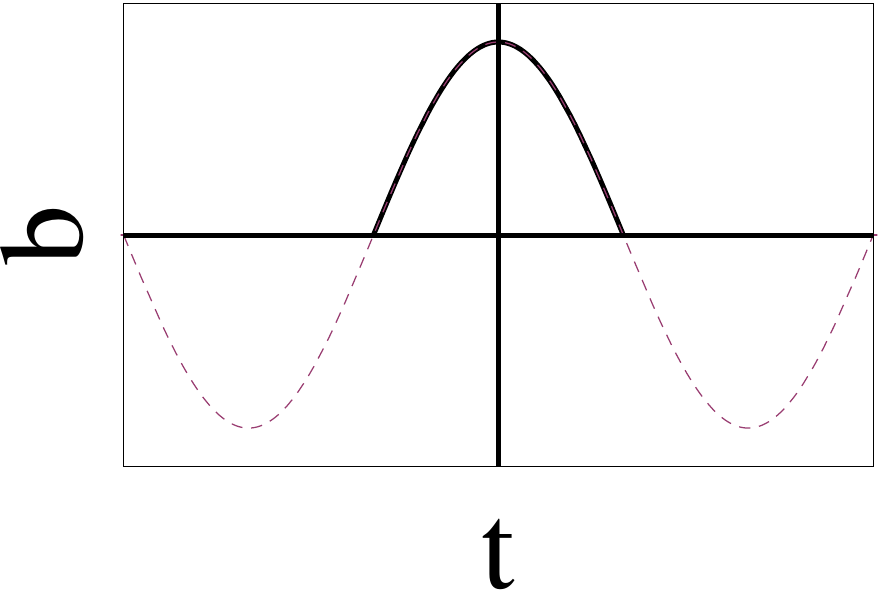}
		\end{tabular}}
}
\caption{Amplitude of the depinning mode in terms of $a(t)$ (upper panels) and $b(t)$ (lower panels) near $r_+$ for (a) $\delta<0$ and (c) $\delta>0$; (b) shows the corresponding bifurcation diagram.  Figure (c) shows three successive nucleation events corresponding to times where $a(t)\to\infty$ (top panel) or $b(t)=0$ (lower panel).
}
\label{fig:asymptoticcf}
\end{figure}

We now discuss the solutions describing a nucleation event near $r_+$, where $\alpha_3>0$; analogous arguments apply in the vicinity of $r_-$, where $\alpha_3<0$.  Within the pinning region, $\delta<0$, a pair of stable and unstable steady state solutions $u_0$ is present, corresponding to the vicinity of a fold on the right of the snaking branch $L_0$ (\cref{fig:asymptoticcf}(b)) and all initial conditions approach the stable state or diverge.  Outside of the pinning region, $\delta>0$, there are no stable solutions and the amplitude $a\to \infty$ for all initial conditions.  The upper panels of \cref{fig:asymptoticcf}(a,c) show typical trajectories $a(t)$ corresponding to the dynamics represented by the arrows in \cref{fig:asymptoticcf}(b), with the right panel showing three successive nucleation events.  We approximate the time between depinning events, i.e. the nucleation time $T^{\text{dpn}}_+$, as the time interval between successive asymptotes where $a$ diverges.

Near the fold of the snaking branch $L_0$ and to its right ($0<\delta\ll 1$) the system undergoes dynamics on the time scale $\delta^{-1/2}$ (\cref{eq:depinningacf}) and thus $T^{\text{dpn}}_+=\mathcal{O}(\delta^{-1/2})$. Upon leaving the vicinity of the fold (i.e. when $a\to \infty$), the system transitions towards the next fold on the snaking branch (\cref{fig:asymptoticcf}) before slowing down again. This transition corresponds to a nucleation event that adds a wavelength to each side of $u_0(x)$ and the process repeats at successive folds. Since the structure of $v_+(x)$ is almost independent of the length $2f$ of the localized state (it is an edge mode) the resulting process is periodic, a fact that can be highlighted by introducing the Riccati variable $b$ defined by $a=-\alpha_1 \dot{b}/\alpha_3 \sqrt{|\delta|} b$. In terms of $b$, \cref{eq:depinningacf} becomes the oscillator equation $\ddot{b}+\delta\Omega_{+}^2 b=0$, where $\Omega_{+}^2\equiv\alpha_2\alpha_3/\alpha_1^2>0$. The lower panels of \cref{fig:asymptoticcf}(a,c) show the oscillator amplitude $b(t)$ corresponding to the $a(t)$ solutions shown just above.   However, care must be taken in the interpretation of this equation, because $b\to 0$ implies $a\to\infty$ while \cref{eq:depinningacf} breaks down already when $a=\mathcal{O}(1)$. Despite this caveat we shall find the use of the variable $b$ useful since it highlights the possibility of a temporal resonance when the system (\ref{eq:SH}) is forced with a time-periodic forcing. A similar discussion applies to annihilation events near $r=r_-$.

\begin{figure}
 \centering

      \subfloat[]{\includegraphics[width=75mm]{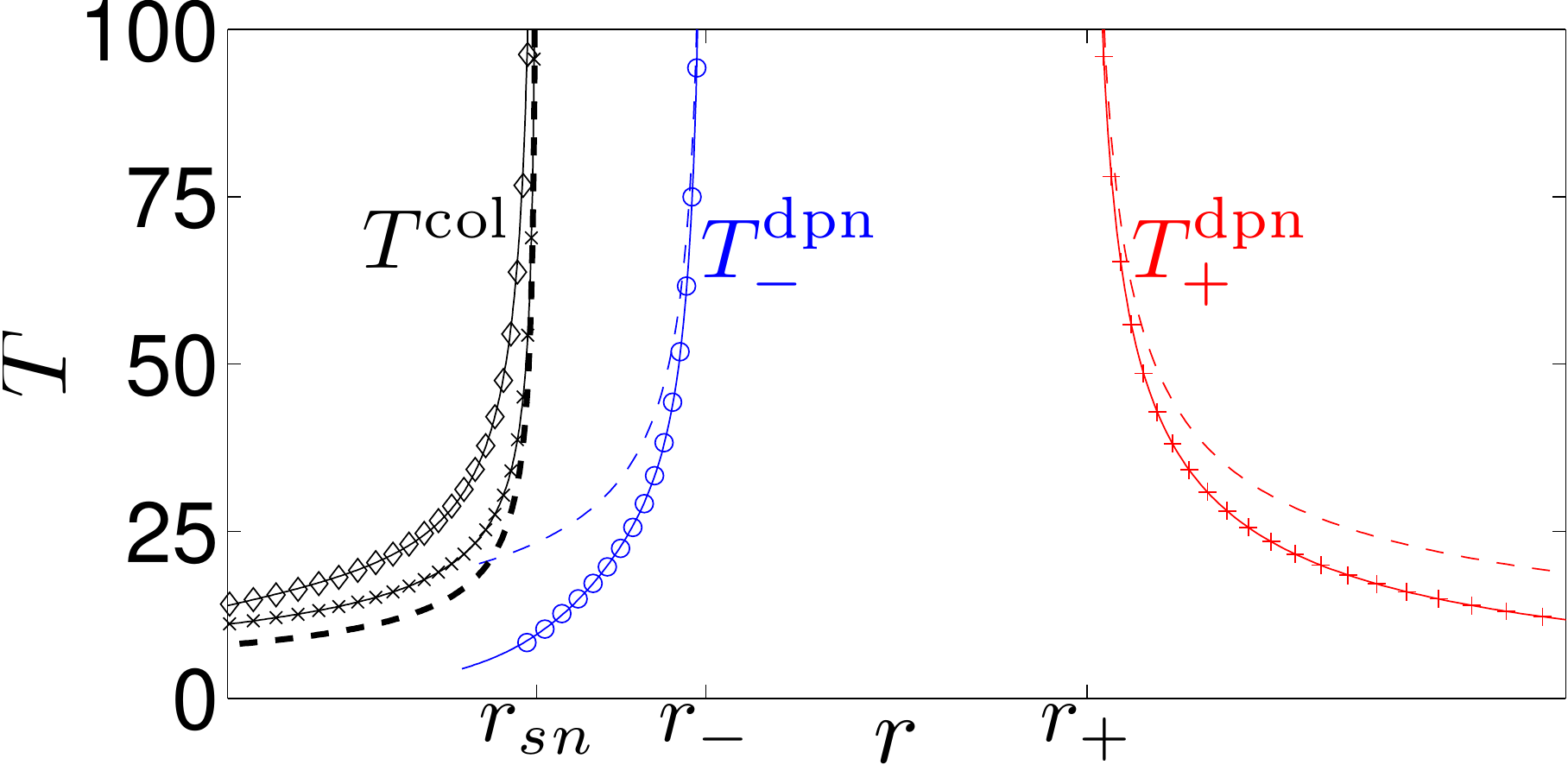}} \\
      \subfloat[]{\includegraphics[width=75mm]{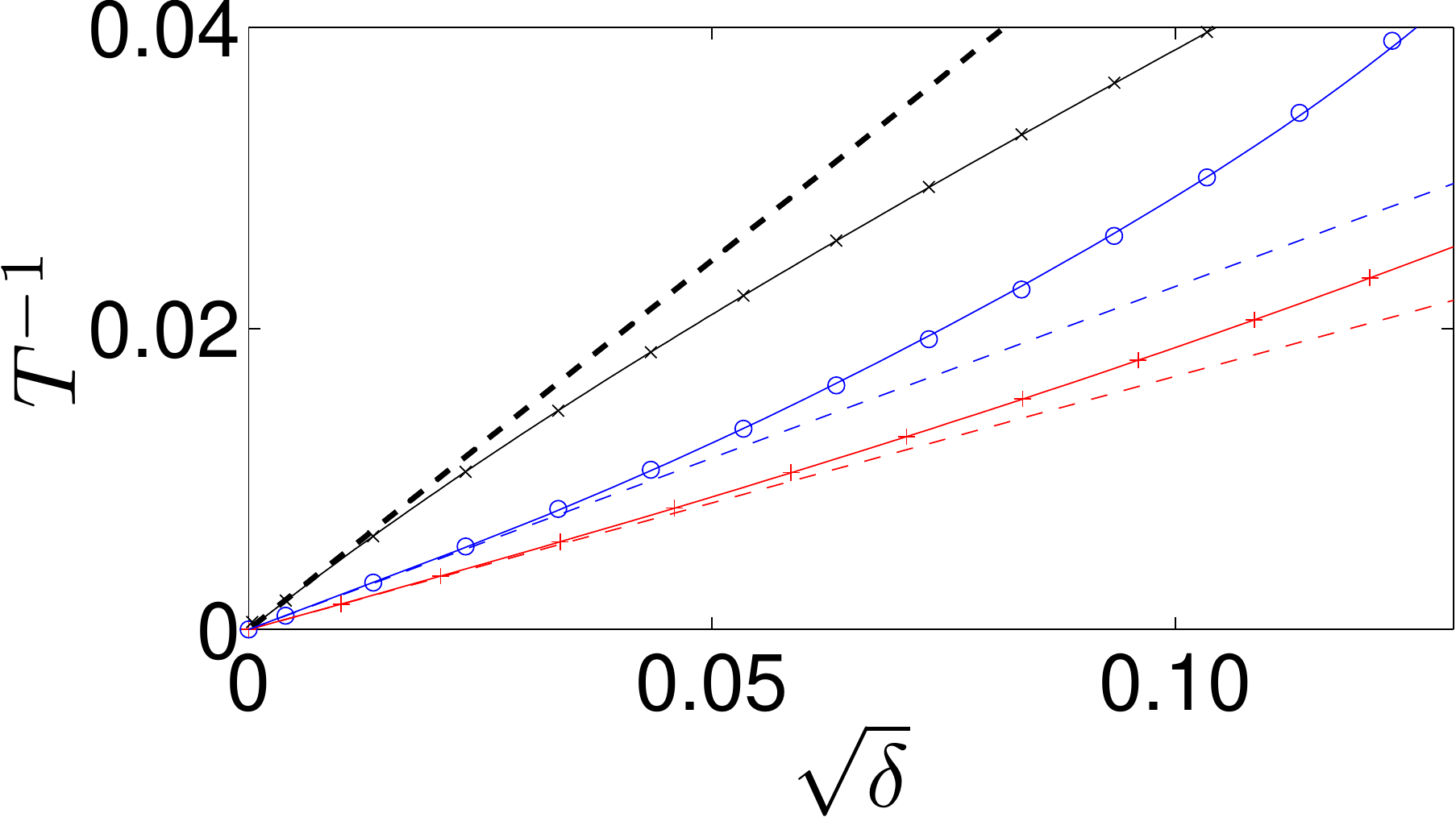} }
      \\
      \subfloat[]{\includegraphics[width=75mm]{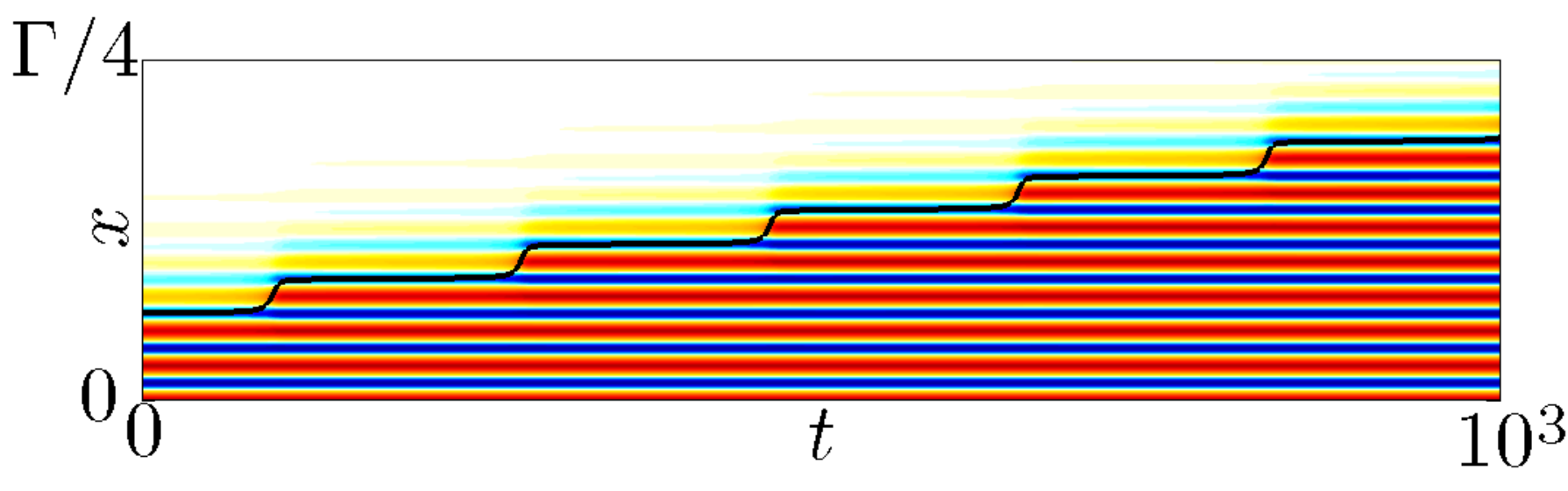}} 
    \caption{(a) The time between nucleation events ($T_{+}^{\mathrm{dpn}}$, red crosses), annihilation events ($T_{-}^{\mathrm{dpn}}$, blue circles) and the time to collapse to the trivial state ($T^{\mathrm{col}}_{\mathrm{per}}$, black crosses) as functions of the parameter $r$, starting from marginally stable $L_{0}$ solutions at $r=r_+$, $r=r_-$ and the periodic state at $r=r_{sn}$, respectively. The symbols show results from direct numerical simulations, the solid lines are fits to this data, and the dashed lines are predictions from the leading order theory in reference~\cite{burke2006}.  The corresponding results for the collapse time for a localized state at $r_-$ are also shown ($T^{\mathrm{col}}_{\mathrm{loc}}$, black diamonds).  (b) Comparison between numerical data (circles/crosses) and the leading order theory (dashed lines), showing $(T_{\pm}^{\mathrm{dpn}})^{-1}$ as a function of the square root of the distance $\delta$ from the pinning region. The corresponding fifth order fits are shown using solid lines. (c) A space-time representation of a simulation at $r\approx-0.2583$ ($\delta\approx 0.001$) initialized using a marginally stable localized solution at $r_+$, with red representing high values and blue low values of $u(x)$. The solid black line shows the instantaneous front position $f(t)$.}
\label{fig:depinningtimecf}
\end{figure}

\Cref{fig:depinningtimecf} compares the leading order theoretical prediction $T^\mathrm{dpn}_{+}=\pi/\sqrt{\delta}\Omega_+$ obtained from \cref{eq:depinningacf} (dashed lines) with numerical simulations of \cref{eq:SH}. The theory works well for $\delta\ll1$ but can be improved by including higher order terms, $(T^{\mathrm{dpn}})^{-1}=\sum \sigma_{n}\delta^{n/2}$, $n \ge 1$, and computing the $\sigma_{n}$ using a least squares fit. Given the results in \cref{fig:depinningtimecf} we determined that a fifth order calculation was sufficiently accurate even when $r\sim r_{sn}$. 
The figure shows the nucleation time in $\mathcal{D}_+$ (red) and annihilation time in $\mathcal{D}_-$ (blue).  The symbols represent results from simulations, the dashed lines represent the prediction from the leading order asymptotic theory, while the solid lines represent the fifth order numerical fit. The times $T^\mathrm{col}$ for a marginally stable periodic state at $r_{sn}$ (black crosses) and a localized state at $r_-$ (black diamonds) to reach the trivial state by amplitude decay in $\mathcal{A}_-$ are shown in black.  The black dashed line represents the leading order asymptotic theory applied to the periodic state near $r_{sn}$. The coefficients $\sigma_{n}$ for both the asymptotic theory and the fifth order numerical fit for our choice of parameters are summarized in \cref{tab:fitconstants}.  We will find that the numerical fit is required for quantitative agreement between the theory presented in section V and numerical simulations presented in section IV. However, the theory cannot be applied to localized states in $\mathcal{A}_-$ since these states undergo both amplitude decay {\it and} depinning.

\begin{table}[h]\footnotesize\centering
\begin{tabular}{c||c|cccccc}
    					 		& $\Omega$ & $\sigma_1$ & $\sigma_2$ & $\sigma_3$ & $\sigma_4$ & $\sigma_5$ \\ \hline\hline
$T^{\mathrm{dpn}}_{+}$   		&0.5285    &  0.1687    & 0.1141     & 0.7709     &  -0.4000   & 0.0803\\
$T^{\mathrm{dpn}}_{-}$ 	 		&0.7519    &  0.2381    & -0.8445    & 33.37      &  -306.4    & 1067\\
$T^{\mathrm{col}}_{\mathrm{per}}$ & 0.7705 &  0.4829    & -1.738     & 10.62      &  -35.00    & 48.31\\       
$T^{\mathrm{col}}_{\mathrm{loc}}$ &  - 	   &  0.2081    & 0.4431     & 2.962      &  -34.15    & 79.52
\end{tabular}
\caption{Values of the coefficients $\sigma_j$ determined from a least squares fit of the depinning/collapse time to simulations with constant forcing of the form $T^{-1}=\sum_{n=1}^{5}\sigma_n |r-r_{\pm,sn}|^{n/2}$. The frequency $\Omega$ is calculated numerically from the integrals in \cref{eq:alpha} in each case. 
}
\label{tab:fitconstants}
\end{table}

\section{The high frequency limit}

We begin our study of the effects of a time-periodic forcing on localized states by considering the limit of fast oscillations.  We first consider the case when the frequency of the forcing cycle is so fast that the motion of the fronts does not permit nucleation/annihilation of additional/existing periods.  We then increase the amplitude of the forcing cycle so that the structure remains unpinned for an appreciable amount of time.

\subsection{The averaged system} 

The qualitative behavior of \cref{eq:SH} is unchanged when the forcing frequency is high enough that insufficient time is spent outside of the pinning region for depinning to occur.  The effect of the periodic forcing in this case is small, producing rapid amplitude fluctuations of the existing localized states. We introduce the effective Maxwell point $\bar{r}_M$ using the relation
\begin{equation}
\langle \mathcal{F}[u]\rangle=0,
\label{eq:rMavg}
\end{equation}
where the brackets indicate an average over the forcing cycle.
We assume that the periodic forcing occurs at a high frequency, $\omega\rightarrow\omega/\epsilon$, where $\epsilon\ll1$ and define a fast timescale $\phi = \omega t/\epsilon$. We seek solutions of \cref{eq:SH} in the form $u(x,t)=u_0(x,t,\phi)+\epsilon u_1(x,t,\phi) + \epsilon^2 u_2(x,t,\phi) + \dots$, satisfying
\begin{equation}
\omega u_{\phi}=\epsilon\left[ (r_0+ \rho \sin\phi) u-\left(1+\partial_{x}^2\right)^2u+b u^2-u^3-u_{t}\right],
\end{equation}
where $t$ is the original timescale on which the averaged dynamics take place, and assume that $\rho,r_0,b,\omega=\mathcal{O}(1)$.
The leading order equation $(u_0)_{\phi}=0$ gives $u_0(x,t,\phi)=A_0(x,t)$. 
At order $\mathcal{O}(\epsilon)$, we obtain
\begin{equation}
\omega \partial_{\phi} u_1= (r_0+ \rho \sin\phi) u_0-\left(1+\partial_{x}^2\right)^2 u_0+ b u_0^2-u_0^3 -\partial_t u_0. \label{eq:SHfastoscu1}
\end{equation}
The solvability condition requires that the integral over a single period of the fast oscillation of the right side of \cref{eq:SHfastoscu1} vanishes. This condition yields the governing equation for $A_0$:
\begin{equation}
\partial_t A_0= r_0 A_0-\left(1+\partial_{x}^2\right)^2A_0+b A_0^2-A_0^3\label{eq:SHfastoscA0}.
\end{equation}
Thus, in the limit of a high frequency forcing cycle with order unity amplitude, the leading order behavior follows the time-independent SHE. Corrections arise at second order, as we now show. 

\Cref{eq:SHfastoscu1,eq:SHfastoscA0} show that the $\mathcal{O}(\epsilon)$ correction to the leading order behavior is given by
\begin{equation}
u_1(x,t,\phi)=  -\tfrac{\rho}{\omega} \cos\phi A_0(x,t) +A_1(x,t).
\end{equation}
At $\mathcal{O}(\epsilon^2)$ we obtain
\begin{equation}
\omega \partial_{\phi}u_2= (r_0+ \rho \sin\phi) u_1-\left(1+\partial_{x}^2\right)^2 u_1+ 2 b u_0 u_1- 3 u_0^2 u_1 -\partial_t u_1
\end{equation}
leading to the solvability condition
\begin{equation}
\partial_t A_1 = r_0 A_1-\left(1+\partial_{x}^2\right)^2A_1+2 b A_0A_1-3A_0^2A_1.
\end{equation}
Similarly, the second order correction to the solution takes the form
\begin{align}
u_2 =  \tfrac{\rho^2}{4\omega^2} \cos2\phi A_0(x,t)
  -\tfrac{\rho}{\omega^2} \sin\phi \left[b A_0(x,t)^2-2 A_0(x,t)^3\right]&\\ \nonumber
  -\tfrac{\rho}{\omega} \cos \phi A_1(x,t)  + A_2(x,t)&,
\end{align}
and we obtain, at $\mathcal{O}(\epsilon^3)$,
\begin{equation}
\omega \partial_{\phi}u_3= (r_0+ \rho \sin\phi) u_2-\left(1+\partial_{x}^2\right)^2 u_2+ 2 b u_0 u_2- 3 u_0^2 u_2 +b u_1^2-3u_0 u_1^2  -\partial_{t}u_2,
\end{equation}
yielding 
\begin{equation}
\partial_t A_2= r_0 A_2-\left(1+\partial_{x}^2\right)^2A_2+b (2 A_0A_2+A_1^2) - 3( A_0^2A_2+A_0A_1^2)
		-\tfrac{1}{2}\left(\tfrac{\rho}{\omega}\right)^2 A_0^3 .
\end{equation}
We can define an averaged variable with error at order $\mathcal{O}(\epsilon^3)$ that describes the dynamics on the long timescale:
\begin{equation}
A \equiv\frac{1}{2\pi}\int_0^{2\pi} \left( u_0+\epsilon u_1 +\epsilon^2 u_2\right)  \;d\phi 
=A_0+\epsilon A_1 +\epsilon^2 A_2.
\end{equation}
On summing the solvability conditions, we obtain the following equation for the dynamics of the averaged variable
\begin{equation}
\partial_t A= r_0 A-\left(1+\partial_{x}^2\right)^2A+b A^2-\left[1+\tfrac{1}{8\pi^2}\left(\rho T\right)^2\right]A^3 +\mathcal{O}(T^3),
\label{eq:AveragedHF}
\end{equation}
where, for clarity, we have introduced the period of the forcing cycle $T \equiv 2\pi \epsilon /\omega$. The result is a SHE with a modified cubic term.  

We find that the averaged Maxwell point of the system, defined by \cref{eq:rMavg}, is in fact the Maxwell point of the averaged system (\ref{eq:AveragedHF}).  This can be checked explicitly by noting that 
\begin{equation}
\bar{\mathcal{F}}[A]=\langle \mathcal{F}_0[u_0+\epsilon u_1+\epsilon^2 u_2 ]\rangle+\mathcal{O}(\epsilon^3),
\end{equation} 
where $\bar{\mathcal{F}}$ is the free energy of the averaged system with periodic forcing, $\mathcal{F}_0$ is the free energy of the system with a constant forcing $r_0$, and the average is over a forcing cycle. Furthermore, 
\begin{equation}
\bar{\mathcal{F}}[A] = \mathcal{F}_0[u_0]+\tfrac{\rho^2 T^2}{32\pi^2\Gamma} \int^{\Gamma/2}_{-\Gamma/2}u_0^4\;dx +\mathcal{O}(T^3),
\end{equation} 
implying that the free energy in the fluctuating system is greater than that of the system with constant forcing $r_0$. We can use this expression to calculate the frequency-induced shift of the Maxwell point explicitly by finding the value of $r$ where $\bar{\mathcal{F}}[A]=0$.  Because the periodic forcing has increased the energy of the spatially periodic state, the Maxwell point of the averaged system necessarily shifts to the right ($\bar{r}_M > r_{M}$) to compensate while the boundaries of the pinning region also shift to the right. Following \cite{burke2006} we obtain
\begin{equation}
\bar{r}_{\pm} = r_{\pm}+\tfrac{\rho^2 T^2}{8\pi^2} \frac{\int^{\Gamma/2}_{-\Gamma/2}u_0^3 v_{\pm}\;dx}{\int^{\Gamma/2}_{-\Gamma/2}u_0 v_{\pm}\; dx}.
\end{equation} 
Here $u_0$ is the marginally stable solution of the constant forcing system at $r_{\pm}$, and $v_{\pm}$ are the eigenmodes of the linearized problem at $r_{\pm}$ responsible for wavelength nucleation/annihilation. Both integrals are positive and we find that
\begin{equation}
2\bar{p}\equiv \bar{r}_+-\bar{r}_-\approx 2 p - 0.0385(\rho T)^2,
\end{equation}
where $p=(r_+-r_-)/2$ is the half-width of the pinning region in the constant forcing case. Thus the introduction of the periodic forcing {\it shrinks} the width of the pinning region and shifts it to the right.

\subsection{Large amplitude forcing}

We may repeat the above calculation in the case $\rho\rightarrow \rho/\epsilon$ so that we are now dealing with fast oscillations with a large amplitude.  This allows the system to spend enough time outside of the pinning region for depinning to take place.  However, in this limit, a large fraction of the forcing cycle is actually spent in the amplitude decay regimes $\mathcal{A}_{\pm}$, and thus the leading order dynamics will not be comprised simply of nucleation and annihilation events. This regime is described by the equation
\begin{equation}
\omega u_{\phi}-\rho \sin(\phi) u= \epsilon\left[r_0 u-\left(1+\partial_{x}^2\right)^2u+b u^2-u^3 -u_{t}\right],
\end{equation}
where $r_0,b,\rho=\mathcal{O}(1)$, and we look for solutions in the form $u=u_0+\epsilon u_1+ \epsilon^2 u_2+\dots$. At leading order we obtain $(\omega u_0)_{\phi}-\rho \sin(\phi) u_0=0$, with solution
\begin{equation}
u_0(x,t,\phi)=e^{-(\rho/\omega) \cos\phi} A_0(x,t).\label{eq:SHfastlargeoscu0}
\end{equation}
At $\mathcal{O}(\epsilon)$, the governing equation becomes
\begin{equation}
\omega \partial_{\phi} u_1-\rho \sin(\phi) u_1= r_0 u_0-\left(1+\partial_{x}^2\right)^2u_0+ b u_0^2-u_0^3 -\partial_{t} u_0
\label{eq:SHfastlargeoscu1}.
\end{equation}
Imposing the requirement that $u_1$ is periodic on the fast timescale leads to the solvability condition $\int \rm{RHS}\, e^{(\rho/\omega) \cos\phi}\;d\phi=0$, where RHS stands for the right-hand-side of \cref{eq:SHfastlargeoscu1}, and an evolution equation for $A_0$:
\begin{equation}
\partial_t A_0= r_0 A_0-\left(1+\partial_{x}^2\right)^2A_0+b I_0(\tfrac{\rho}{\omega}) A_0^2-I_0(\tfrac{2\rho}{\omega})A_0^3\label{eq:SHfastlargeA0}.
\end{equation}
Here $I_0(x)$ is the modified Bessel function of the first kind. Thus the slowly evolving amplitude $A_0$ of the leading order solution $u_0$ satisfies a SHE with modified coefficients. 
A higher order calculation shows that there is no additional correction at $\mathcal{O}(\epsilon)$ and the averaged dynamics follows SHE with the modified nonlinear coefficients of \cref{eq:SHfastlargeA0} up to  $\mathcal{O}(\epsilon^2)$. 

In contrast to the $\rho\sim\mathcal{O}(1)$ case, the cubic nonlinearity is now dramatically increased relative to the quadratic one. This results in a rapid decrease in the region of bistability as $\rho/\omega$ increases.  Indeed, at  $\rho/\omega\approx 7.02$ the region of bistability disappears in a codimension two point where the bifurcation that creates the periodic state transitions from subcritical to supercritical.  

\section{Intermediate Frequencies: Breathing localized structures}

We now move away from the high frequency limit and investigate parameter combinations that permit depinning. For this purpose we consider parameter excursions that allow the system to traverse $\mathcal{P}_{\pm}$ and spend a significant time in both $\mathcal{D}_+$ and $\mathcal{D}_-$, i.e., we take $r_-<r_0<r_+$, and $\rho> p$, where $p\equiv (r_+-r_-)/2$ is the half-width of the pinning region. The resulting structures oscillate in width and amplitude and we refer to them as ``breathing" localized structures.

\subsection{The fate of stable localized initial conditions}

\begin{figure}
\centering
\subfloat[$T=50$]{\includegraphics[width=60mm]{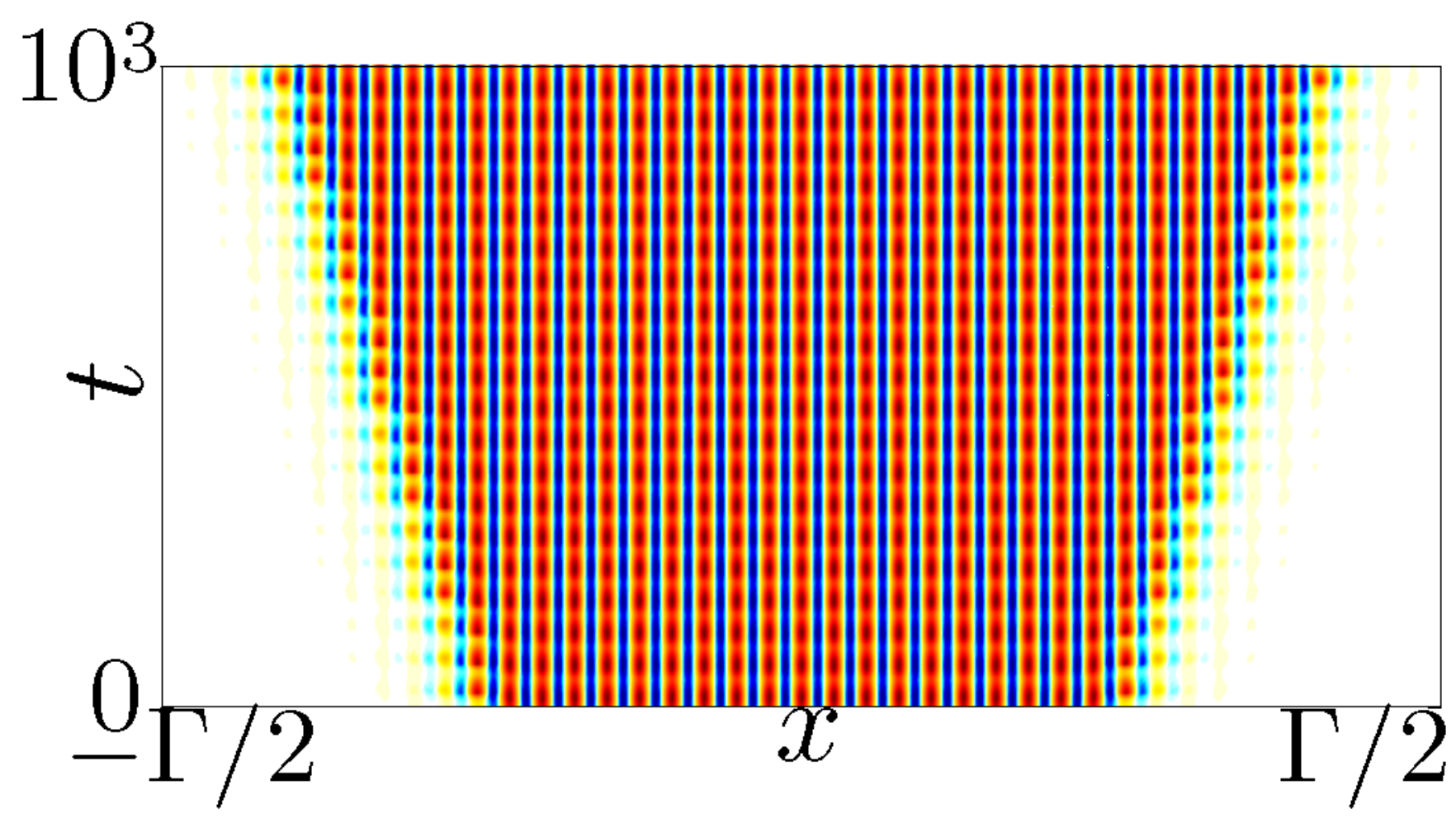}\quad
			\includegraphics[width=60mm]{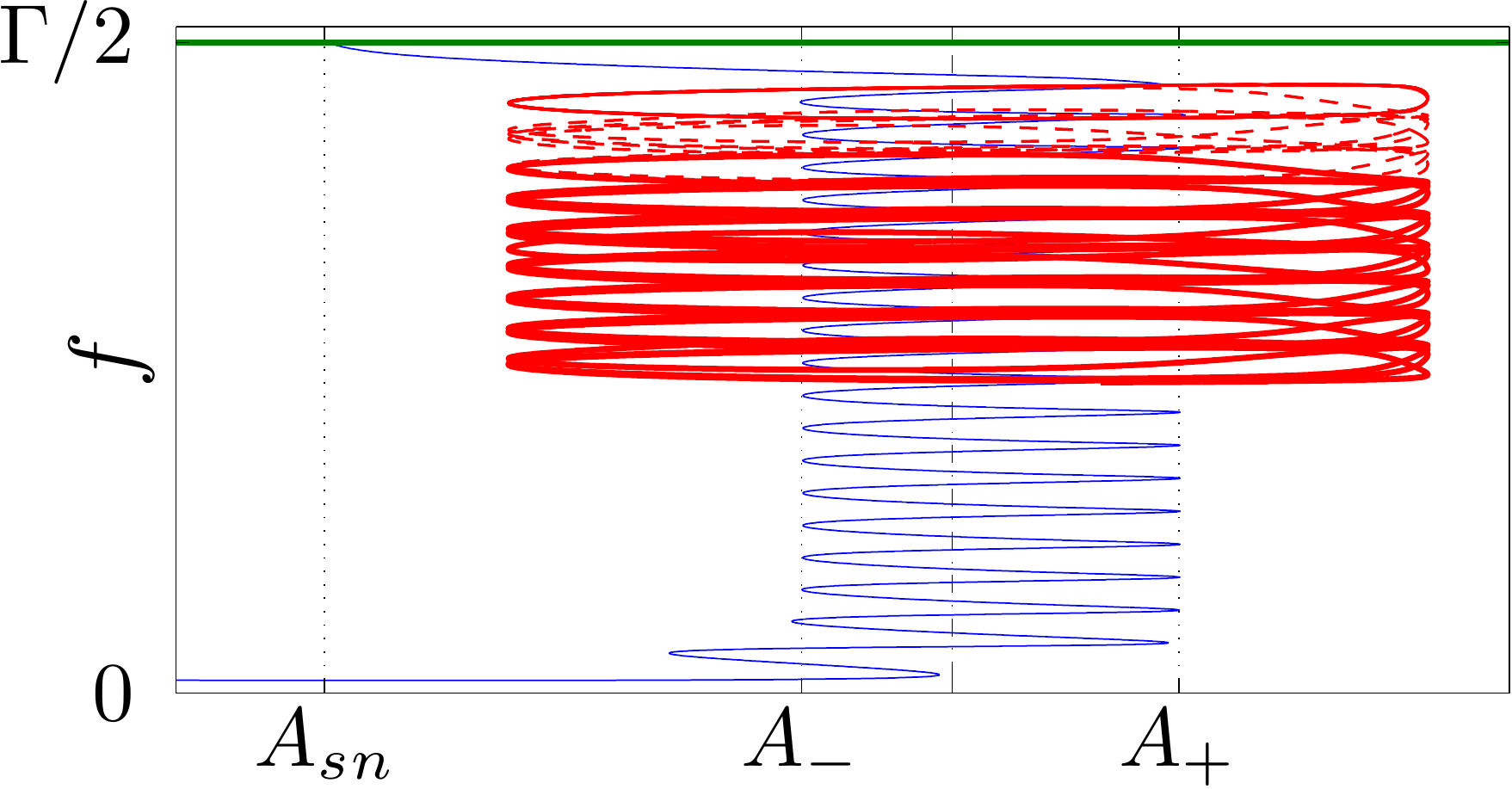}}
 \\
\subfloat[$T=150$]{\includegraphics[width=60mm]{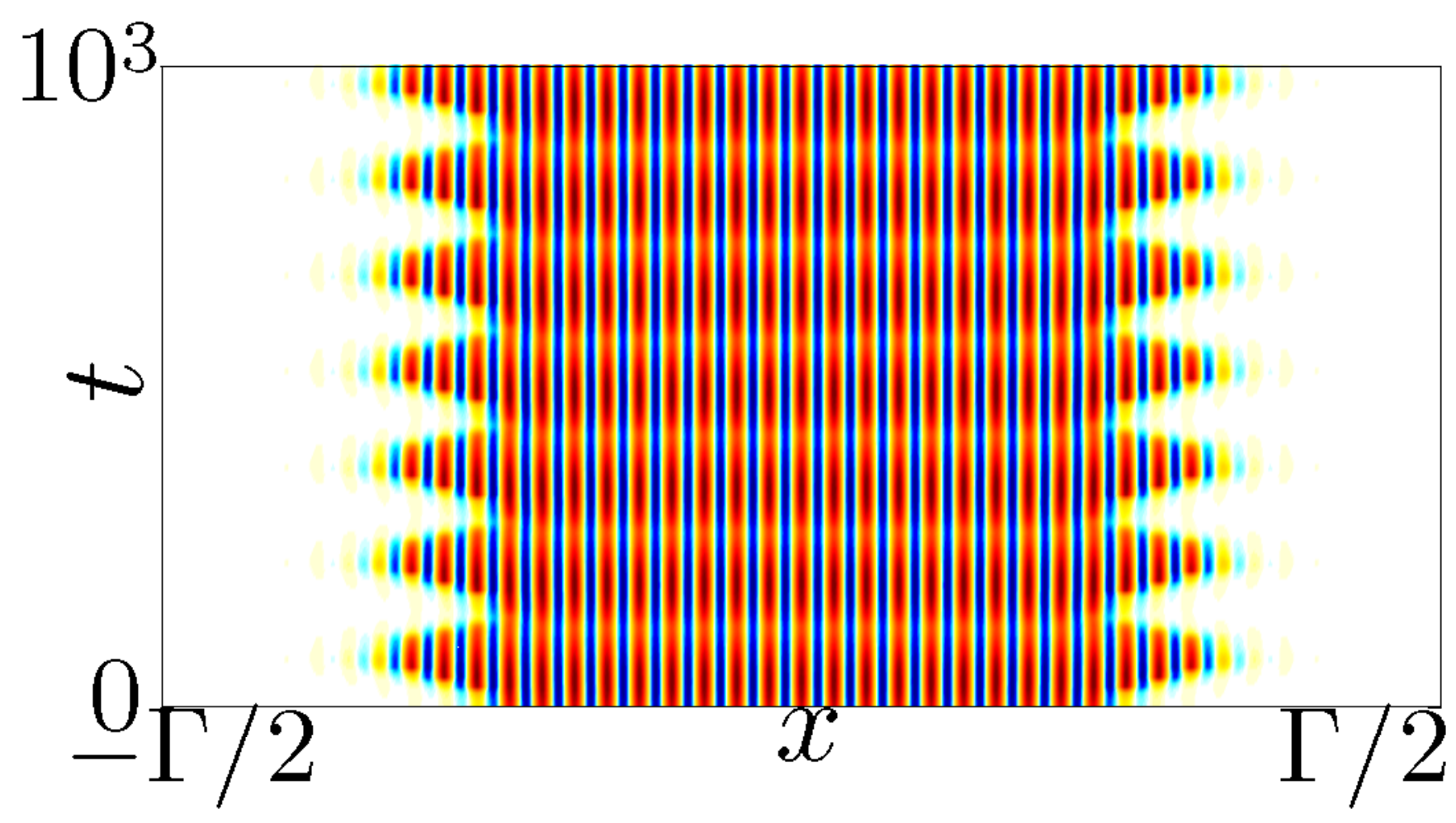}\quad
			\includegraphics[width=60mm]{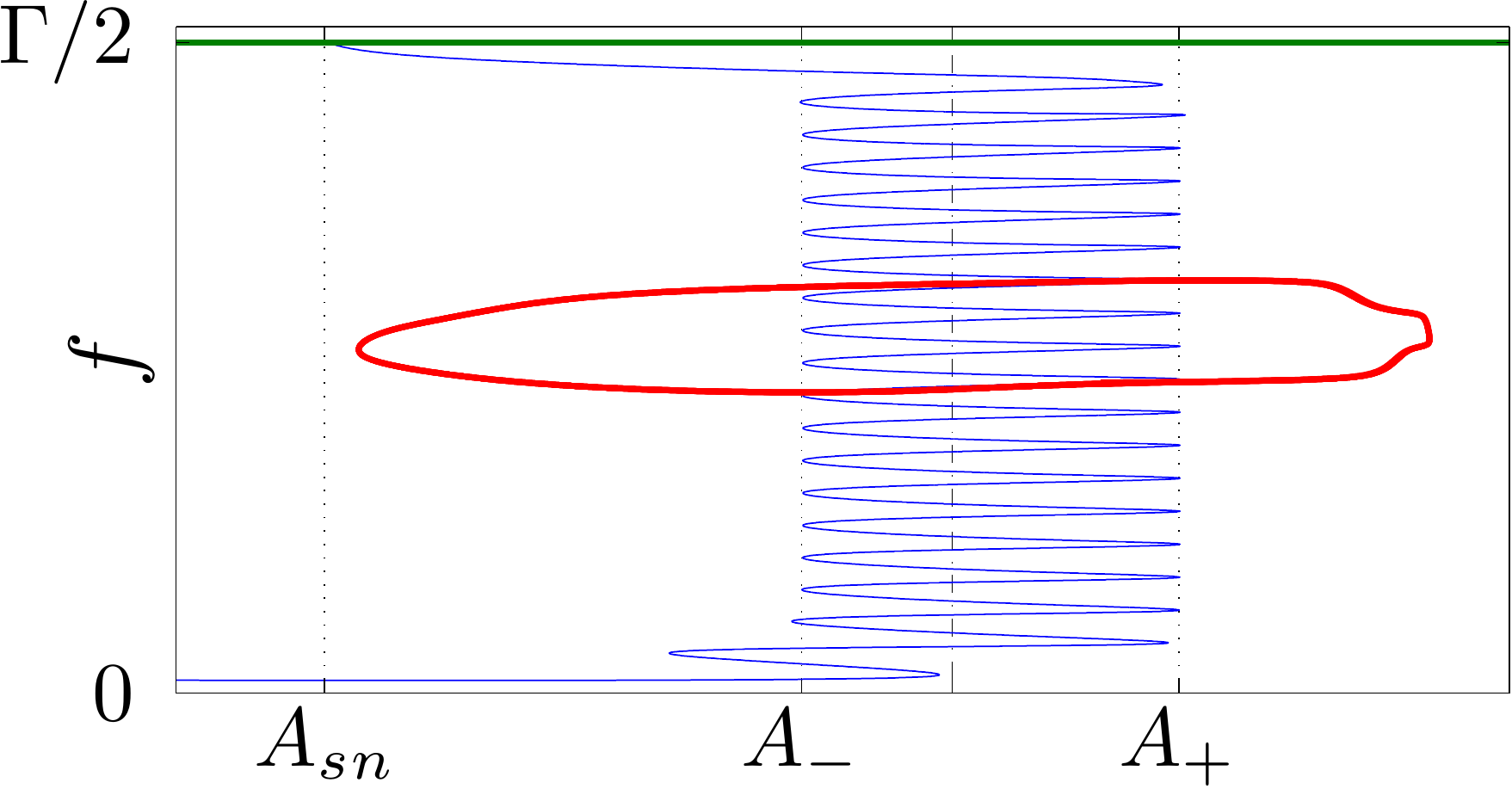}} 
\\
\subfloat[$T=250$]{\includegraphics[width=60mm]{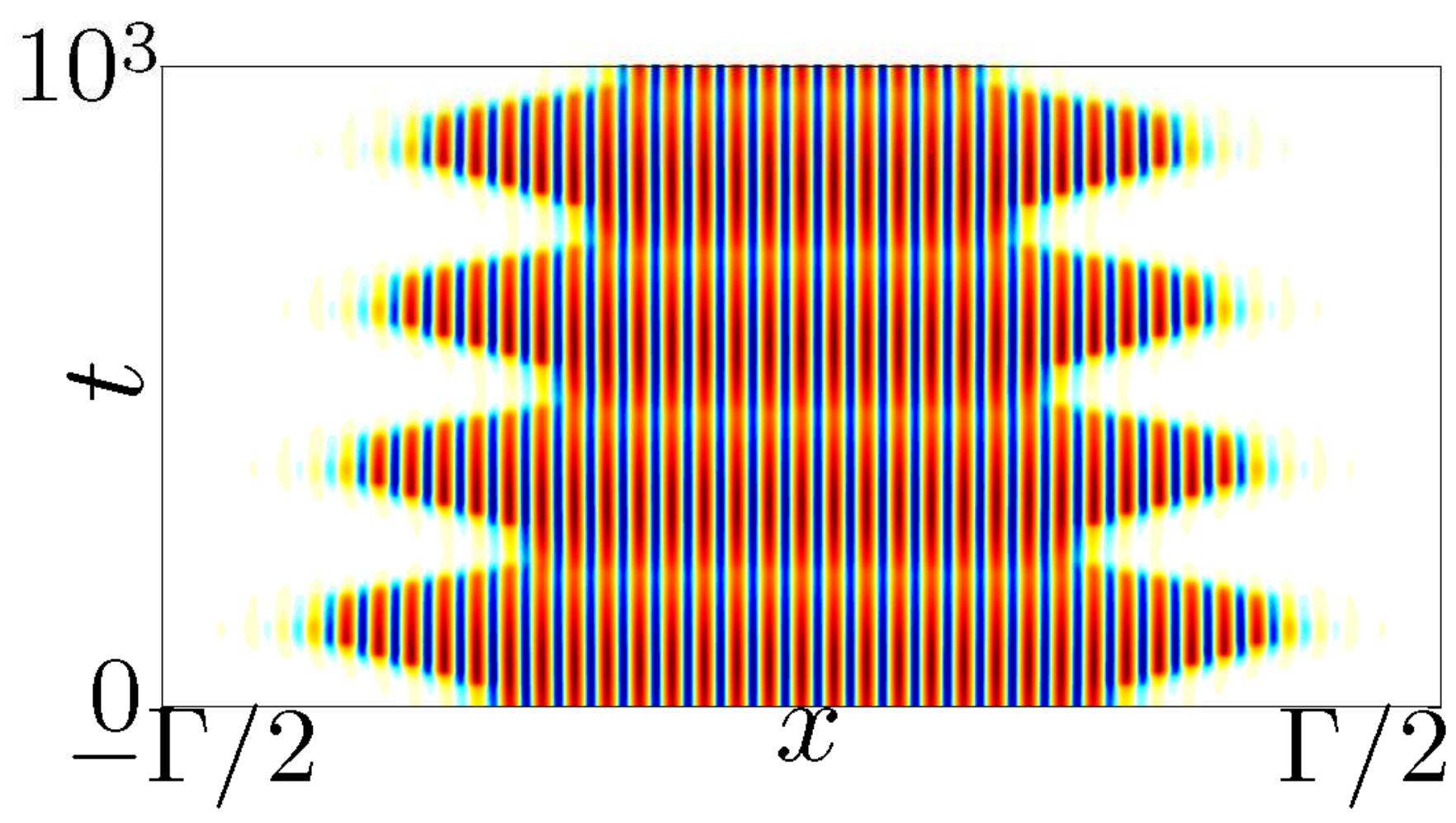}\quad
			\includegraphics[width=60mm]{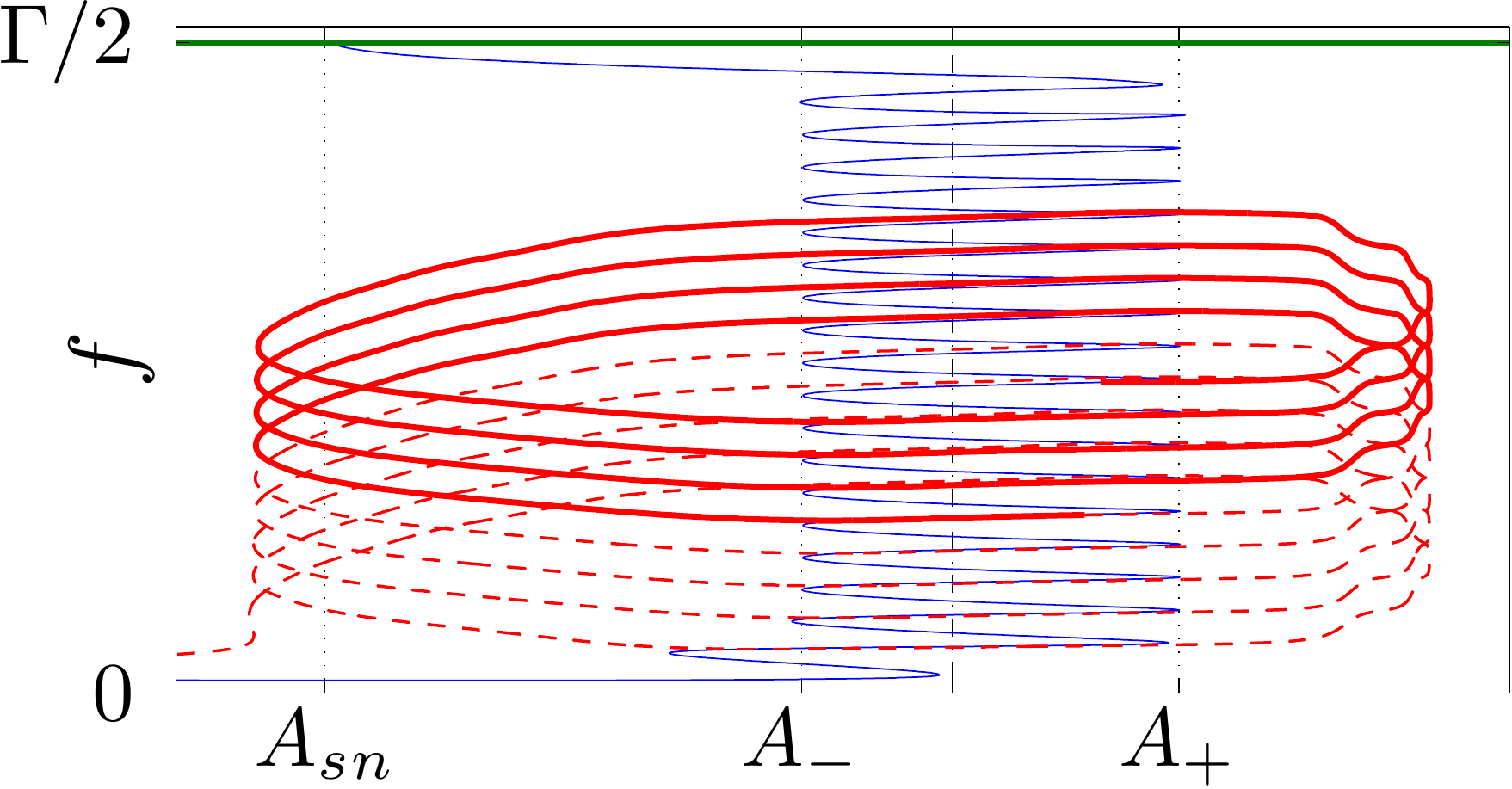}}
\\
\subfloat[$T=350$]{\includegraphics[width=60mm]{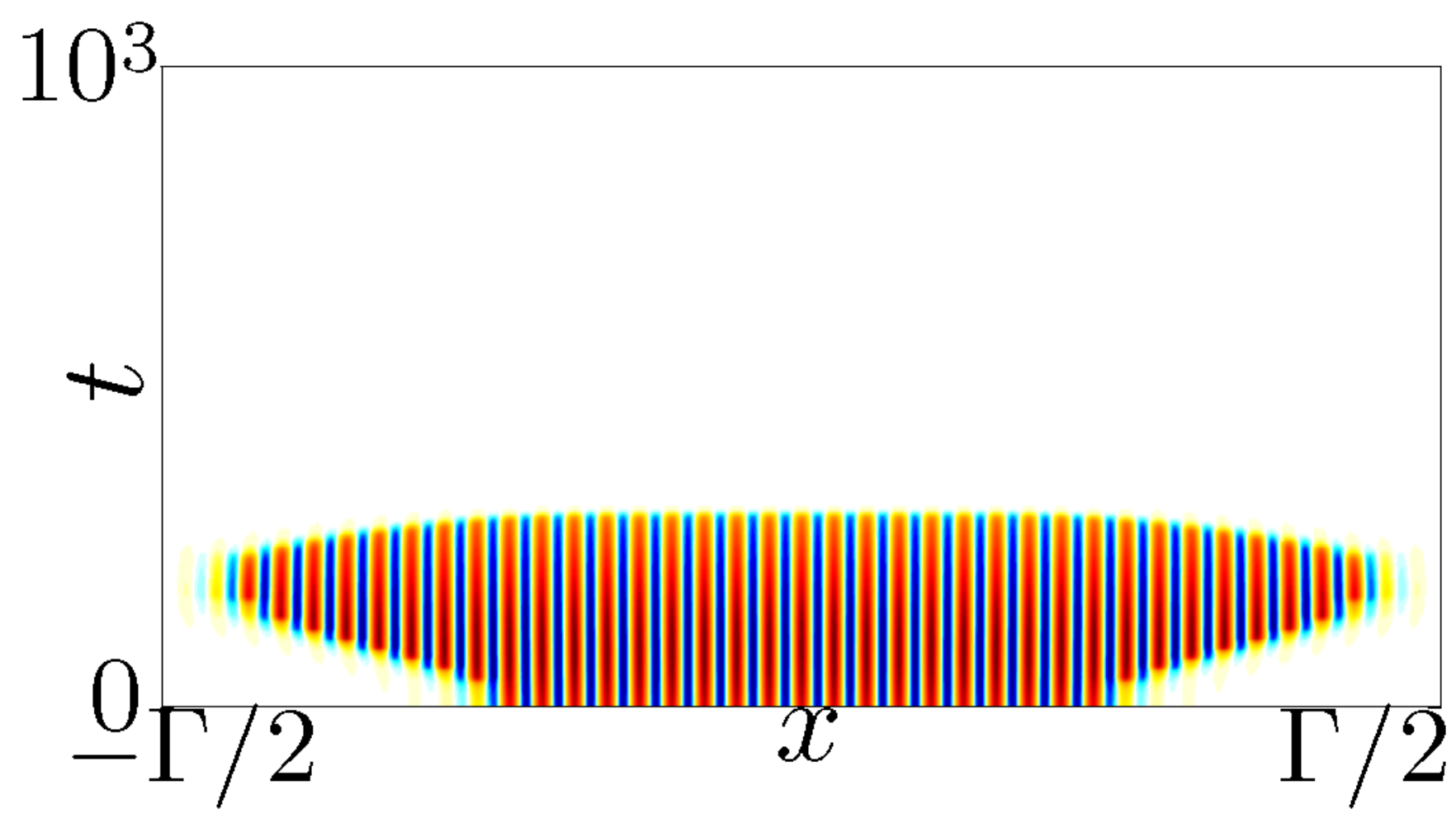}\quad
			\includegraphics[width=60mm]{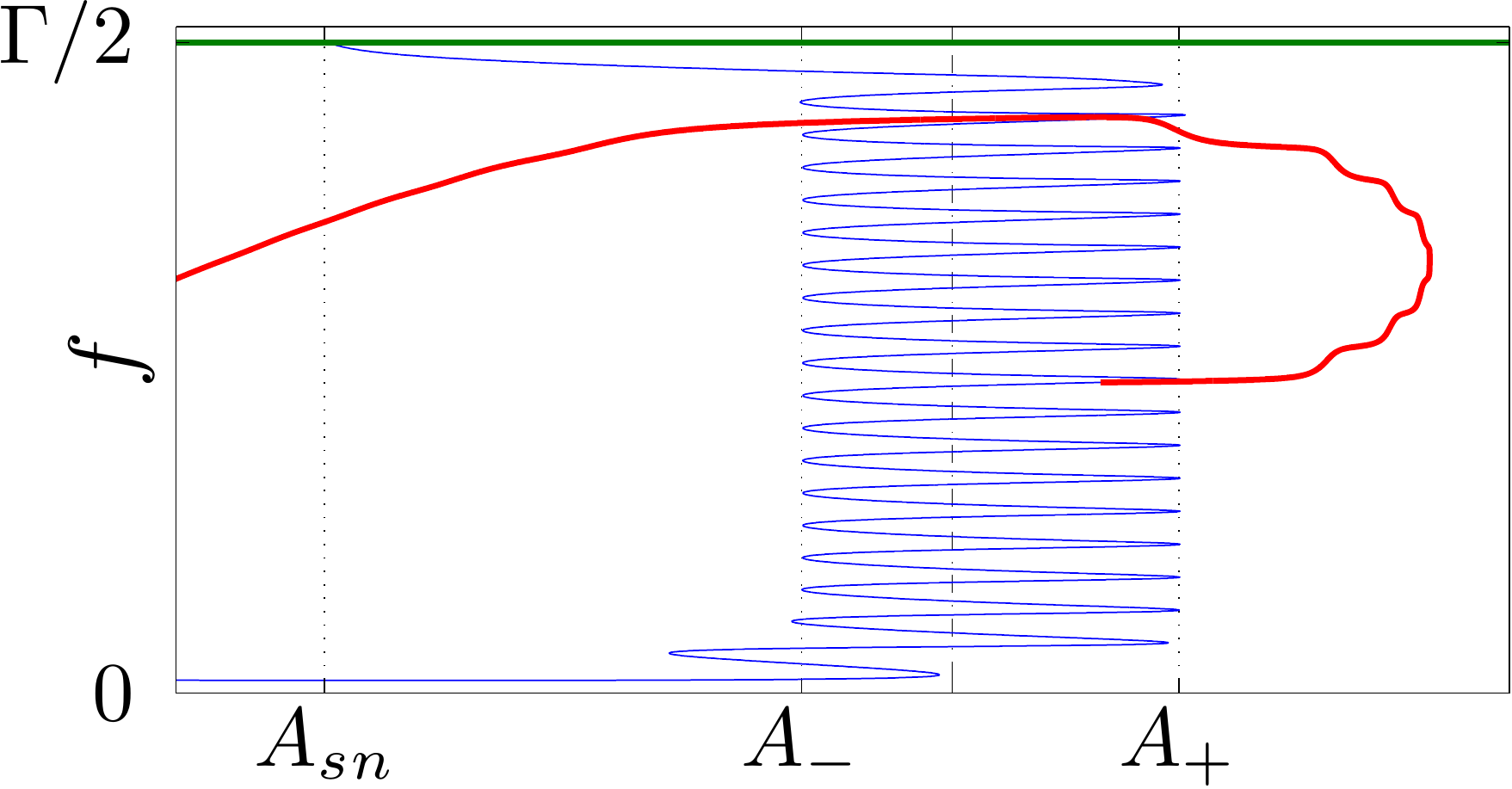}}
\caption{Space-time plots (left panels) and the corresponding phase space trajectories (right panels) for solutions of \cref{eq:SH} with $r(t) = -0.28 + 0.1 \sin (2 \pi t / T)$, $b=1.8$, initialized using an $L_{0}$ solution at $r=-0.28$. The red dashed lines in the right panels correspond to evolution past the time window represented in the left panels, while the green lines represent spatially periodic solutions of the time-independent case. The period $T$ is indicated below each plot. The trajectory in (a) terminates on a time-periodic defect state.}
\label{fig:preview}
\end{figure}
\Cref{fig:preview} shows sample results for $\rho = 0.1$, $r_0 = -0.28$, in each case starting from the same stable spatially localized $L_{0}$ solution of the time-independent problem $r\equiv r_0$. The figure shows that, depending on $T$, the solution can undergo growth/decay through a depinning-like process (\cref{fig:preview}(a,c)), decay to the trivial state via an amplitude mode (\cref{fig:preview}(d)), or take the form of a periodic orbit corresponding to a localized solution with no net motion of the fronts (\cref{fig:preview}(b)). Moreover, the growth/decay of new wavelengths can occur regularly from one period of the forcing to the next, or in a seemingly irregular way.
In particular, \cref{fig:preview}(a) shows a growth scenario for $T=50$ in which the solution grows in length by one wavelength on each side after approximately three cycles of the forcing. This process is irregular in the sense that the number of nucleation and decay events is not constant from one period of the forcing to the next. It is also interesting to note that this simulation does not reach the spatially periodic state, but instead approaches an oscillating state with a defect at the edge of the periodic domain. In contrast, \cref{fig:preview}(c), obtained for $T=250$, shows a very regular pattern of five nucleations events followed by six decay events during the course of each forcing cycle, resulting in an overall decay of the state. Finally, \cref{fig:preview}(d) for $T=350$ shows an initial growth phase followed by abrupt amplitude decay to the trivial state.

Although the wavelength of a localized solution depends on the forcing parameter $r$, it is always near the preferred wavelength $2\pi$, and thus $f$ undergoes abrupt jumps by approximately $2\pi$ whenever the fronts depin. These jumps are most evident during the growth phase since the time between nucleations is longer than the time between annihilations. The results shown in \cref{fig:preview} are independent of the length of the initial stable state selected for the simulation, provided that $6\pi \lesssim f \lesssim \Gamma/2-6\pi$ throughout the simulation, i.e., provided the structure remains well localized. We likewise report that the initial phase of oscillation does not affect the dynamics.

We distinguish periodic orbits from growing and decaying orbits using the instantaneous front velocity $V_f$ defined by the relation $V_f\equiv\dot{f}$.  We look at the averaged front velocity $\langle V_f\rangle$ over a cycle period (calculated for oscillation periods for which $6\pi \lesssim f \lesssim \Gamma/2-6\pi$ and disregarding the first oscillation period), and consider an orbit periodic if $|\langle V_f\rangle|<5\times 10^{-3}$ (corresponding to no net nucleation or annihilation event within a time period of about $3\times 10^{4}$ units). For decaying and growing orbits, the average change in the front position $\langle \Delta f \rangle$ over one cycle helps distinguish the regular behavior in \cref{fig:preview}(c) from the irregular dynamics in \cref{fig:preview}(a): regular dynamics translate into $\langle \Delta f \rangle$ close to an integer number of nucleation/decay events ($\approx 2 n \pi$).

\subsection{Spatially localized periodic orbits}

\begin{figure}
\centering
\includegraphics[width=120mm]{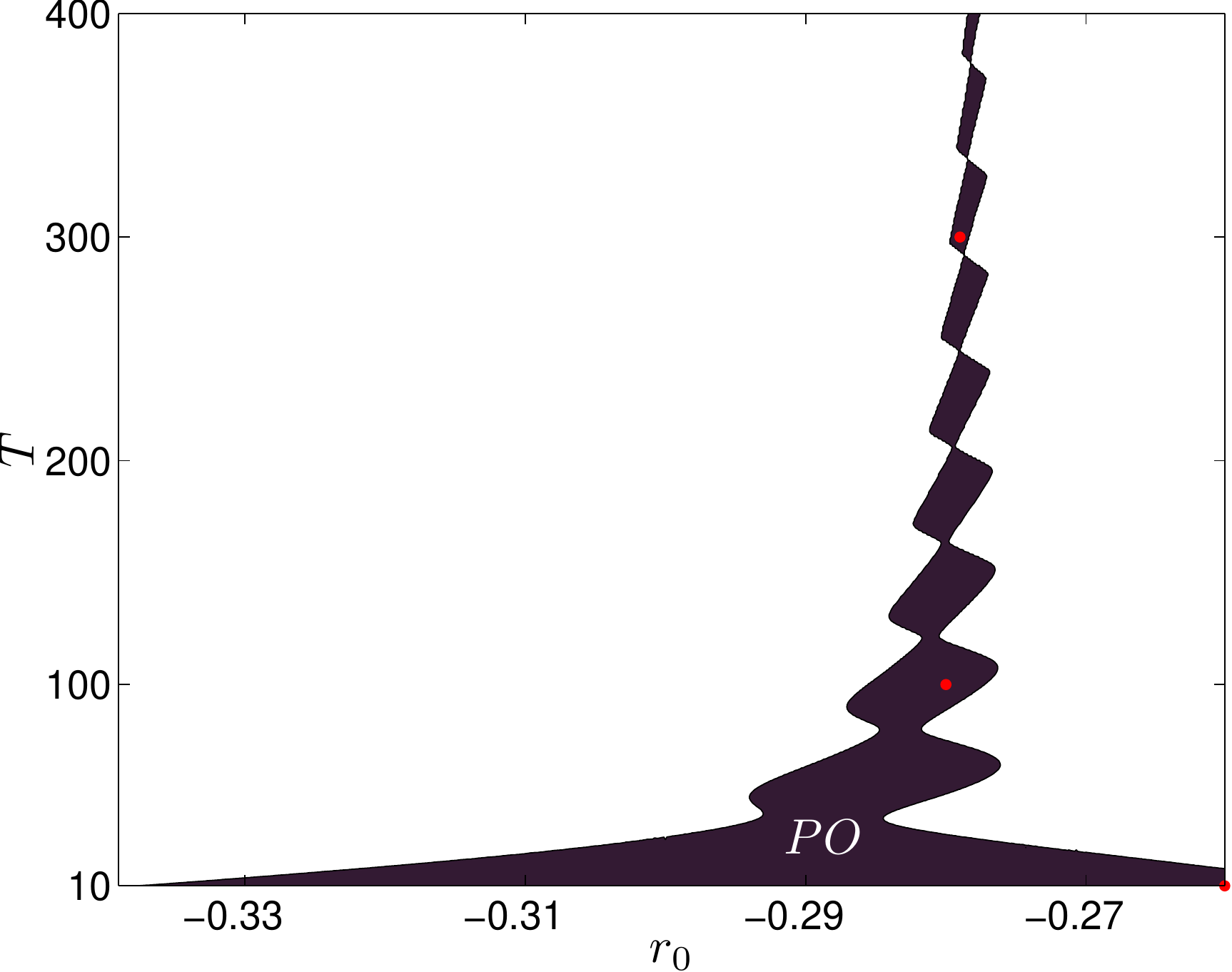}
\caption{\label{fig:vcmstable}Diagram showing the region of existence of periodic orbits $PO$ (shaded region) for oscillation amplitude $\rho=0.1$. The simulations are initialized with a $L_{0}$ solution that is stable for constant forcing $r_0$. The range of $r_0$ shown corresponds to the pinning interval $(r_-,r_+)$ in the time-independent case. The three dots (lower right corner, 2nd and 7th sweet spots from bottom) indicate parameter values for the periodic orbits shown in \cref{fig:convergedperiod}.
}
\end{figure}

We now investigate the existence of periodic orbits like the state exemplified in \cref{fig:preview}(b). For $\rho=0.1$, we do a parameter scan varying the mean forcing amplitude $r_- \le r_0 \le r_+$ in steps of $\Delta r_0 = 10^{-4}$ and the oscillation period $10 \le T \le 400$ in steps of $\Delta T=1$. At each point a simulation is run to calculate $\langle V_f\rangle$ initialized with a steady state localized solution at $r\equiv r_0$.  In most cases the simulations were run for 2000 units of time (4000 units were necessary for the longer oscillation periods). The results are shown in \cref{fig:vcmstable}.
The region where $\langle V_f\rangle=0$ is labeled $PO$ and corresponds to parameter values at which periodic orbits are found. For short periods $T$ for which there is insufficient time for nucleation or annihilation within a cycle, the region of periodic orbits spans nearly the whole pinning region (bottom part of \cref{fig:vcmstable}). With increasing $T$ the range of existence of periodic orbits narrows as predicted by the theory in section III for fast oscillations but does not do so monotonically. The figure reveals that {\it sweet spots} where the range is larger than in the {\it pinched regions} above and below occur at regular intervals of the forcing cycle period. For $\rho = 0.1$, the pinched regions are separated by $\Delta T\approx 43$.  The region of existence of periodic orbits is asymmetric owing to major differences in the depinning dynamics in regimes $\mathcal{D}_-$ and $\mathcal{D}_+$. Moreover, the region slants to higher values of the forcing as the period $T$ increases, a property that is related to the additional time spent in regime $\mathcal{A}_-$ during the decay phase. Region $PO$ eventually asymptotes to $r_0\approx -0.2744$ as $T\to \infty$, the threshold for entering regime $\mathcal{A}_-$, where amplitude decay takes over from depinning as the leading mode of decay. In contrast to the high frequency case, here the Maxwell point determined from the time-averaged free energy moves to lower values of $r$ and is no longer a good predictor of the region of periodic orbits. 
\begin{figure}
\centering
   	\subfloat[$T=10$, $r_0=-0.26$]{
   		\includegraphics[width=50mm]{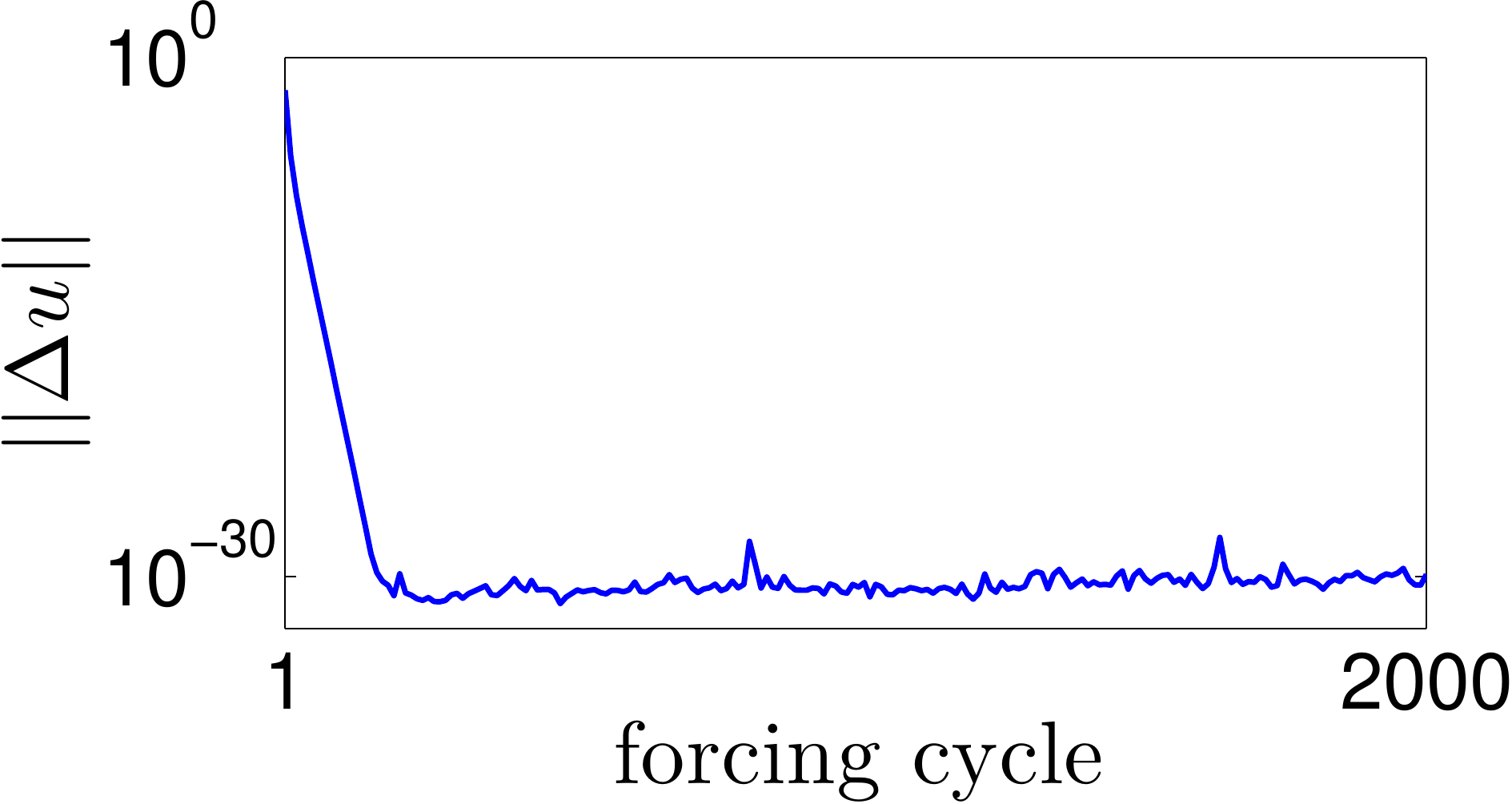}
		\includegraphics[width=50mm]{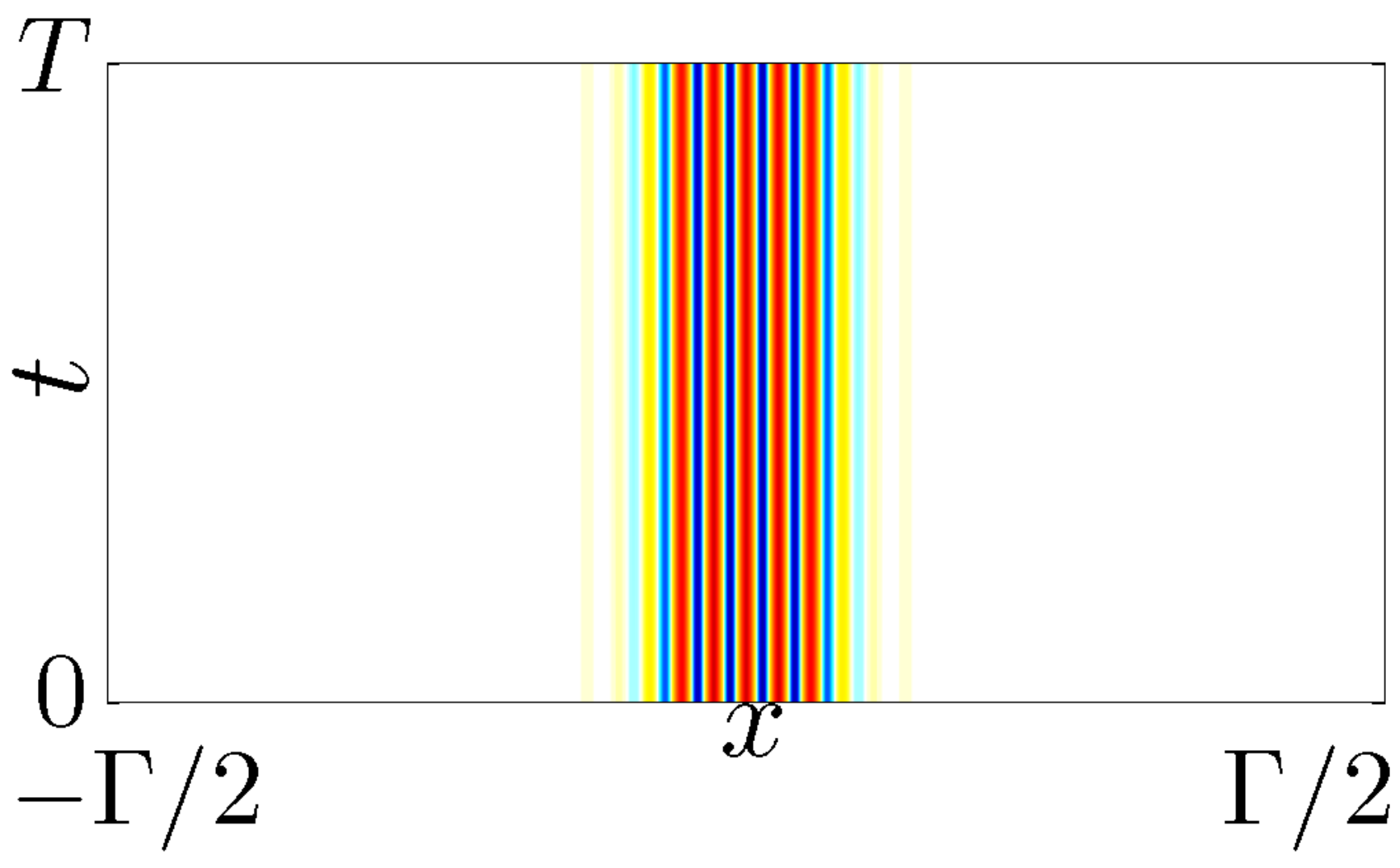}
		\includegraphics[width=50mm]{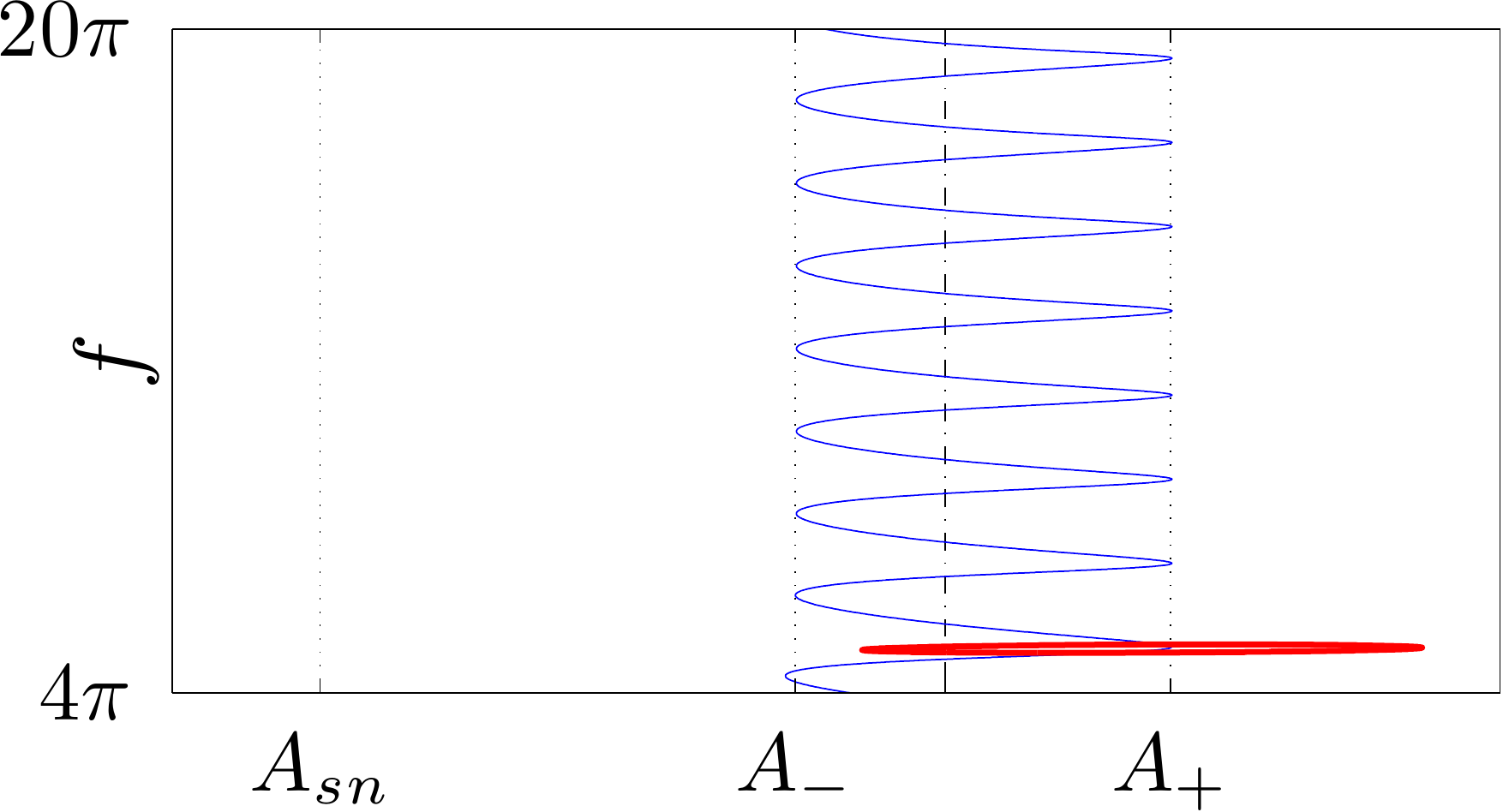}}
\\
   	\subfloat[$T=100$, $r_0=-0.28$]{
   		\includegraphics[width=50mm]{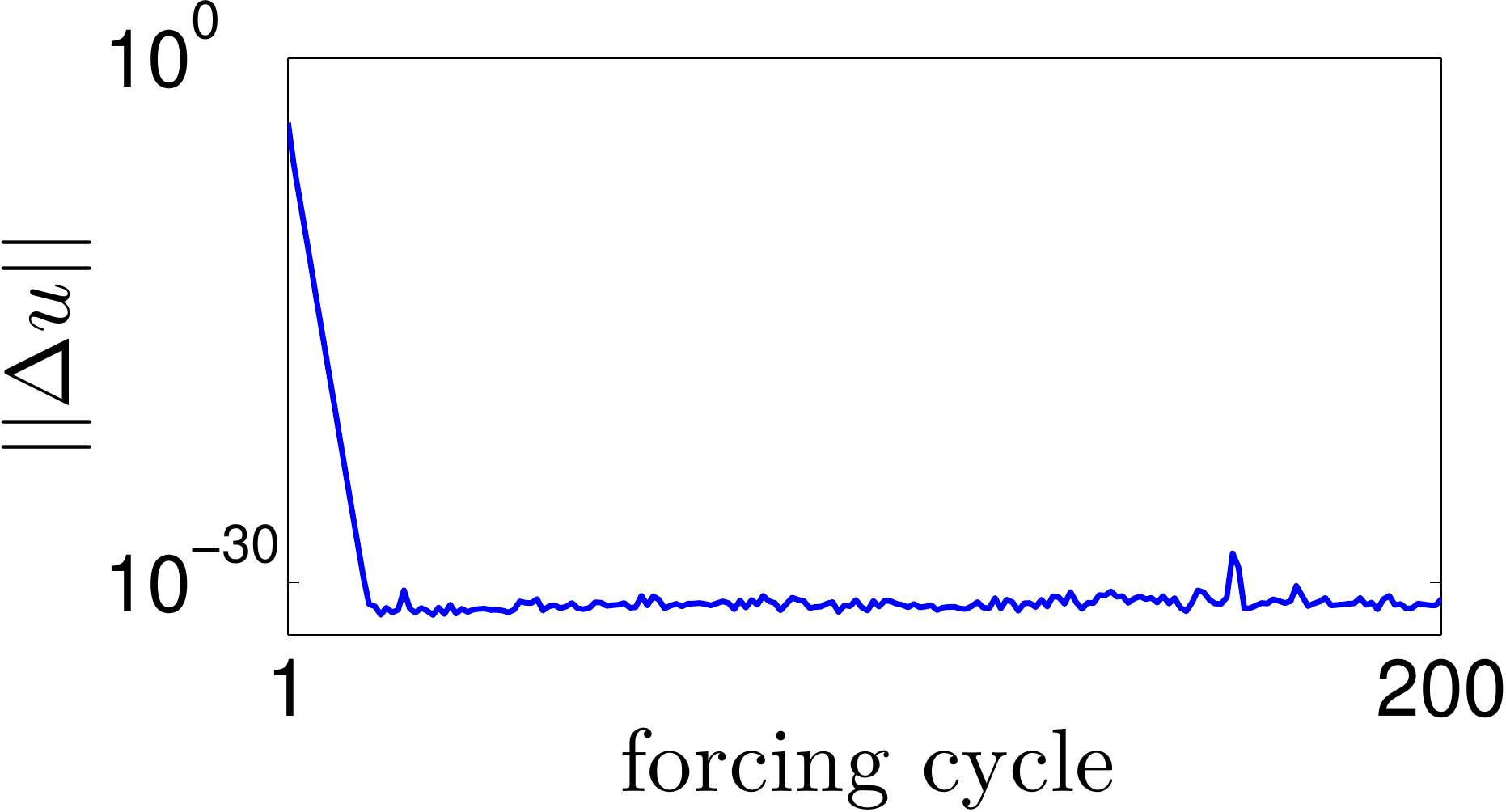}
		\includegraphics[width=50mm]{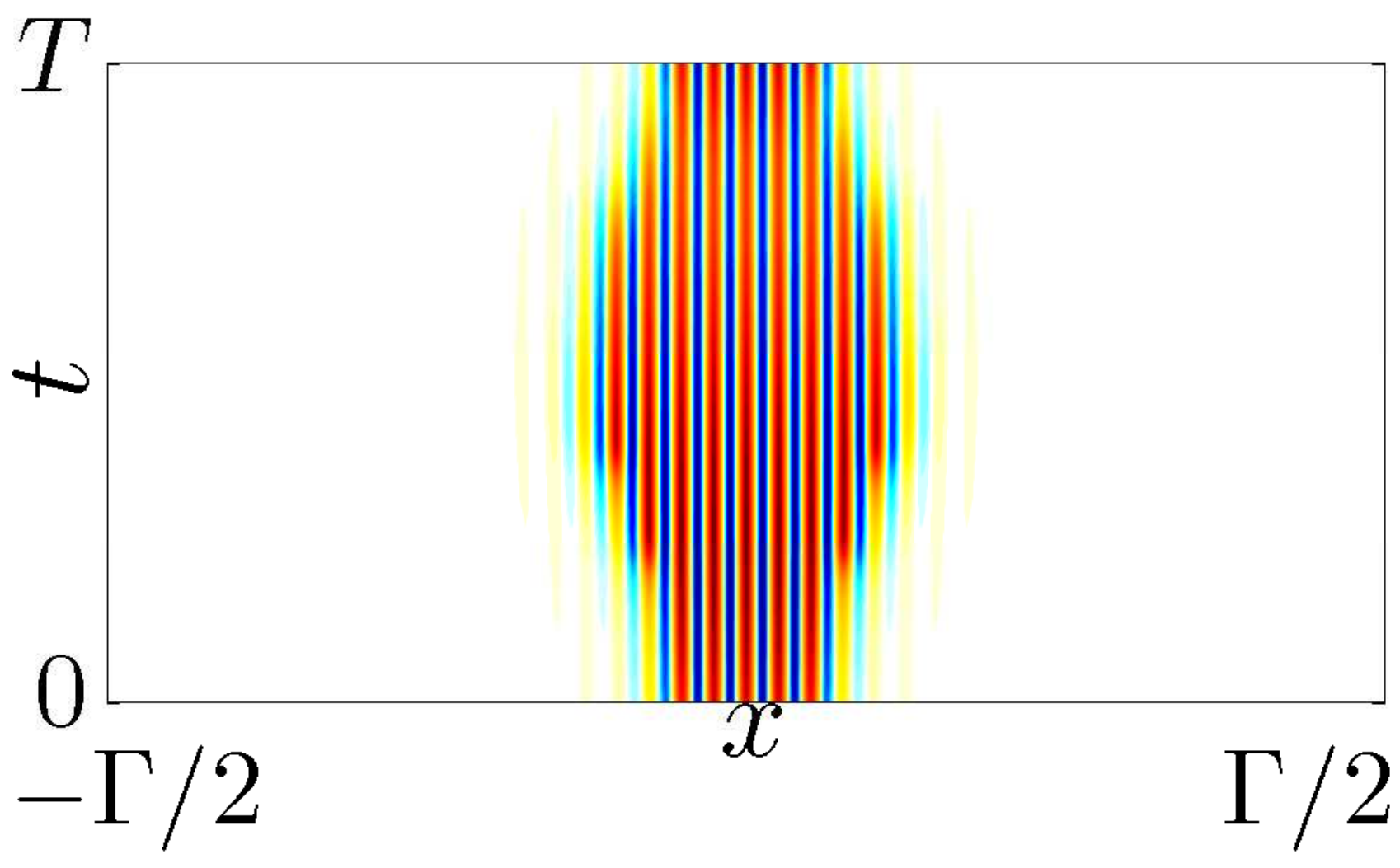}
		\includegraphics[width=50mm]{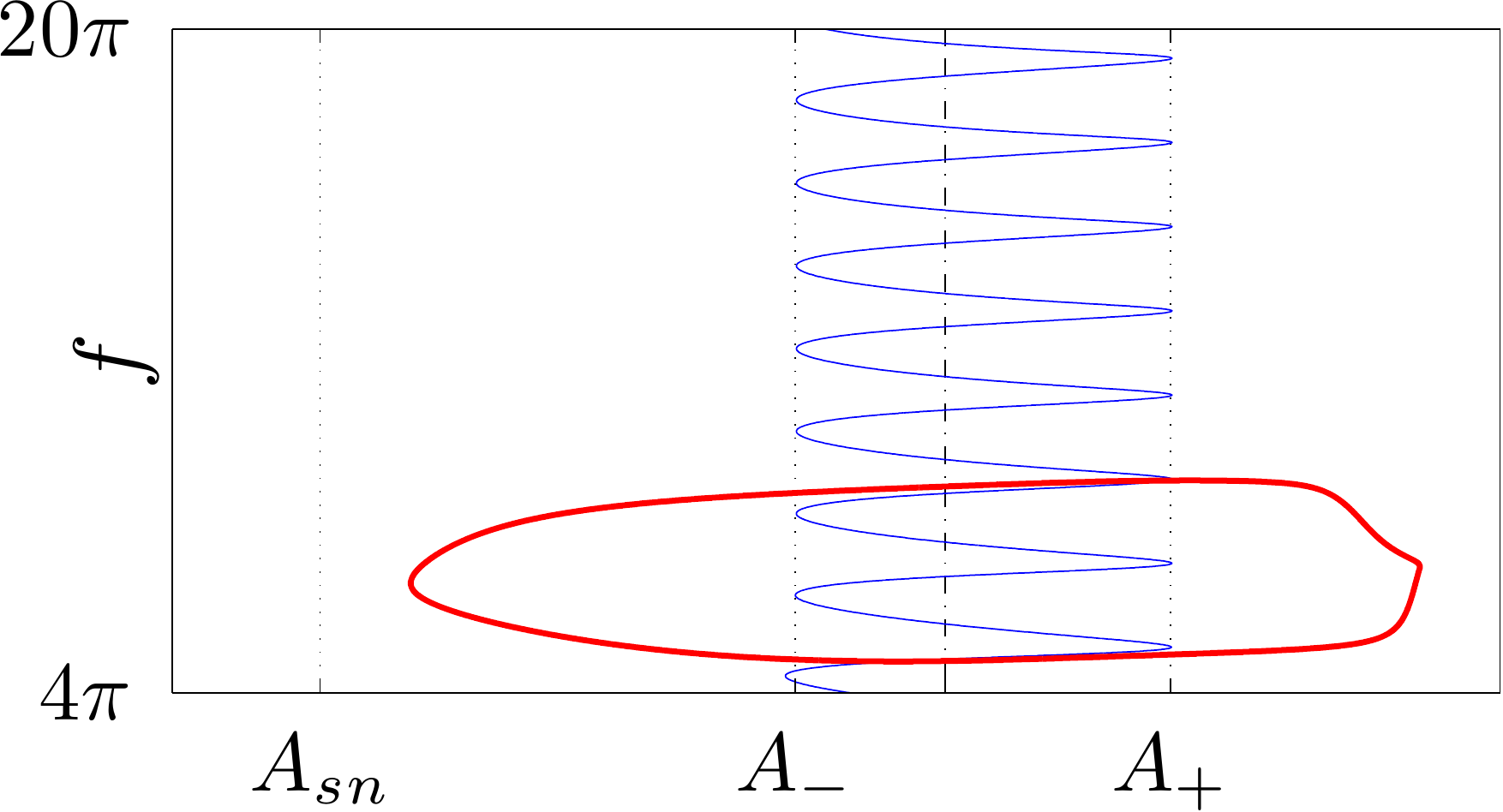}}
\\
   	\subfloat[$T=300$, $r_0=-0.279$]{
   		\includegraphics[width=50mm]{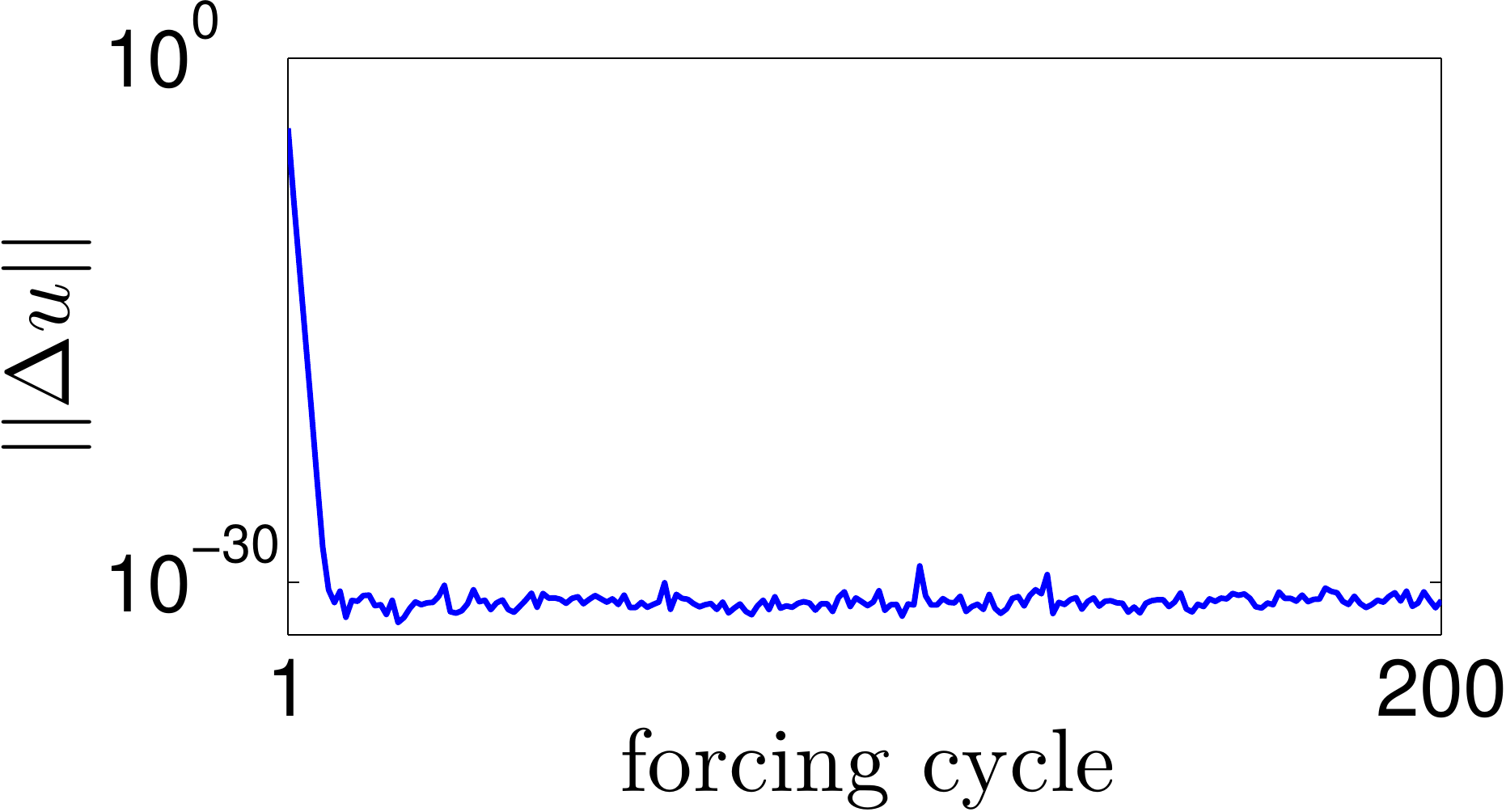}
		\includegraphics[width=50mm]{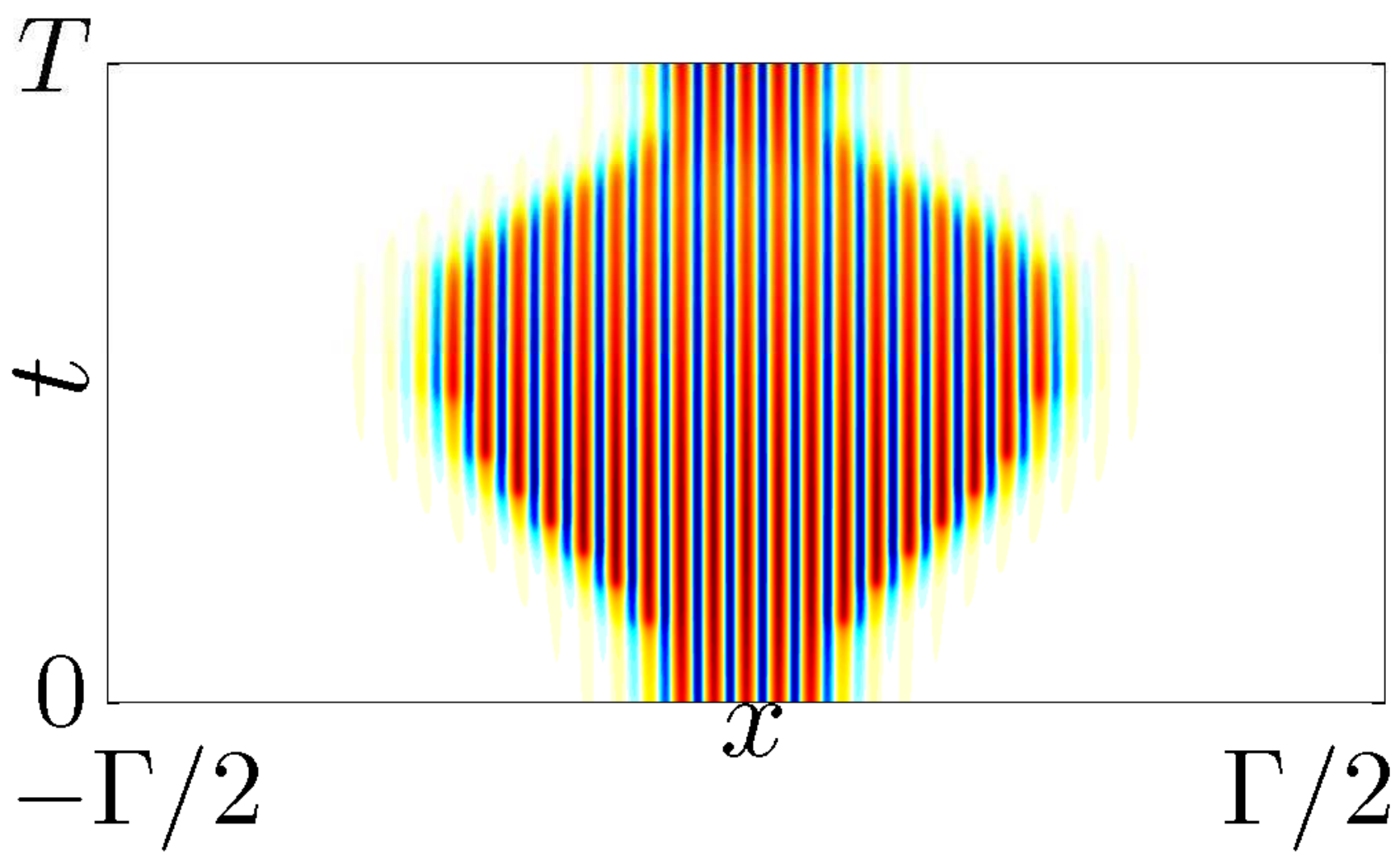}
		\includegraphics[width=50mm]{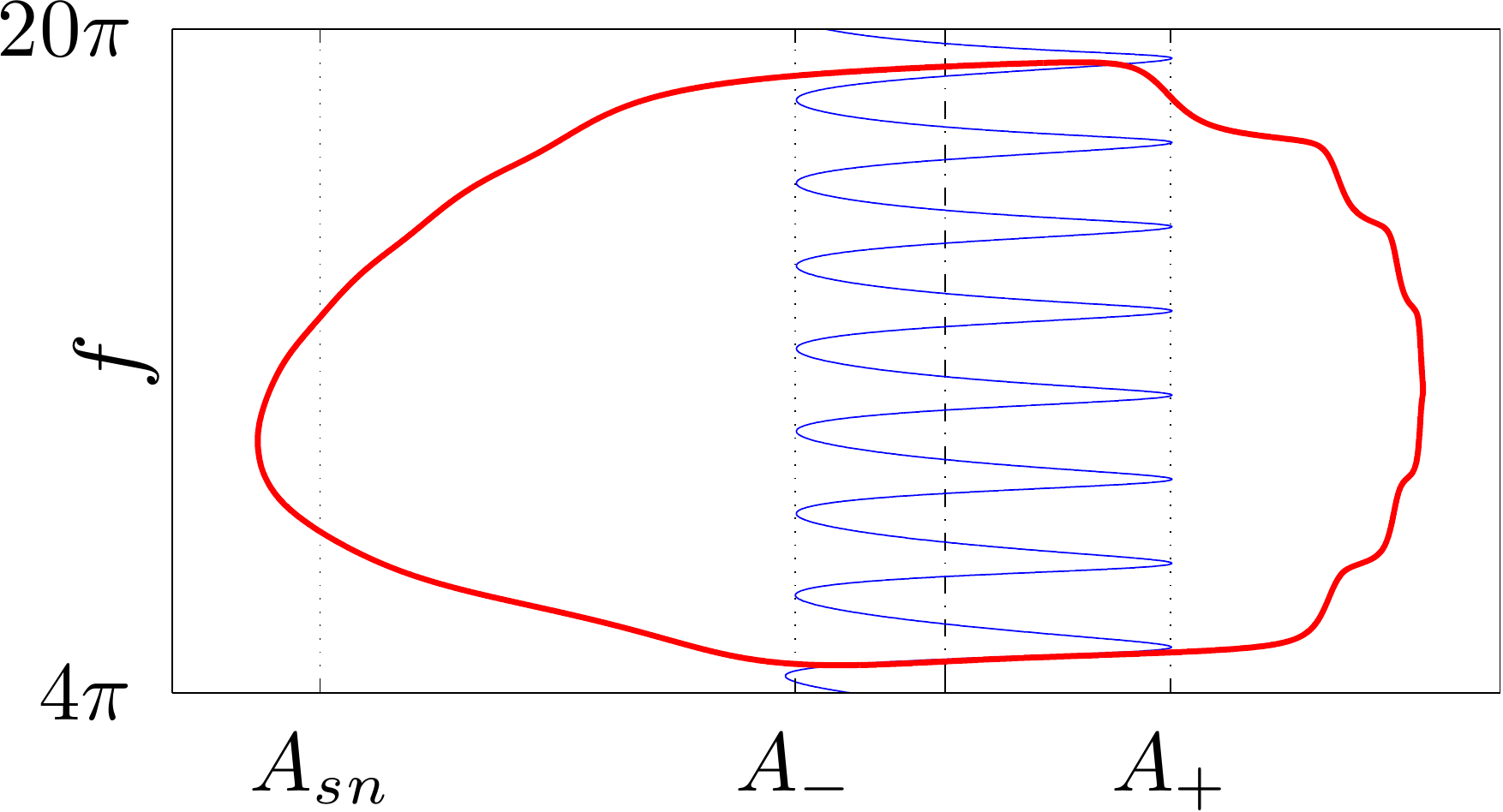}}
    \caption{Periodic orbits for three parameter combinations: The normalized $L_2$ norm of the difference of two solutions exactly one period apart showing convergence to machine precision (left panels), space-time plots of the corresponding converged solution over one cycle period (middle panels), and the $(A,f)$ trajectory of the converged solution (right panels).  }
    \label{fig:convergedperiod}
\end{figure}

\Cref{fig:convergedperiod} shows three different stable periodic orbits from different sweet spots, corresponding to $T=10$, 100 and 300. The left panels indicate that these solutions are converged to machine precision and do not seem to suffer from slow instabilities, while the remaining panels provide insight into the balance between growth and decay over the course of the cycle period. \Cref{fig:convergedperiod}(a), for $T=10$, shows a periodic pulsation in amplitude but no front motion. \Cref{fig:convergedperiod}(b), for $T=100$, reveals a periodic orbit characterized by both amplitude and front oscillation as does \cref{fig:convergedperiod}(c) for $T=300$.  The $T=100$ orbit, which is located in the \textit{second} fully formed sweet spot from the bottom of \cref{fig:vcmstable}, undergoes \textit{two} nucleation events followed by \textit{two} decay events during the course of each forcing cycle. The $T=300$ example is from the 7th sweet spot and undergoes 7 nucleation/decay events per cycle; the example shows that the nucleations occuring during the growth phase of the forcing between $t \approx 50$ and $t \approx 150$ are significantly slower than the decay between $t \approx 200$ and $t \approx 250$. Note that periodic orbits are present despite entering $\mathcal{A}_-$, something that is only possible because of the short amount of time spent in this regime.

\begin{figure}
\centering
    \mbox{
\subfloat[$\rho=0.06$]{\includegraphics[width=50mm]{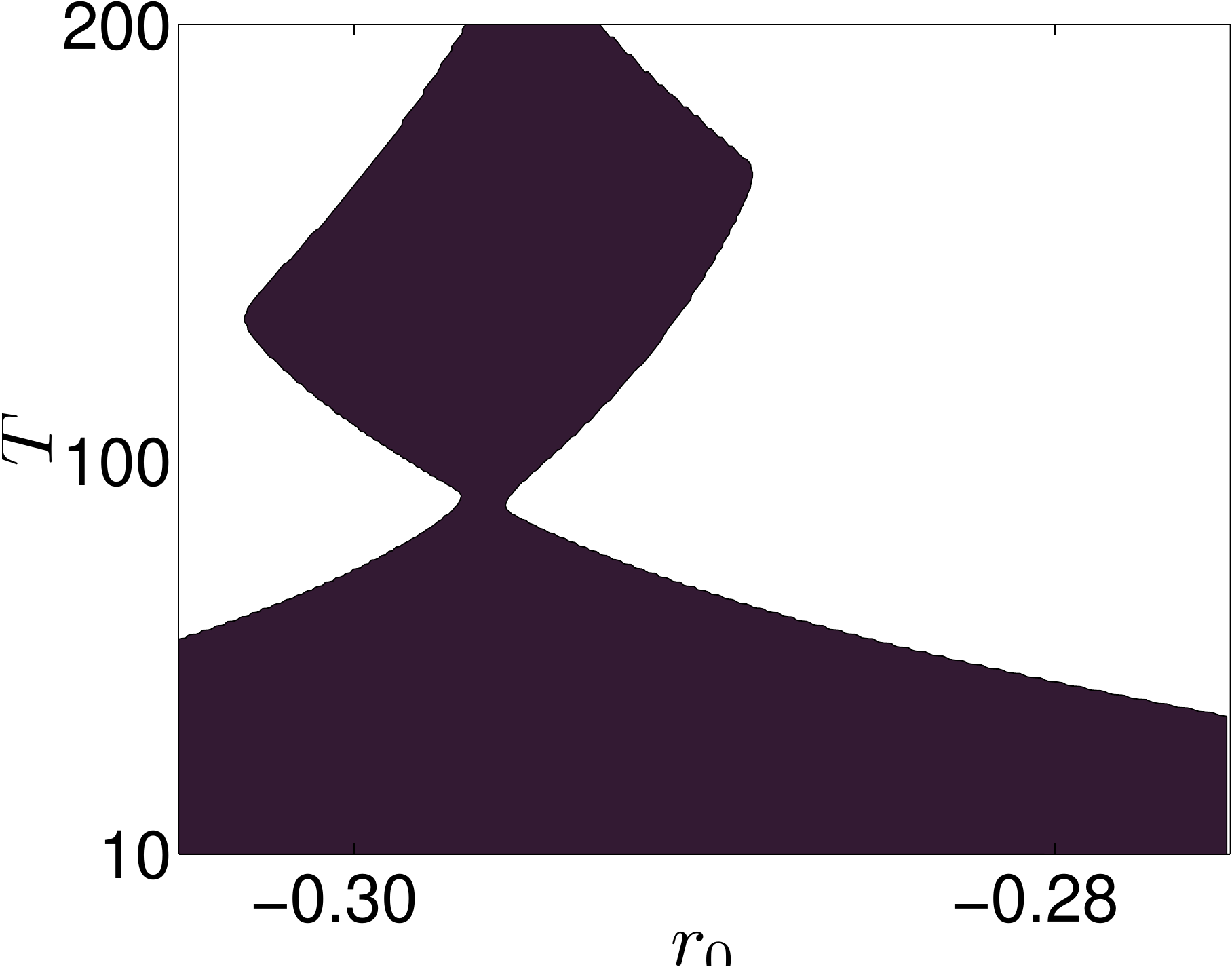}}\;
\subfloat[$\rho=0.08$]{\includegraphics[width=50mm]{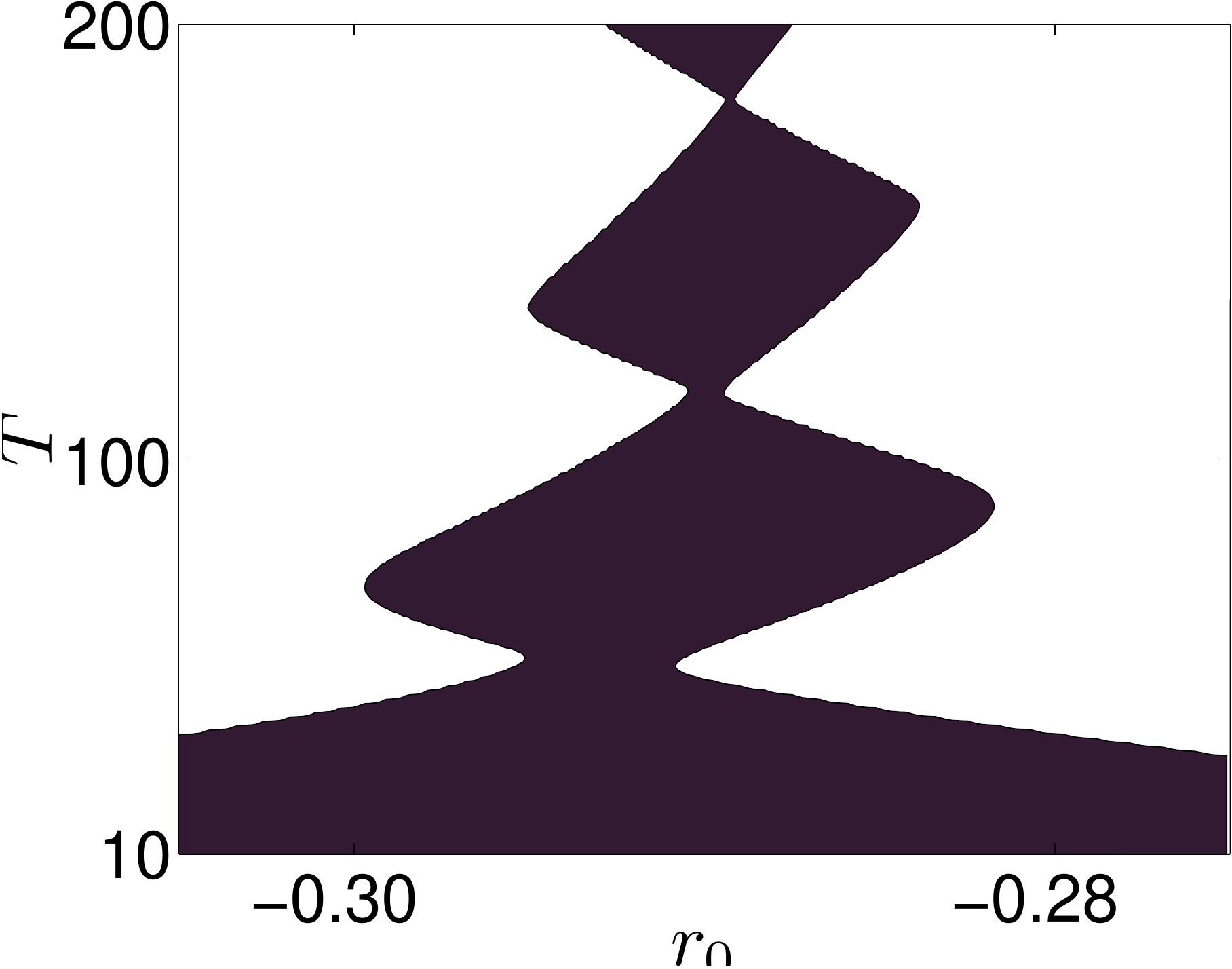}}\;
\subfloat[$\rho=0.10$]{\includegraphics[width=50mm]{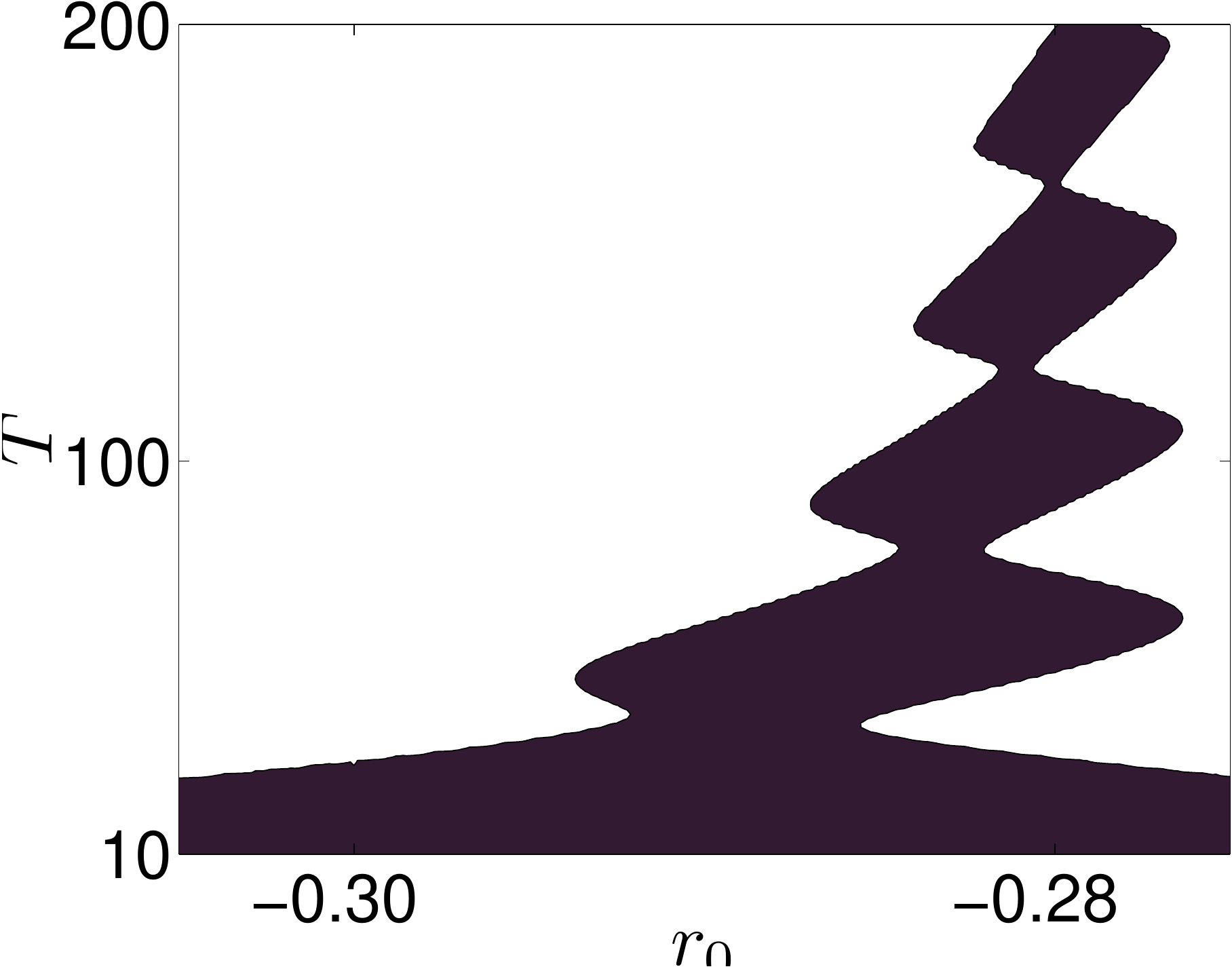}}
}
\caption{Region of existence of periodic orbits when (a) $\rho=0.06$, (b) $\rho=0.08$, (c) $\rho=0.1$, using the same color code as in \cref{fig:vcmstable}.}
    \label{fig:vcmcompare}
\end{figure}

We also examined the dependence of the region $PO$ on the amplitude of oscillation, $\rho$. The results for $T \le 200$ (\cref{fig:vcmcompare}) show that as $\rho$ increases, the sweet spots span an increasingly smaller interval in the period $T$ and thus a larger variety of periodic orbits can be observed within a given range of $T$ as $\rho$ increases. In addition, the whole sweet spot structure asymptotes more quickly towards $r_0=-0.2744$ as $\rho$ increases.

\subsection{Structures undergoing net growth or decay}

\begin{figure}
\centering
\includegraphics[width=120mm]{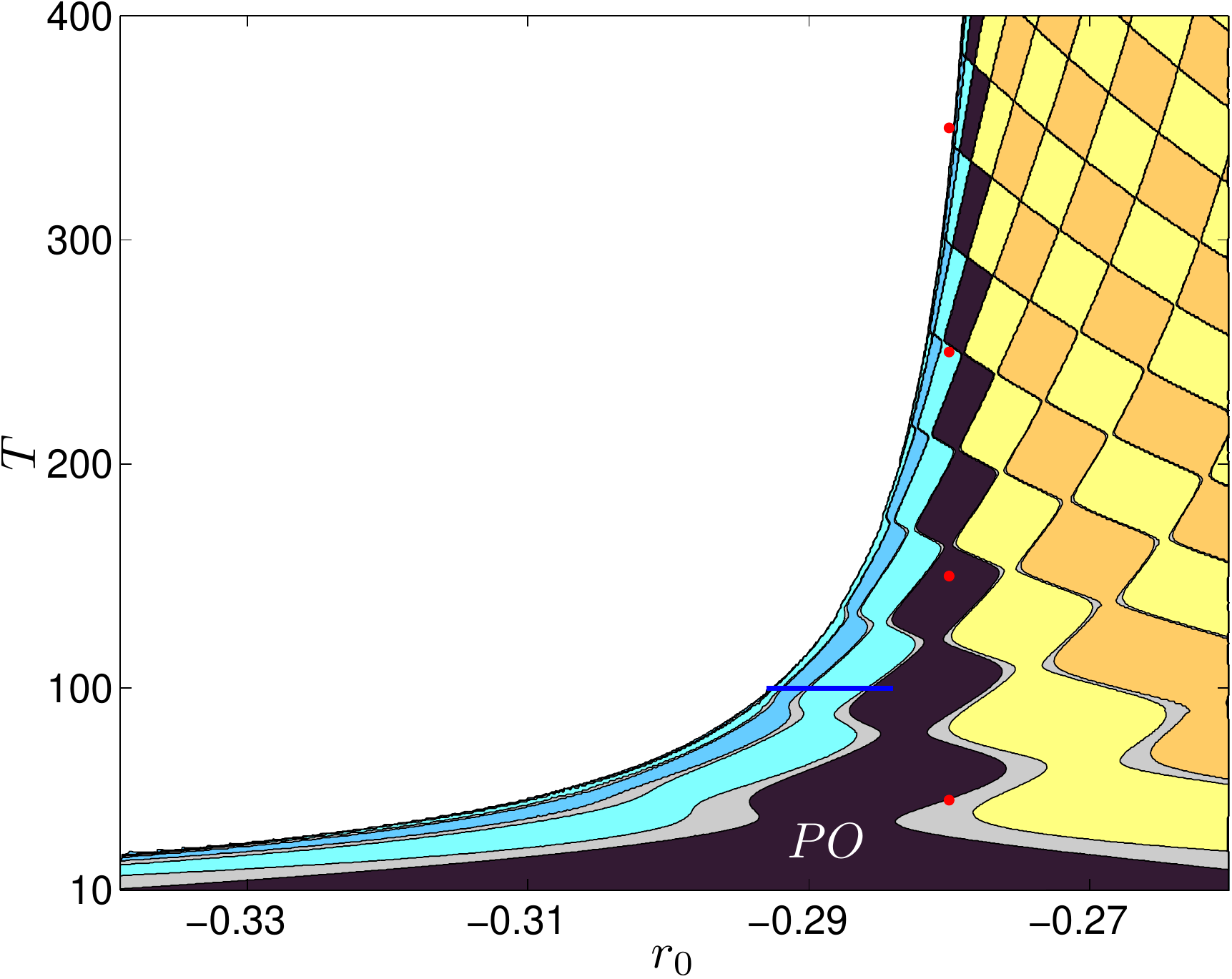}
\caption{\label{fig:wholediag}The number of spatial periods added/lost per cycle for an oscillation amplitude $\rho=0.1$. The simulations were initialized with $L_0$ localized solutions that are stable at $r_0$ in the time-independent system. The central black region corresponds to the $PO$ region (cf. the shaded region in \cref{fig:vcmstable}). The light blue region to the left corresponds to decay by one wavelength on each side of the localized state per cycle, the next to decay by two wavelengths per cycle etc. The regions to the right of $PO$ correspond instead to net growth by one wavelength, two wavelengths etc.  on each side of the localized state per cycle. The large white region to the left indicates the location of decay to the trivial state within one cycle period. Transition zones where irregular behavior is observed are shown in gray. The dots indicate the location in parameter space of the solutions plotted in \cref{fig:preview} while the horizontal line refers to a region that is studied in \cref{fig:cliff}.}
\end{figure}

The existence of the periodic orbits discovered above is closely related to the dynamics of the fronts connecting the localized patterned state to the background state, suggesting that we can use the net displacement $\langle\Delta f\rangle$ of the fronts within a forcing cycle to classify the growing/decaying solutions.  We therefore calculated $\langle\Delta f\rangle$ on the same grid as that used to find the periodic orbits $PO$, starting from narrower localized states to the right of $PO$ and broader localized states to the left of $PO$, all stable. In some cases (e.g. for $T>200$), a domain of twice and sometimes four times the size used in the $PO$ calculations was necessary to capture enough oscillations.

\begin{figure}
\centering
    \mbox{
\subfloat[]{\includegraphics[width=50mm]{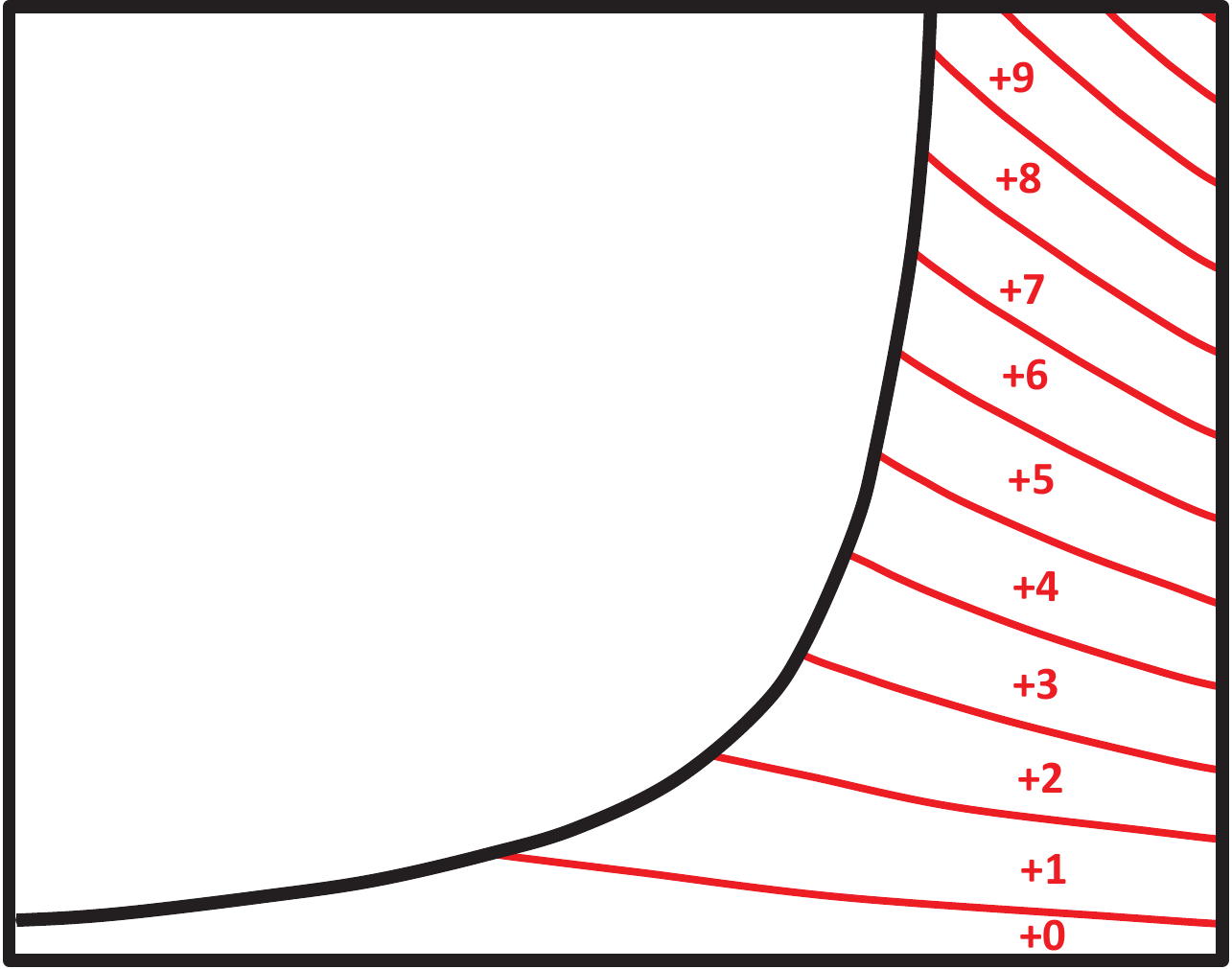}}
\subfloat[]{\includegraphics[width=50mm]{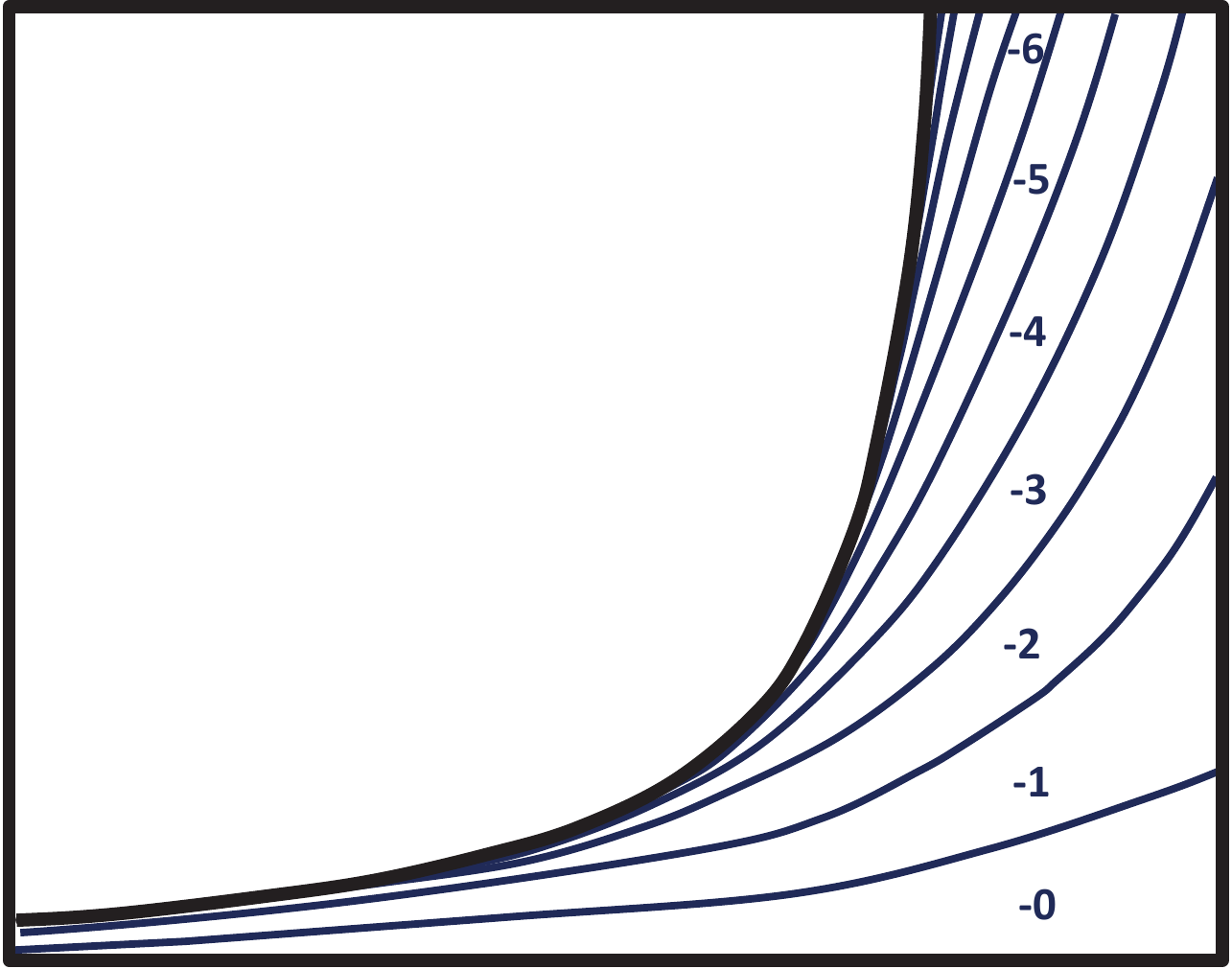}}
}\\
\subfloat[]{\includegraphics[width=100mm]{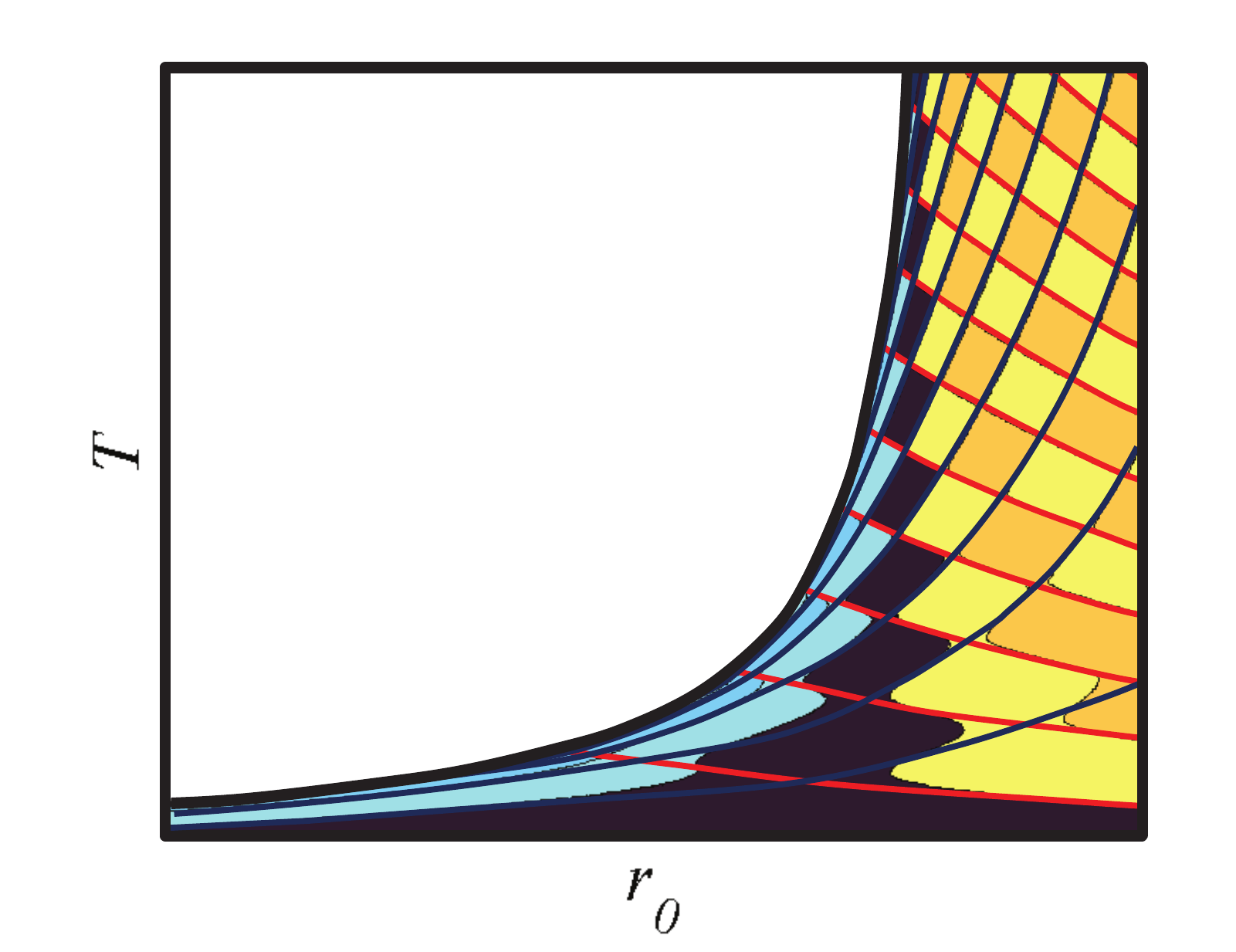}}

\caption{Sketch of the sweet spot classification scheme. The lines indicate transitions between the number of wavelengths gained in (a) and lost in (b) on each side of the localized pattern during one cycle period. These lines are superimposed over the data from numerical simulations in (c) as a means to classify the regions of growth and decay. 
}
    \label{fig:classify}
\end{figure}

The results are summarized in \cref{fig:wholediag}. The different colored regions are determined by the conditions $(n-0.25)2\pi <\langle\Delta f\rangle < (n+0.25)2\pi$, $n=\pm1,\pm2,\dots$ and represent regions where regular behavior is observed. The zones between these regions (shown in gray) are ``transition zones" that will be discussed below. The figure shows that the region of existence of the periodic orbits is surrounded by regions of decay (to the left) and growth (to the right). Beginning in the periodic orbit region $PO$ and moving to the right (increasing $r_0$), the first region encountered ($O_{+1}$) corresponds to growth by one wavelength on either side of the pattern per cycle. The next region ($O_{+2}$) corresponds to growth by two wavelengths on either side, and so on for the subsequent regions which we refer to as $O_{n}$, where $n$ is a positive integer. The regions to the left of $PO$ correspond to decay instead of growth. The closest region to $PO$, $O_{-1}$, exhibits one wavelength decay on either side of the pattern per cycle and so on for $O_{n}$, $n<-1$. Each of these regions is separated from its neighbor by a transition zone where irregular dynamics are observed and displays the same sweet spot--pinched structure as the $PO$ region: the regions expand and contract successively as $T$ increases. Some insight into this structure can be gained by looking at the number $q$ ($m$) of wavelengths gained (lost) on each side of a localized structure during an excursion of the trajectory into regime $\mathcal{D}_+$ ($\mathcal{D}_-$).  A sketch of the corresponding results in \cref{fig:classify} shows that areas corresponding to the gain (loss) of a fixed number of wavelengths during a $\mathcal{D}_+$ ($\mathcal{D}_-$) excursion form bands, and that the intersections of these bands define subregions labeled $^{-m}O_n^{+q}$, where $n=q-m$, corresponding to net gain or loss of $n$ wavelengths per cycle resulting from the annihilation of $m$ wavelengths followed by the nucleation of $q$ wavelengths (\cref{fig:classify}(c)). This procedure allows us to assign a unique label to each subregion in the parameter plane (excluding the transition zones in between). 
Spending more time or going farther into $\mathcal{D}_+$ ($\mathcal{D}_-$) will result in more nucleations (annihilations) over a forcing cycle because more time is spent outside of the pinning region.  This explains the evolution of the $^{-m}O_n^{+q}$ structure as $r_0$ increases: the time spent in $\mathcal{D}_+$ increases and the time spent in $\mathcal{D}_-$ decreases.  Similarly, as the period $T$ of the forcing increases, more time is spent in both $\mathcal{D}_+$ and $\mathcal{D}_-$, resulting in an increase of both $q$ and $m$. This translates into larger oscillations in the location of the fronts of the localized structures for longer periods.
Finally we can gain intuition about the cliff beyond which localized solutions collapse to the trivial state within a single forcing cycle by considering the length of time spent in the regime $\mathcal{A}_-$.  This region is characterized by the time required for solutions to decay to the homogeneous state. As the cycle period increases, the center of oscillation that allows the system to just reach this threshold time is pushed further to the right.  So the edge of the cliff moves to increasing values of $r_0$ as the cycle period $T$ increases.

\begin{figure}
\centering 
\includegraphics[width=120mm]{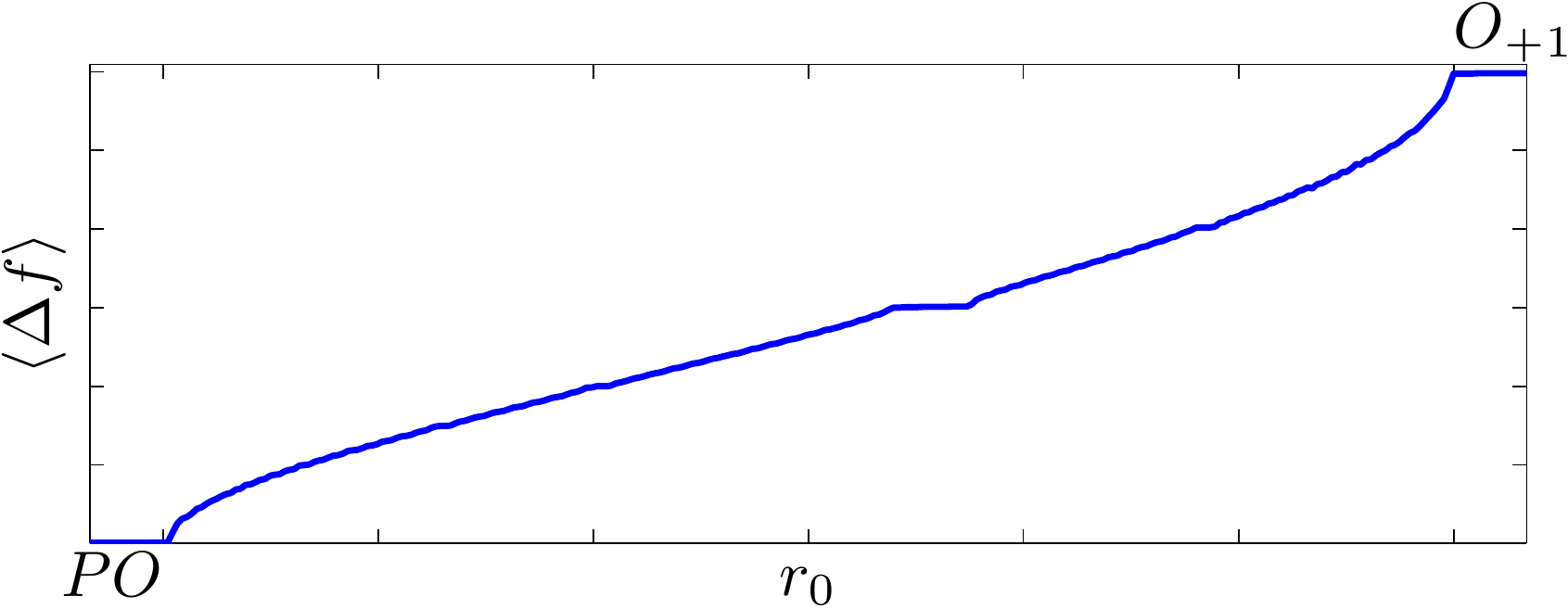}
\caption{The quantity $\langle \Delta f\rangle$ in the transition zone between $PO$ and $O_{+1}$ at $T=80$ exhibits a structure characteristic of a devil's staircase. }    
    \label{fig:transition}
\end{figure}

The transition zones narrow as the period $T$ increases, and a closer look reveals a complex structure resembling a devil's staircase, a characteristic of mode-locking. \Cref{fig:transition} shows $\langle f \rangle$ within the transition zone between $PO$ and $O_{+1}$ at $T=80$ calculated on a domain of 160 spatial periods using a grid of $r_0$ values with spacing $\Delta r_0=10^{-5}$. The results reveal the presence of increasingly thin regions in which $n$ wavelengths are gained/lost from either side of the localized structure within $N$ cycles. These regions thus correspond to fractional growth/decay of the solutions suggesting a complex structure on all scales.  Whether there are regions of nonperiodic dynamics corresponding to irrational numbers cannot be determined through simulations, but the asymptotic results of section IVE seem to indicate that they form a dense subset of the transition zone. 

\subsection{Amplitude decay}

In addition to the depinning-like dynamics observed in the colored regions outside of $PO$ in \cref{fig:wholediag}, amplitude decay occurs in the white region. In this region the initial localized solution collapses to the trivial state within a single forcing cycle.  The boundary of this region is formed by the accumulation of the depinning bands identified in \cref{fig:classify}(b).
\begin{figure}
\centering
    \mbox{
\subfloat[]{\includegraphics[width=75mm]{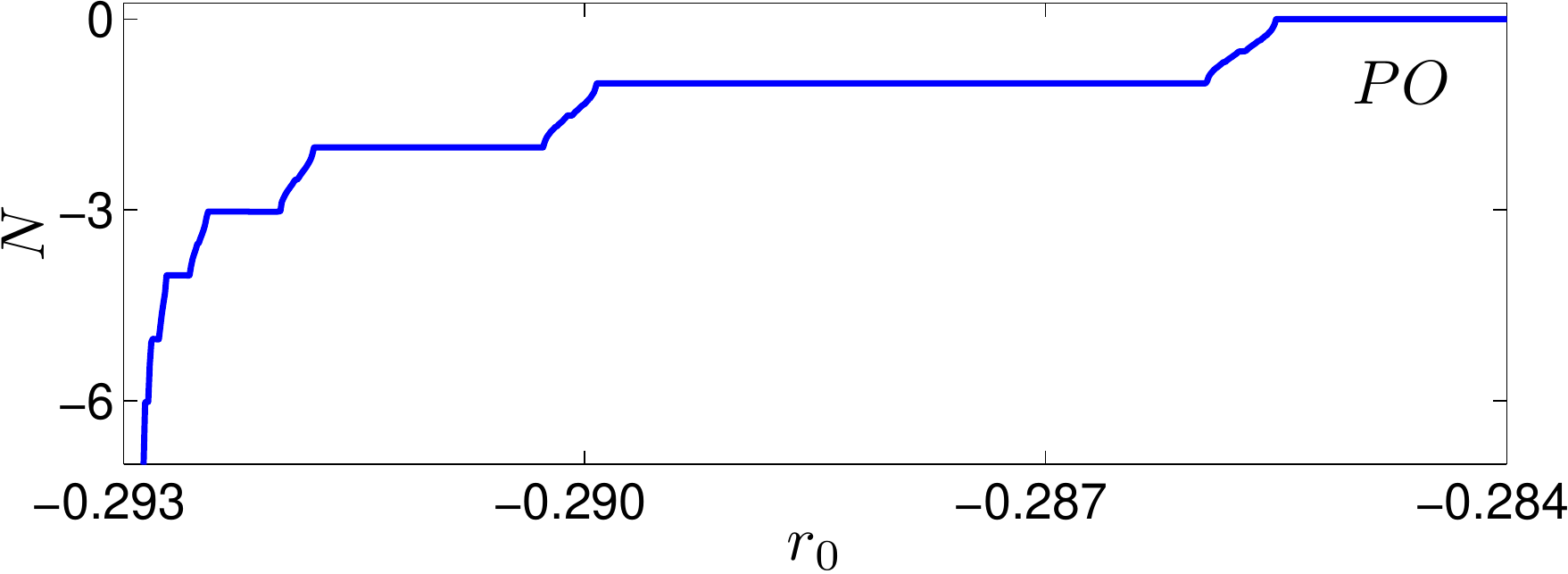}}\quad
\subfloat[]{\includegraphics[width=75mm]{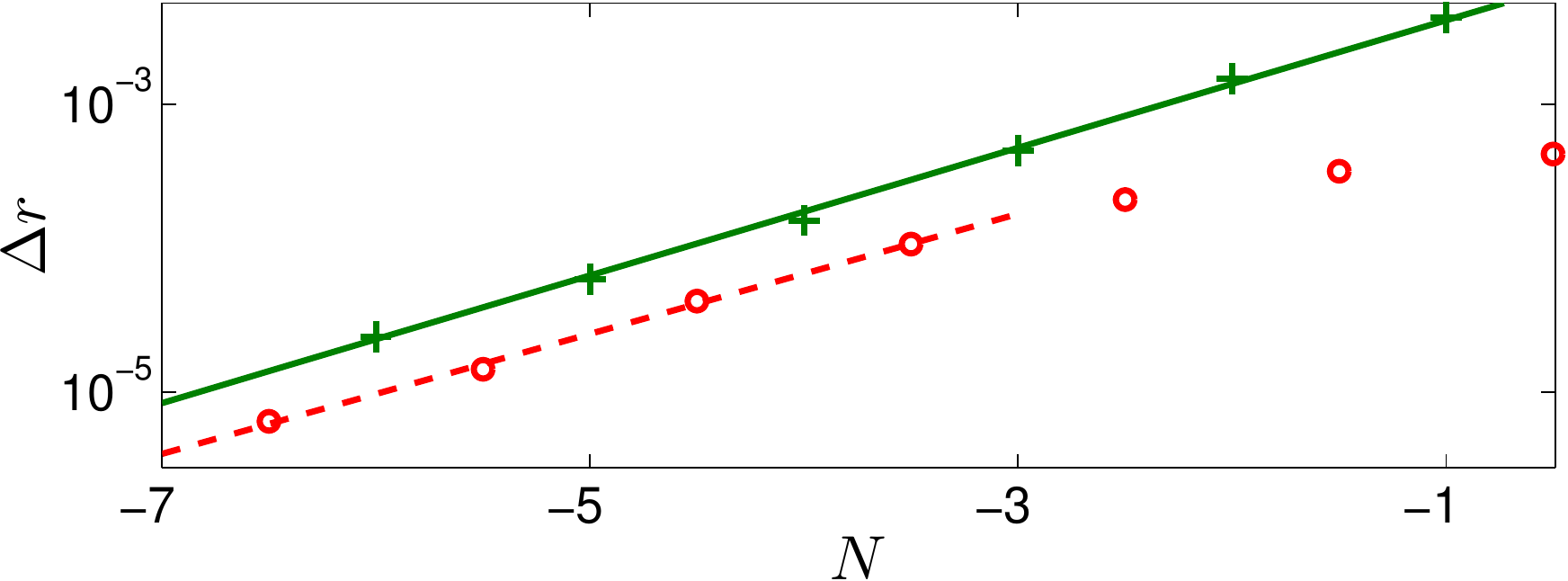}}
}
\caption{(a) The quantity $N\equiv\langle \Delta f  \rangle/ 2 \pi$, representing the shift in the front location averaged over a cycle period, as a function of $r_0$ in regime $D_-$ when $T=100$ (\cref{fig:wholediag}, horizontal line). (b) Length of the plateaus (green crosses) and of the transition regions between them (red circles) as determined from (a), shown in a semilog plot, together with linear approximations to the data (straight lines, given in the text). The plateaus are plotted at integer values of $N$ while the transition zones between plateaus $N=n$ and $N=n+1$ are taken to correspond to $N=n+0.5$. }
    \label{fig:cliff}
\end{figure}

We look more carefully at the accumulation point of the decay bands in \cref{fig:cliff}. The plateaus correspond to the loss of integer numbers of wavelengths per forcing cycle. \Cref{fig:cliff}(a) shows that the width of the plateaus as well as of the transition zones between them decreases as one approaches the accumulation point, while \cref{fig:cliff}(b) shows the width $\Delta r$ of the plateaus and the transition zones as a function of $N$, the number of wavelengths lost per cycle ($N<0$). The data obtained shows that both these widths decrease exponentially with increasing $|N|$ and are consistent with the fits
\begin{eqnarray}
\Delta r_P (N_P) = 0.010557 e^{1.0188 N_P},\\
\Delta r_T (N_T) = 0.0033069 e^{0.9584 N_T},
\end{eqnarray}
where $\Delta r_P$ (resp. $\Delta r_T$) denotes the width of the plateau corresponding to the loss of $N_P$ wavelengths per cycle (resp. width of the transition zone between the pair of closest integers to $N_T$). To obtain these fits we used all the data in \cref{fig:cliff}(b) on the plateau widths but only the transition zones between $N=-6.5$ and $N=-3.5$. To consider even smaller values of $N$ would have required considerably more numerical effort without improving substantially the accuracy of the fit, while values of $N$ closer to $0$ lead to departures from the asymptotic regime. Both formulas show similar exponential decrease, thereby confirming the presence of an abrupt ``cliff'' at the accumulation point (\cref{fig:cliff}(a)).  Furthermore, we see that the width of the transition zones tends to about $1/3$ of that of the plateaus as $|N|$ increases. 

\subsection{Asymptotic theory: small oscillations}

To understand the structure of the parameter plane in \cref{fig:wholediag} we need to understand the process of depinning in the time-dependent system. For this purpose we will consider parameter excursions that take the system outside of the pinning region long enough for a nucleation or annihilation event to occur. We therefore suppose that $r\rightarrow r_{+}+ \epsilon^2(\delta+\rho\sin\epsilon\omega t)$ for which the oscillation period is of the same order as the nucleation time.  An analogous calculation near $r_-$ would produce similar results. In this regime the problem is governed by the equation 
\begin{equation}
u_t= \left(r_++ \epsilon^2(\delta+ \rho \sin\epsilon\omega t)\right) u-\left(1+\partial_{x}^2\right)^2u+b u^2-u^3.\label{eq:SHsmallslow}
\end{equation}
Since the dynamics takes place on an $\mathcal{O}(\epsilon^{-1})$ timescale we define the slow timescale $\tau=\epsilon t$ and write $\partial_t\rightarrow\epsilon\partial_{\tau}$. We look for a solution in the form $u=u_0+\epsilon u_1+\epsilon^2 u_2+\dots$, obtaining, at leading order, 
\begin{equation}
r_+u_0-\left(1+\partial_{x}^2\right)^2u_0+b u_0^2-u_0^3=0.
\end{equation}
As a result we pick $u_0$ to be a localized solution at a saddle-node bifurcation of the snaking branch in the time-independent case.  In this case $u_0$ is stationary but only marginally stable.  At $\mathcal{O}(\epsilon)$, we obtain
\begin{equation}
\partial_{\tau} u_0=r_+u_1-\left(1+\partial_{x}^2\right)^2u_1+2b u_0u_1-3u_0^2u_1.\label{eq:SHsmallsloweps}
\end{equation}
Since $u_0$ is stationary, $u_1$ must be of the form of a zero eigenvector of the Swift--Hohenberg equation linearized about the saddle-node solution.  The relevant eigenvector $v_+$ corresponds to wavelength addition and is symmetric with respect to $x\to -x$. Since we focus on states that do not drift we can ignore the marginal but antisymmetric eigenvectors corresponding to translation and phase. Thus $u_1=a(\tau)v_{+}$.

To determine the amplitude $a$ we must go to $\mathcal{O}(\epsilon^2)$.  At this order, the equation is
\begin{equation}
\partial_{\tau} u_1=r_+u_2-\left(1+\partial_{x}^2\right)^2u_2+2b u_0u_2-3u_0^2u_2 +(\delta+\rho \sin\omega \tau )u_0 +bu_1^2- 3u_0 u_1^2.
\end{equation}
The solvability condition for $u_2$ is \cite{burke2006}
\begin{equation}
\alpha_1 \dot{a} = \alpha_2 (\delta+\rho\sin\omega  \tau)+\alpha_3 a^2,
\label{eq:smallosca}
\end{equation}
with the coefficients $\alpha_j$ calculated from the integrals defined in \cref{eq:alpha}.

We can turn this equation into a Mathieu equation using the Riccati transformation $a=-\alpha_1 \dot{b}/\alpha_3 b$, obtaining
\begin{equation}
\ddot{b} = -\Omega_+^2 \left(\delta+\rho\sin\omega  \tau \right)b,\qquad \delta>0,
\label{eq:depinosc}
\end{equation}
where $\Omega_+^2\equiv \alpha_2 \alpha_3 / \alpha_1^2>0$ with $\alpha_j$ evaluated at $r_+$ (cf. section IIB). Thus $\Omega_+\approx 0.5285$.
The same procedure at the left boundary of the pinning region leads to an equation for the dynamics of the annihilation mode amplitude as a function of the distance $\delta$ from the boundary, $r_0=r_-+\delta$:
 \begin{equation}
\ddot{b} = \Omega^2_- \left(\delta+\rho\sin\omega  \tau \right)b,\qquad \delta<0,
\label{eq:depinosc-}
\end{equation}
where we have set $a=\alpha_1\dot{b}/\alpha_3 b$. The integrals $\alpha_j$ are now evaluated at $r = r_-$ and $\Omega_-=\sqrt{-\alpha_2\alpha_3/\alpha_1^2}\approx 0.7159$.

\begin{figure}
\centering
\subfloat[]{\includegraphics[width=120mm]{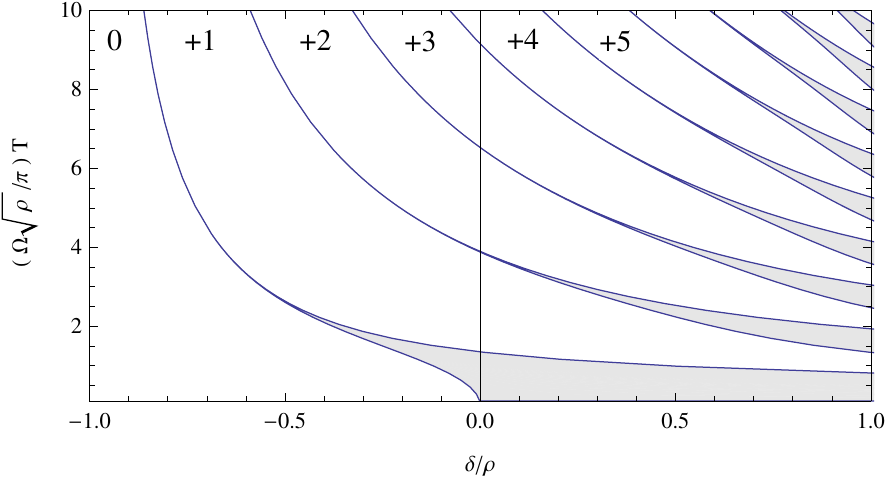}}\\
\mbox{
\subfloat[]{ \includegraphics[width=40mm]{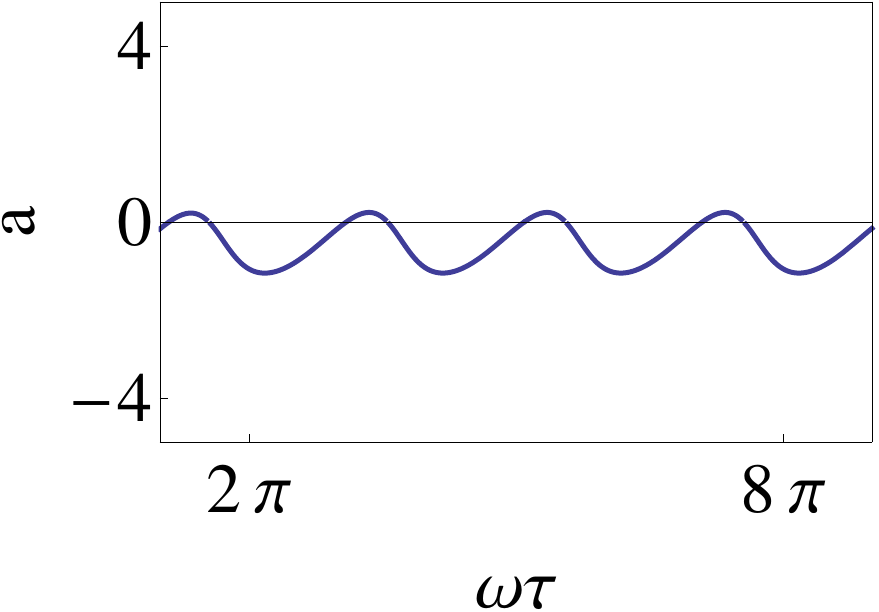}}\;
\subfloat[]{ \includegraphics[width=40mm]{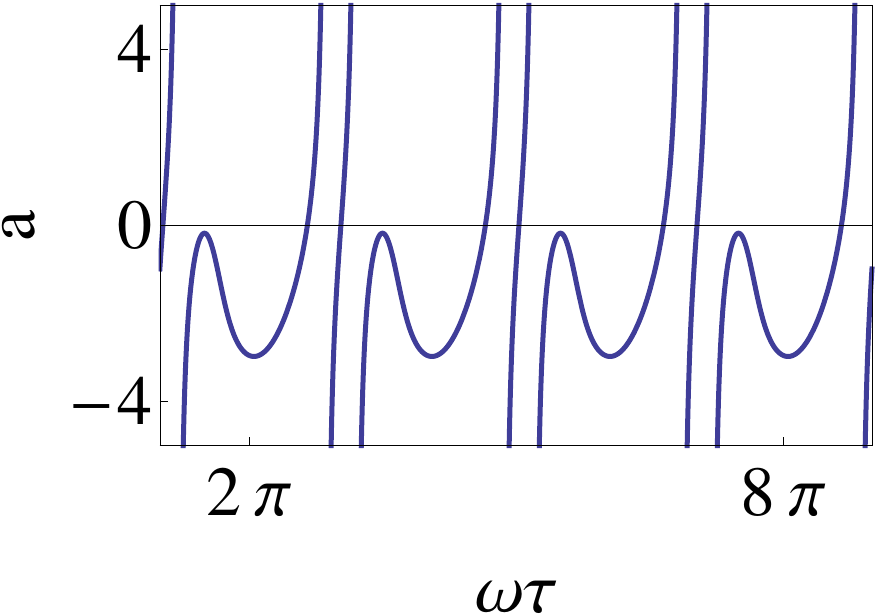}}\;
\subfloat[]{ \includegraphics[width=40mm]{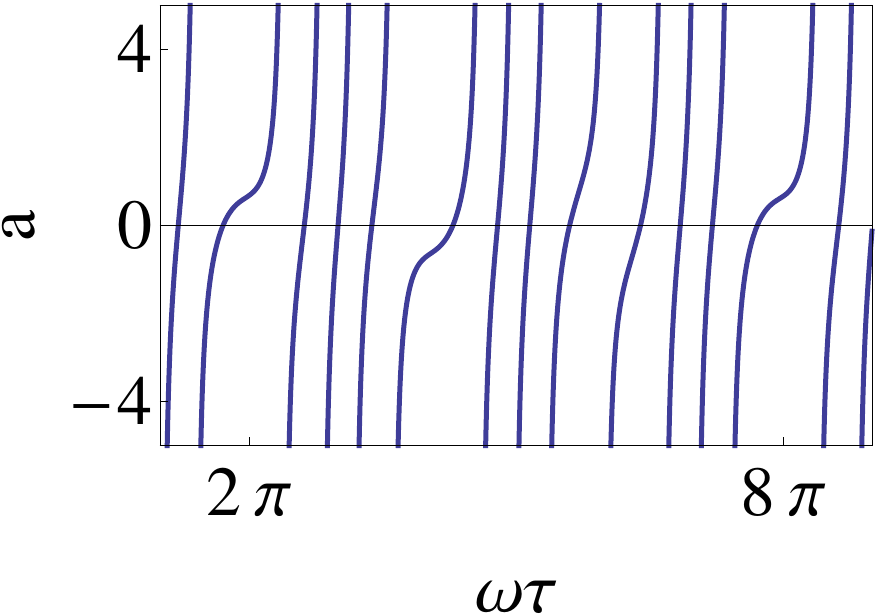}}
}
\caption{(a) The stability diagram for \cref{eq:depinosc}.  The white bands correspond, from left to right, to states that undergo exactly 0,1,2,3,... nucleation events per forcing cycle.  A non-integer number of nucleations per cycle occurs in the gray transition zones in between. (b) Sample solution $a(\tau)$ within the $0$ region when $\delta=-\rho/2$,  $T=2\pi/\Omega_+\sqrt{\rho}$. (c) Sample solution $a(\tau)$ within the $+2$ region when $\delta=0$, $T=6\pi/\Omega_+\sqrt{\rho}$. (d) Sample solution $a(\tau)$ within the transition region between regions $+3$ and $+4$ when $\delta=\rho$, $ T=4\pi/\Omega_+\sqrt{\rho}$. Nucleation events correspond to divergences in $a(\tau)$. 
}
\label{fig:mathieu}
\end{figure}

We can make use of the known properties of the solutions of the Mathieu equation to understand the origin of the resonances between the forcing frequency $\omega$ and the characteristic depinning frequency $\sqrt{\delta}\Omega_+$. The properties of \cref{eq:depinosc} are summarized in the standard stability diagram for the Mathieu equation~\cite{mclachlan1951theory} shown in \cref{fig:mathieu} in terms of the scaled distance from the boundary of the pinning region $\delta/\rho$ and the scaled oscillation period $\Omega_+\sqrt{\rho} T/\pi$. The shaded zones indicate that the solutions of (\ref{eq:depinosc}) are bounded for all time, while the solutions are unbounded in the white bands. In terms of the amplitude $a$ in \cref{eq:smallosca}, the shaded areas correspond to transition zones where a non-integer number of nucleation events occurs during each cycle of the forcing.  In fact, nonperiodic dynamics occur for irrational values of the associated Mathieu characteristic exponent within these zones.   The first white band on the far left corresponds to stable periodic orbits that do not undergo nucleation.  The state undergoes one nucleation per oscillation in the white band immediately to the right, and the number of nucleations per oscillation increases by integer values within each subsequent white band. 

We remark (cf. section IIB) that care must be taken in interpreting the solutions to \cref{eq:depinosc,eq:depinosc-} since the zeros of $b(\tau)$ correspond to solutions of \cref{eq:smallosca} that diverge to $\pm\infty$. During this process higher order nonlinearities enter \cref{eq:smallosca} with the result that the Riccati transformation no longer yields a linear equation. Thus the solutions of \cref{eq:depinosc,eq:depinosc-} in fact fail to describe the depinning process near the zeros of $b(\tau)$, and the corresponding solution $a(\tau)$ is determined by ``gluing'' together a series of individual nucleation events. However, as suggested by the description in \cref{eq:depinosc,eq:depinosc-}, the resulting nucleation process is indeed periodic, albeit in the frame moving with the front $x=f(\tau)$.

\subsection{Asymptotic theory: large oscillations}

\begin{figure}
\centering

\subfloat[]{ \includegraphics[width=75mm]{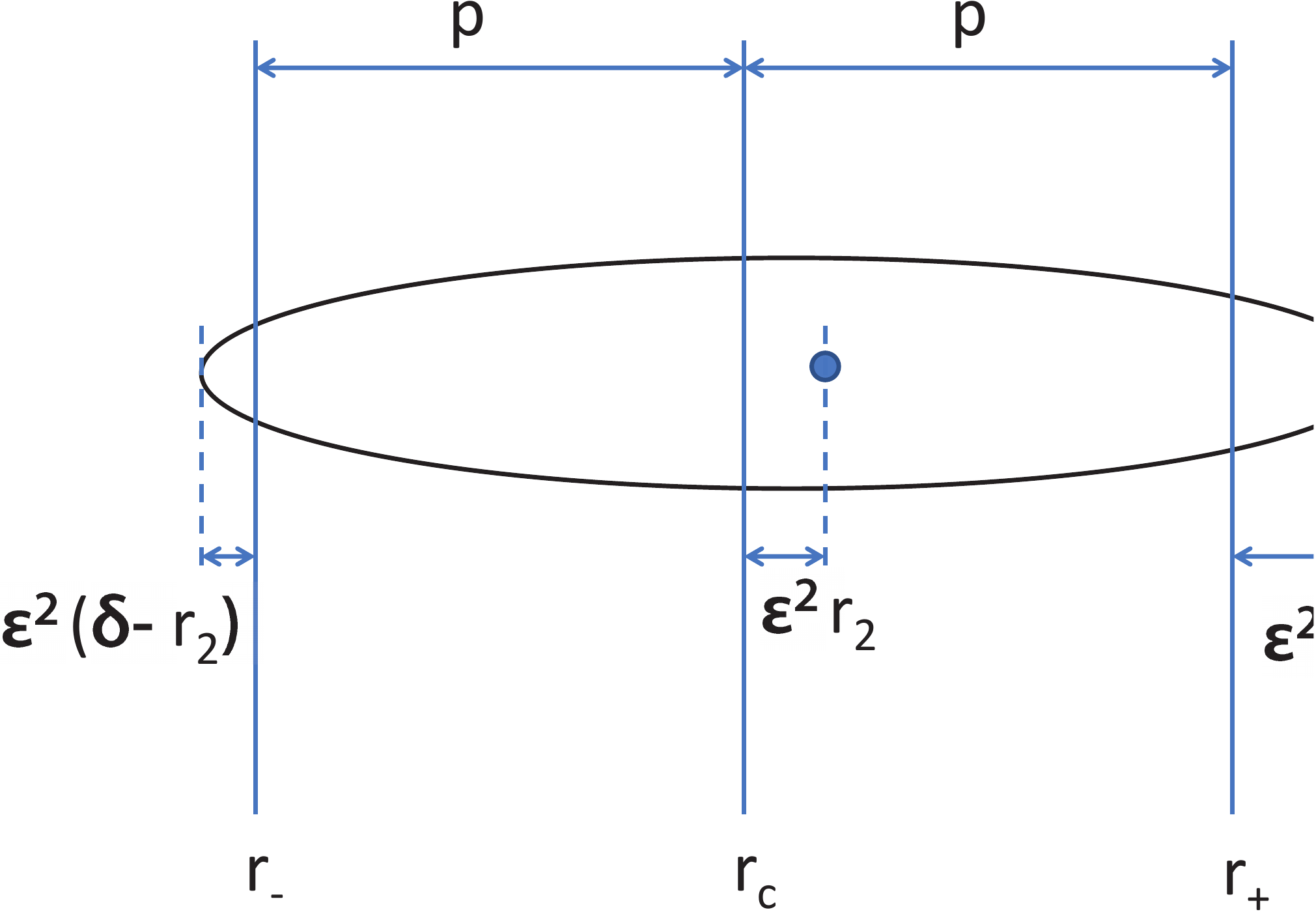}}\\
\subfloat[]{ \includegraphics[width=75mm]{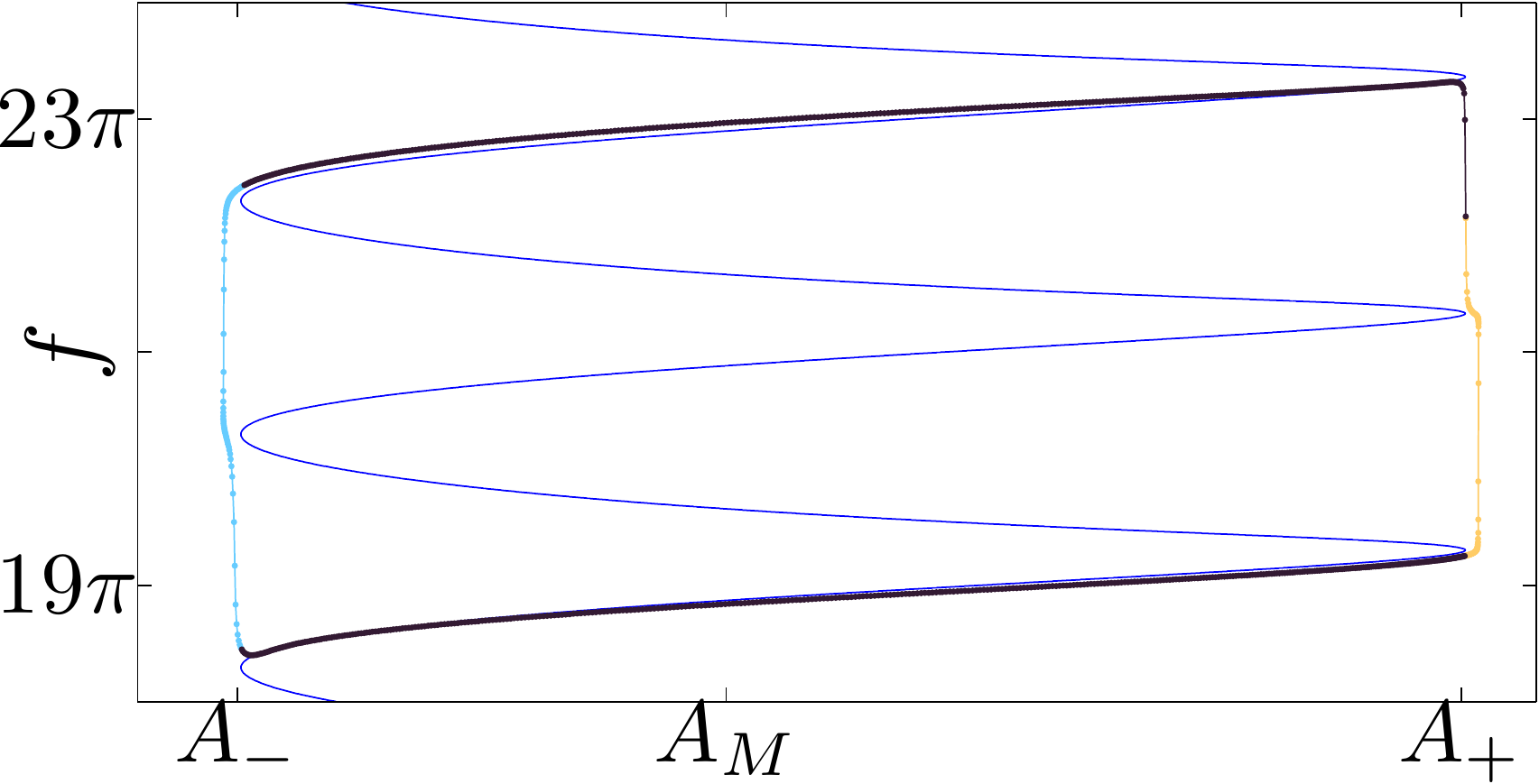}}

\caption{ (a) A schematic of the forcing function $r(t)$ used in the asymptotic theory. (b) A periodic orbit with $\rho=p+10^{-3}$, $T=5000$, and $r_0= -0.299$. The orbit is colored by the magnitude of $r$: purple corresponds to $\mathcal{P}_{\pm}$, orange to $\mathcal{D}_{+}$, and blue to $\mathcal{D}_{-}$.}  
    \label{fig:asymptoticschematic}
\end{figure}

With intuition gained from the small amplitude theory we now consider the case of large parameter oscillations that take the system just outside of the pinning region, but with a long enough period that there is time for depinning to occur. Because the system spends only a small fraction of the forcing cycle outside of $\mathcal{P_{\pm}}$, we require the forcing cycle period to be yet longer, $T=\mathcal{O}(\epsilon^{-2})$, in order that depinning takes place. We therefore define the slow timescale $\mathcal{T} = \epsilon^2 t$ and write the forcing parameter in the form
\begin{equation}
r=r_c +\epsilon^2 r_2 + (p+\epsilon^2 \delta)\sin \left( \epsilon^2\omega t\right),
\label{eq:asymptoticlimitlarge}
\end{equation}
where $r_c\equiv (r_++r_-)/2$ corresponds to the center of the pinning region, and $p\equiv (r_+-r_-)/2$ is its half-width. This choice allows for the oscillations to take the system just outside of the pinning region on both sides where depinning can be described quantitatively; the small offset represented by $r_2$ is included for greater generality, as is illustrated in the schematic in \cref{fig:asymptoticschematic}(a). A periodic orbit from a simulation, colored according to the value of the forcing parameter $r$ is also shown.

We anticipate that in the above setup nucleation will occur on the faster timescale $\tau=\mathcal{O}(\epsilon^{-1})$ and so look for solutions in the form $u=u_0+\epsilon u_1+\epsilon^2 u_2+...$, where $u_j\equiv u_j(x,{\tau},{\mathcal{T}})$. Writing $\partial_t = \epsilon\partial_{\tau}+\epsilon^2\partial_{\mathcal{T}}$, we obtain at leading order
\begin{equation}
\left[r_c+p \sin \left( \omega \mathcal{T} \right)-(1+\partial_x^2)^2\right]u_0+bu_0^2-u_0^3=0.
\end{equation}
Thus we can choose $u_0(x,\mathcal{T})$ to be a stable localized solution of the time-independent Swift--Hohenberg equation within the pinning region, with $\mathcal{T}$ determining the value of the forcing parameter within this region. The solutions follow the corresponding segment of the $L_0$ branch (\cref{fig:asymptoticschematic}(b)) as long as $\pi(n-1/2) <\omega \mathcal{T}< \pi(n+1/2)$, for any integer $n$.  As we will see, special care must be taken near the extrema of the forcing cycle when the system leaves the pinning region and the dynamics take place on the faster timescale $\tau$. 

The $\mathcal{O}(\epsilon)$ correction reads
\begin{equation}
\label{eq:Largeoeps}
\left[r_c+p \sin (\omega \mathcal{T})-(1+\partial_x^2)^2+2bu_0-3u_0^2\right]u_1 = 0.
\end{equation}
When $\omega \mathcal{T} =(2n+1/2)\pi $ (resp. $ (2n+3/2)\pi$), the quantity $u_1$ solves the linearized Swift--Hohenberg equation at $r_{+}$ (resp. $r_-$) with symmetric solution $v_+$ (resp. $v_-$), i.e. the depinning mode responsible for growth (resp. decay) of the localized pattern. In contrast, when $\pi(n-1/2) <\omega \mathcal{T}< \pi(n+1/2)$, no reflection-symmetric marginally stable modes are present.
To examine the dynamics near $r_{+}$ we take the slow time to be $\omega \mathcal{T}= \pi/2 +\epsilon \theta$; a similar procedure can be carried out near $r_-$ by taking $\omega \mathcal{T}= 3\pi/2 +\epsilon \theta$, and subsequent cycles of the forcing can be handled in the same way. The time derivative now becomes $\partial_t=\epsilon\omega\partial_{\theta}$ and the $\mathcal{O}({\epsilon})$ equation (\ref{eq:Largeoeps}) becomes
\begin{equation}
\label{eq:Largeoepspeak}
\left[r_+ -(1+\partial_x^2)^2+2bu_0-3u_0^2\right]u_1 = \omega \partial_{\theta} u_0.
\end{equation}
Since $u_0$ is the marginally stable localized solution at $r_+$ it follows that $\partial_{\theta} u_0=0$ and hence that $u_1=a(\theta)v_+(x)$.

At $\mathcal{O}(\epsilon^2)$ we obtain 
\begin{equation}
\label{eq:Largeoeps2peak}
\left[r_+ -(1+\partial_x^2)^2+2bu_0-3u_0^2\right]u_2 = \omega \partial_{\theta} u_1-(b-3u_0)u_1^2-(r_2+\delta-\tfrac{1}{2}p\theta^2)u_0,
\end{equation}
for which the solvability condition is 
\begin{equation}
\alpha_1\omega a' = \alpha_2 (r_2+\delta-\tfrac{1}{2}p\theta^2)+\alpha_3 a^2,
\label{eq:asymptotica}
\end{equation}
where the prime denotes the $\theta$ derivative and the coefficients $\alpha_j$ are determined by the integrals in \cref{eq:alpha}.

Using the transformation $a=-\alpha_1\omega b'/ \alpha_3 b$, we obtain a linear oscillator problem with a time-dependent frequency,
 \begin{equation}
b'' = -\frac{p\Omega_+^2}{2\omega^2} \left(\theta_+^2-\theta^2\right)b,
\end{equation}
where $\theta_+^2=2(r_2+\delta)/p$ and, as before, $\Omega_+^2=\alpha_2\alpha_3/\alpha_1^2$. The system exits $\mathcal{P}_+$ when $r_2+\delta>0$, and in this case $[- \theta_+,\theta_+]$ corresponds to the time interval spent in $\mathcal{D}_+$. We now use a matching procedure to connect this solution to the case when $\omega \mathcal{T}\neq \pi/2$, noting that $u_1(x,\mathcal{T})\to 0$ as $\omega\mathcal{T}\to \pi/2$, so that the solution remains stable as it approaches the boundary of the pinning region. Since the leading order solution for large $|\theta|$ is given by $a(\theta)\approx \sqrt{p\alpha_2 /2\alpha_3}|\theta|$ 
we require that $a(\theta)\to \sqrt{p\alpha_2 /2\alpha_3}\theta<0$ as $\theta\to -\infty$. The solution of \cref{eq:asymptotica} satisfying this requirement can be written in terms of parabolic cylinder functions~\cite{abramowitz1972handbook}, $b=D_{\nu}(z)$, where 
\begin{equation}
\nu=\frac{\sqrt{p}\Omega_+\theta_+^2}{2\sqrt{2} \omega}-\frac{1}{2},\qquad  z=-(2p)^{1/4}\sqrt{\frac{\Omega_+}{\omega}}\theta. 
\end{equation}
Each zero $z=z_0$ of $b=D_{\nu}(z)$ corresponds to a nucleation event, since $a$ diverges to $\pm \infty$ as $z\to z_0$. To determine the outcome of such an event we must consider the limit as $\theta\to \infty$ to match the $\dot{r}>0$ phase of the forcing cycle with the $\dot{r}<0$ phase that follows. This will tell us what branch the solution follows for $\pi/2<\omega \mathcal{T}< 3\pi/2$. The parabolic cylinder function behaves like 
\begin{equation}
D_{\nu}(z) \to \tfrac{\sqrt \pi}{\Gamma[-\nu]
} |z|^{-(\nu+1)}e^{z^2/4}\label{eq:Dliminf}
\end{equation}
for real $z\to -\infty$, where $\Gamma[\nu]$ is now the Gamma function. Examining the sign of $\Gamma[\nu]$ shows that as long as $\nu$ is not a positive integer, $a\to \sqrt{p\alpha_2/2\alpha_3}\theta>0$ in this limit, and the solution does indeed settle on the nearest stable localized solution. The special cases for which $\nu$ is a positive integer correspond to the solution landing exactly on an unstable solution.

We can get a very simple expression for the number of nucleation events that occur near $r_+$ by counting the number of real zeros of $D_{\nu}(z)$ for a given $\nu$. We start by noting that for $z\to \infty$ ($\theta\to -\infty$), $D_{\nu}(z)$ approaches zero from above. When $\nu<0$, there are no real zeros and $D_{\nu}(z)>0$ for all $z$.  \Cref{eq:Dliminf} shows that in the limit that $z\to -\infty$, the sign of $D_{\nu}(z)$ depends on the sign of $\Gamma[-\nu]$. For $\nu<0$, the sign is positive and $D_{\nu}(z)\to +\infty$  as $z\to -\infty$ without crossing zero. At $\nu=0$, $\Gamma[-\nu]=\infty$ and $D_{\nu}(z)\to 0$ from above; there are still no zero crossings. For $0<\nu<1$, there will be one zero crossing as $D_{\nu}(z)\to -\infty$ as $z\to -\infty$.  The number of zeros continues to increase by one each time there is a sign change in $\Gamma[-\nu]$ so that for $n-1<\nu<n$, there will be $n>0$ zeros of $D_{\nu}(z)$.  Therefore there will be $n_+$ nucleation events if 
\begin{equation}
n_+-\tfrac{1}{2}<\frac{\Omega_+ T}{2\pi\sqrt{2p} }(r_0+\rho-r_+)<n_++\tfrac{1}{2},
\label{eq:nPasymptotic}
\end{equation}
where we have re-expressed the condition in terms of the amplitude $\rho\equiv p+\epsilon^2\delta$, offset $r_0\equiv r_c+\epsilon^2 r_2$, and the period $ T\equiv 2\pi/\epsilon^2\omega$. 
A similar relation applied for the number of annihilations $n_-<0$:
\begin{equation}
n_--\tfrac{1}{2}<\frac{\Omega_- T}{2\pi\sqrt{2p} }(r_0-\rho-r_-)<n_-+\tfrac{1}{2}.
\label{eq:nMasymptotic}
\end{equation}

\begin{figure}
\centering
    \mbox{
		\subfloat[$\nu=-0.5$]{\begin{tabular}[c]{c}
			\includegraphics[width=40mm]{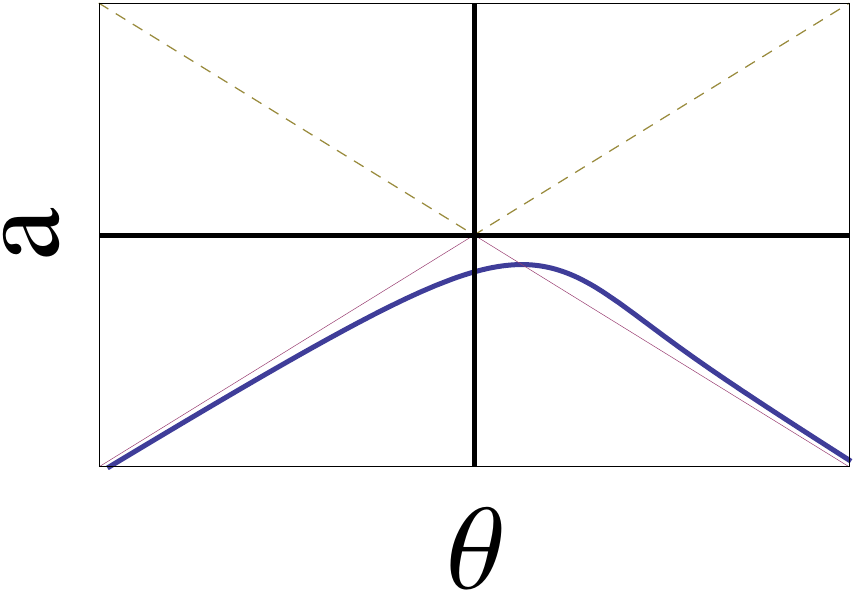}\\
		\includegraphics[width=40mm]{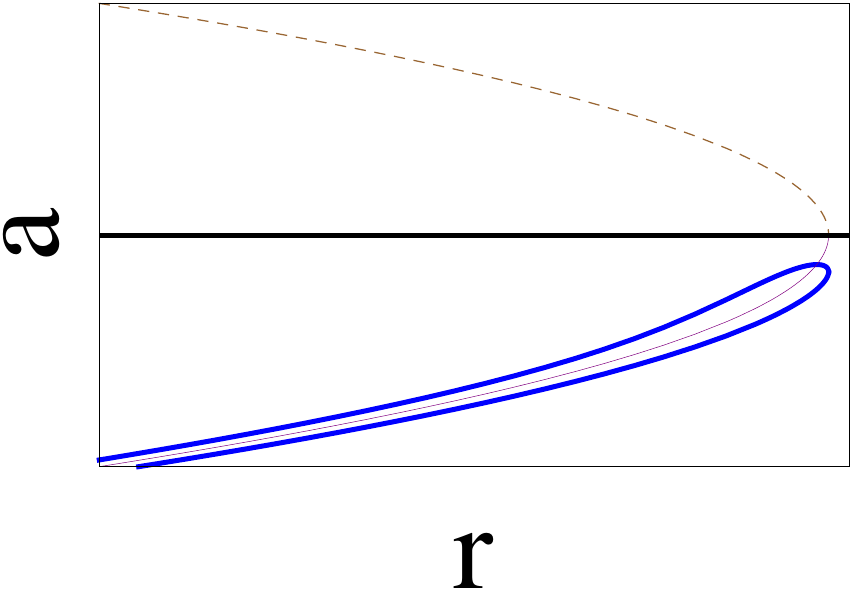}
		\end{tabular}
		}
		
		\subfloat[$\nu=-0.1$]{\begin{tabular}[c]{c}
			\includegraphics[width=40mm]{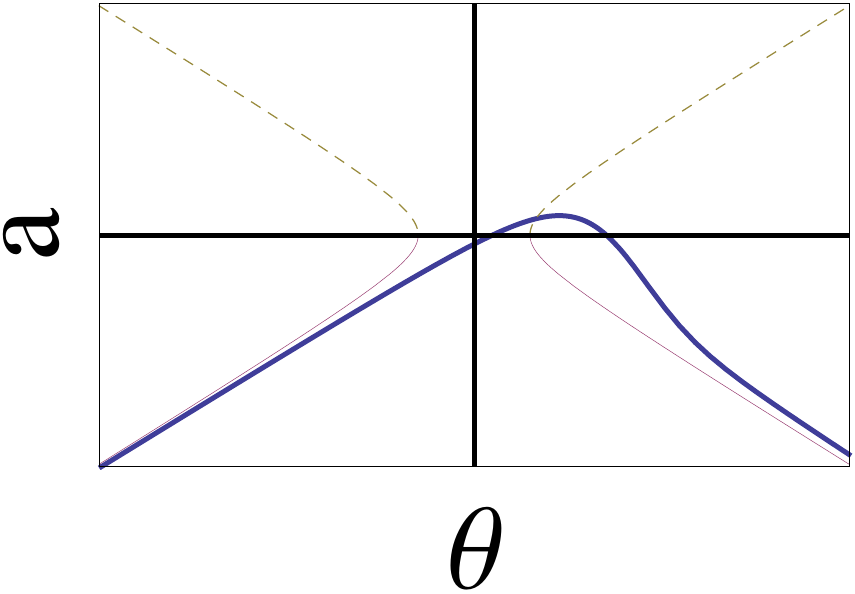}\\
		\includegraphics[width=40mm]{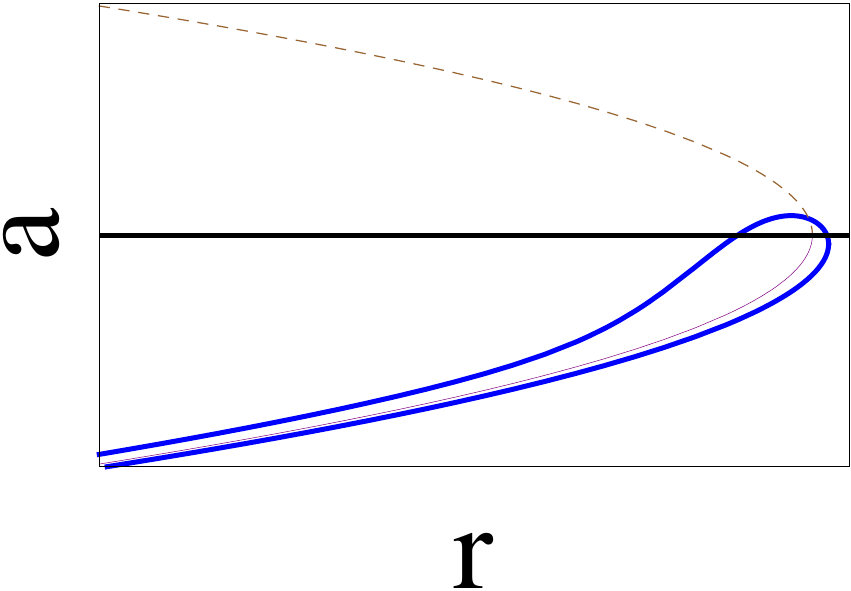}
		\end{tabular}		
		}
		\subfloat[$\nu=0.1$]{\begin{tabular}[c]{c}
\includegraphics[width=40mm]{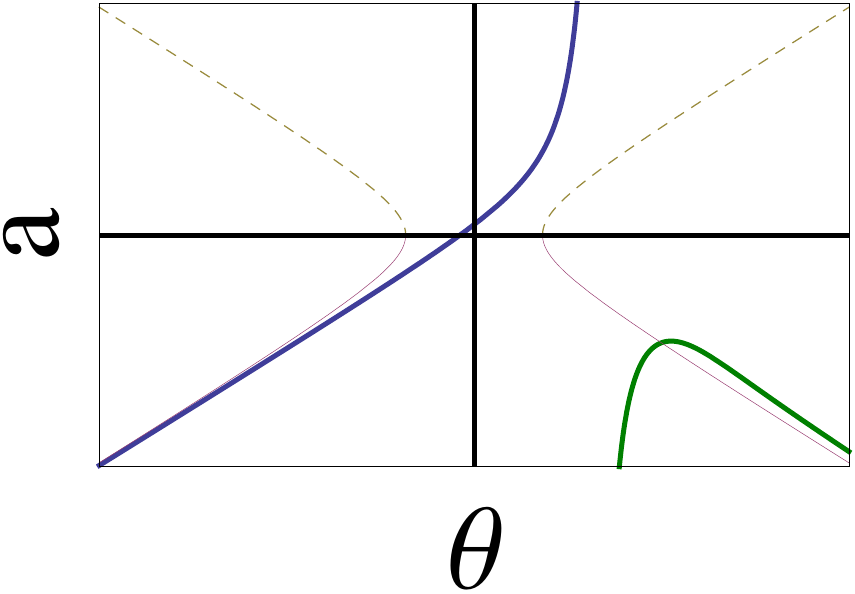}\\
		\includegraphics[width=40mm]{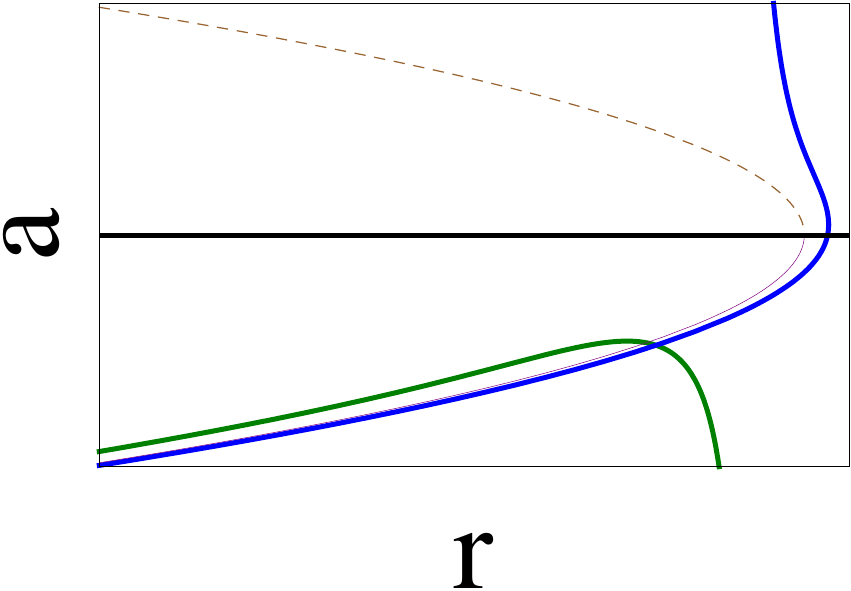}
		\end{tabular}		
		}
}
\caption{A plot of the amplitude $a(\theta)$ of the $\mathcal{O}(\epsilon)$ correction $u_1(x)$ to the solution that is marginally stable at $r=r_+$ as a function of a scaled $\mathcal{O}(\epsilon)$ time near the boundary of the pinning region for different values of $\nu=\frac{\sqrt{p}\Omega_+\theta_+^2}{2\sqrt{2} \omega}(r_0-\rho-r_-)-\frac{1}{2}$. The time $\theta=0$ corresponds to the peak of the forcing cycle where ``inertial" effects are expected.  The thin solid (dashed) line shows the amplitude $a$ for the stable (unstable) localized solution as functions of $r$ but replotted in terms of the time $\theta$.  Below each frame is a schematic representation of the trajectory of the amplitude $a$ as a function of the forcing parameter $r$.  The stable (unstable) steady state branches of the constant forcing case are shown in solid (dashed) lines for reference.  (a) $\nu=-0.5$: the system does not leave the pinning region, but there are still deviations from the stable state. (b) $\nu=-0.1$: the system exits the pinning region, but not far enough for nucleations to occur. (c) $\nu=0.1$: the system penetrates into $\mathcal{D}_+$ past the threshold for a nucleation to occur (represented by a discontinuous jump).  }
    \label{fig:asymptoticinertia}
\end{figure}

The above conditions also reveal the presence of bifurcation delay, as expected of a nonautomous bifurcation problem. This delay manifests itself in the shift of the critical value $r_2+\delta=0$ for the presence of a fold to the threshold value determined by $\nu=0$, viz., $r_2+\delta=\omega/\sqrt{2p}\Omega_+$: the system enters $\mathcal{D}_+$ by as much as $\omega/\sqrt{2p}\Omega_+$ without triggering a nucleation event. \Cref{fig:asymptoticinertia}(b,c) shows the amplitude $a$ as a function of the scaled time $(2p)^{1/4}\Omega_+\theta/\omega$ just before and after this threshold. The discontinuous jump in \cref{fig:asymptoticinertia}(c) represents a nucleation event and is obtained by gluing together two separate asymptotic calculations near different, but adjacent saddle-nodes on the same snaking branch. The same ``inertial" effect is observable even when the system does not leave $\mathcal{P}_+$, i.e., $\delta+r_2<0$. Using the property
\begin{equation}
D_{\nu}(z)\to \sqrt{2^{\nu}\pi}\left(\frac{1}{\Gamma [(1-\nu )/2]}-\frac{\sqrt{2}z}{\Gamma [-\nu /2]}- \frac{(1+2\nu )z^2}{4 \Gamma [(1-\nu)/2]}\right), \qquad z\to 0,
\end{equation}
we see that even when the system just barely reaches the boundary of the pinning region, $r_2+\delta=0$, the perturbation $a(\theta)$ remains finite (\cref{fig:asymptoticinertia}(a)). Indeed the minimum value of $a(\theta)$ occurs for $\theta>0$ instead of $\theta=0$. In fact $a(\theta)$ can be calculated explicitly in terms of parabolic cylinder functions using the relation $\frac{d D_{\nu}(z)}{dz}=\tfrac{1}{2} z D_{\nu}(z)-D_{\nu+1}(z)$ (\cref{fig:asymptoticinertia}).   

In the parameter regime analyzed here, the system tracks a given stable localized state with a delay in $r$ of up to $\mathcal{O}(\epsilon^2)$ for most of the forcing cycle. All of the interesting dynamics occur within a small time interval when the system visits the vicinity of the boundary of the pinning region. If it ventures far enough outside of this boundary, nucleation/annihilation events begin to take place after a delay.  Once the system reenters the pinning region, the system settles on the nearest stable but longer/shorter localized structure.  The settling process also happens within the vicinity of the boundary of the pinning region and there may or may not be an additional nucleation/annihilation event during this settling, depending on where in the process the system was upon re-entering the pinning region $\mathcal{P}_{\pm}$.

\begin{figure}
\centering
\includegraphics[width=120mm]{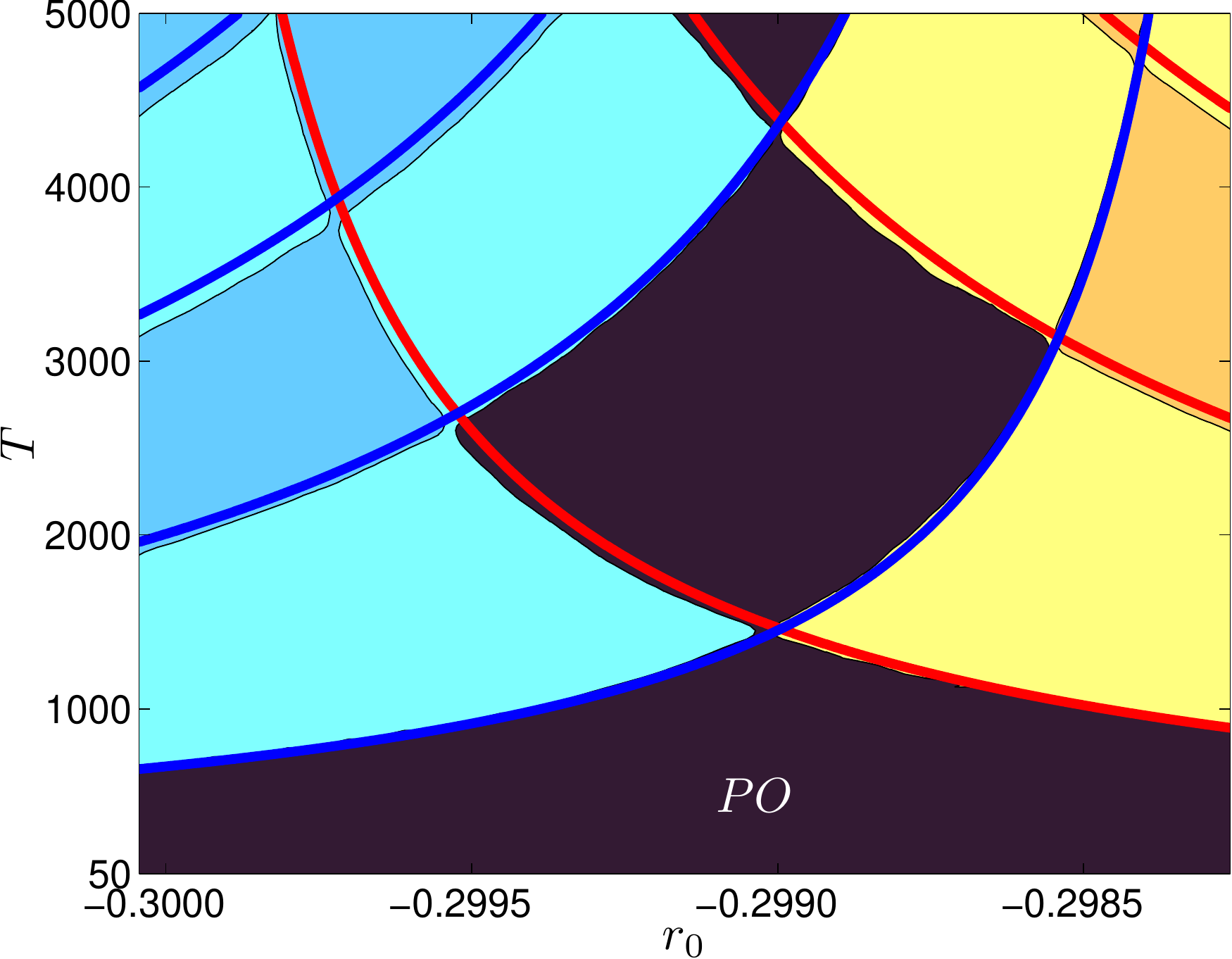}
\caption{A comparison of the asymptotic theory (red/blue lines) with numerical simulations (colors) for $\rho=p+10^{-3}\approx 0.04$.  The dark region corresponds to the region $PO$. The red (blue) lines indicate transitions in the number of nucleations (decays) that occur during one forcing cycle. }
    \label{fig:asymptotic}
\end{figure}

To understand the structure of growing, steady-state, and decaying solutions in this limit, we need only compare the growth from depinning that occurs near $\omega \mathcal{T}=\pi/2$ to the decay from depinning that occurs near $\omega\mathcal{T}=3\pi/2$. The formation of the pinched zones and the sweet spot structure of the stationary solutions can be predicted by balancing the growth on the right of the pinning region to the decay on the left as we shall now see. The resulting prediction is compared with numerical simulation in \cref{fig:asymptotic} for $\rho=p+10^{-3}$, and shows excellent quantitative agreement. Specifically, the colored regions are determined by numerical simulation and these match the coding scheme of \cref{fig:wholediag} while the red and blue lines are predictions for the transitions between the various regions of the classification scheme detailed in \cref{fig:classify}. Note, however, that the values of $r_0$ in \cref{fig:asymptotic} span only about $1/40$th of the pinning region of the constant forcing system:  the sweet spot--pinched structure here is asymptotically small as a result of our choice of $\rho$.  

\begin{figure}
\centering
    \mbox{
\subfloat[$T=100$]{\includegraphics[width=75mm]{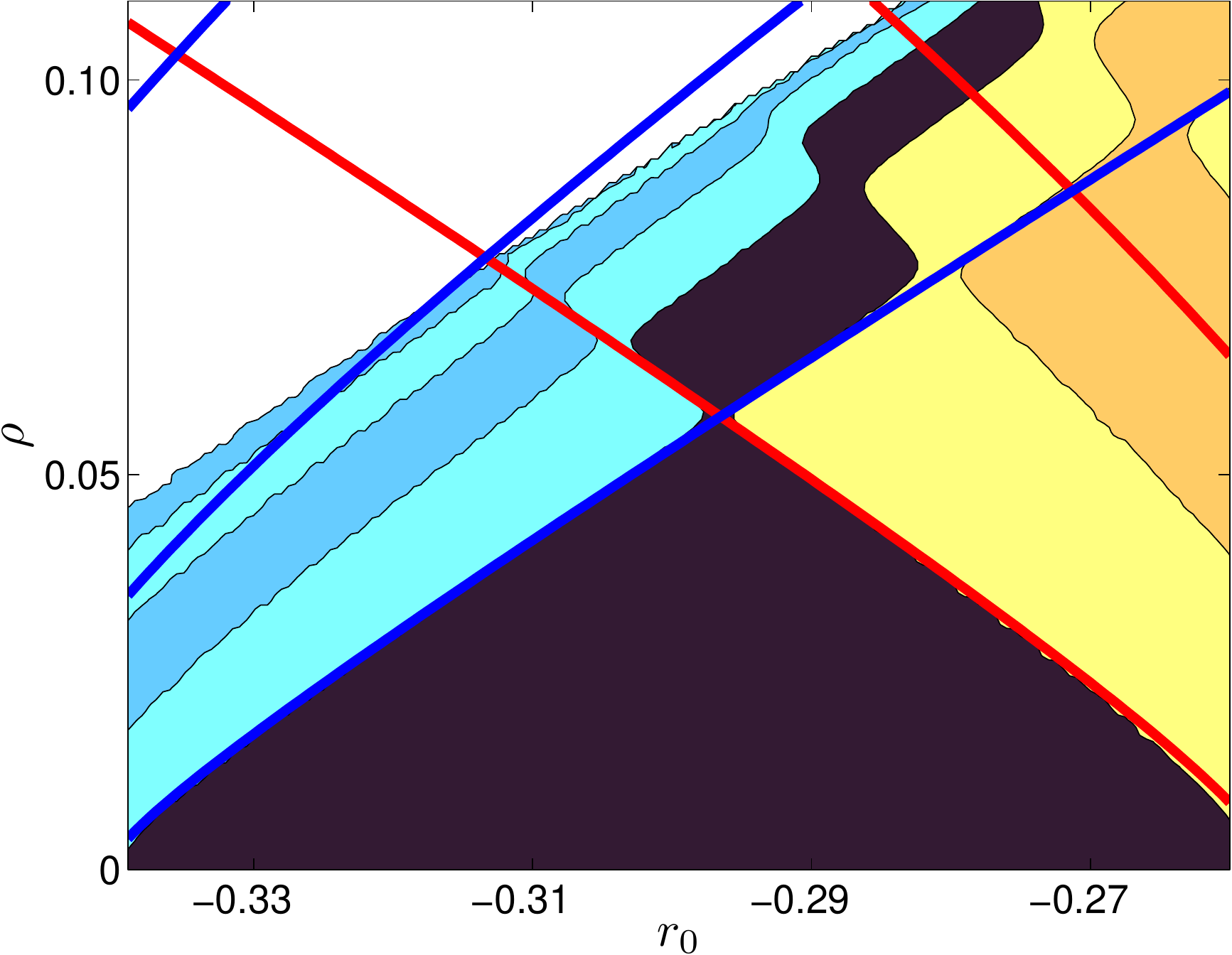}}\quad
\subfloat[$T=200$]{\includegraphics[width=75mm]{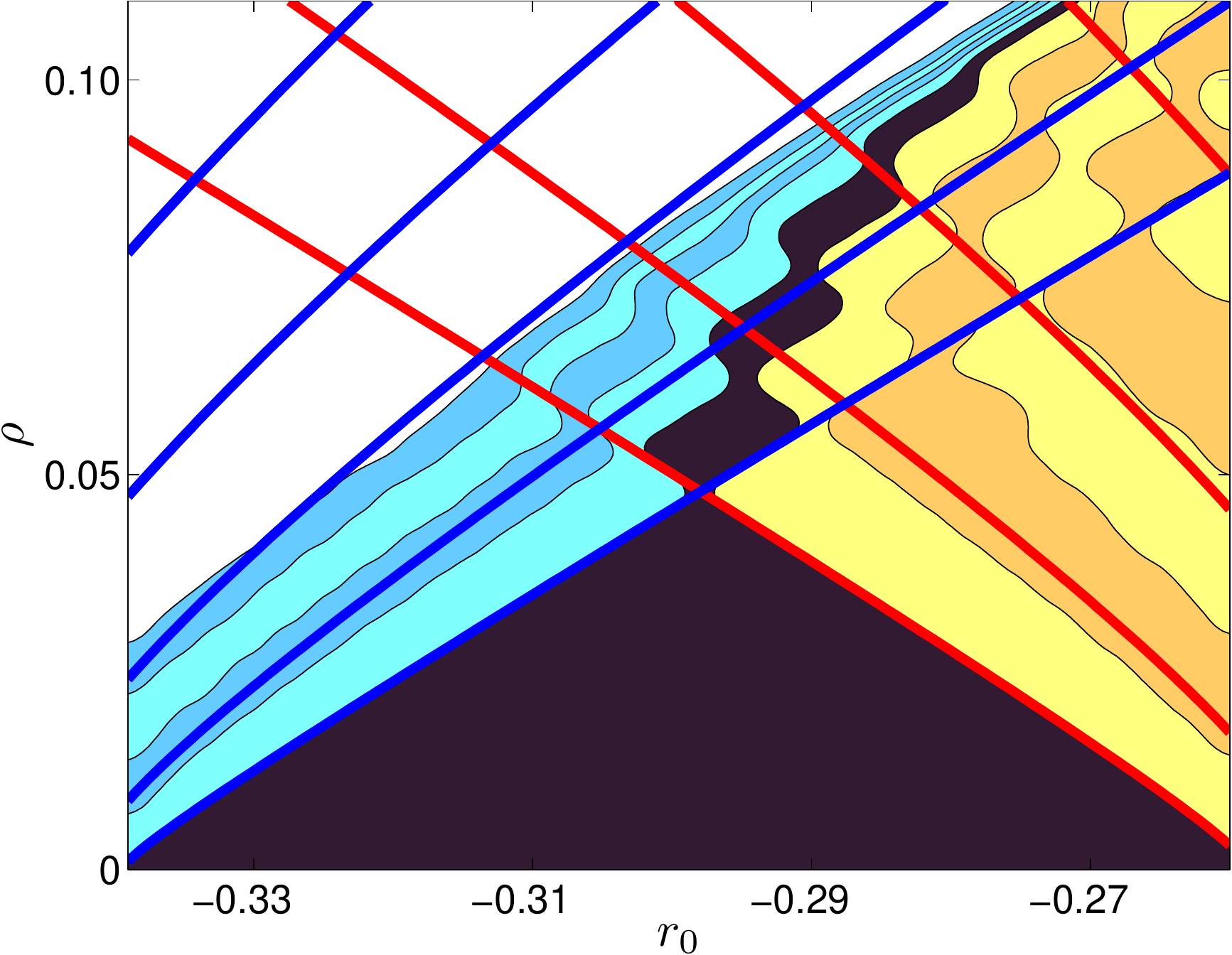}}
}
\caption{The number of spatial periods gained/lost in one forcing cycle when (a) $T=100$ and (b) $T=200$. All simulations were initialized with stable $L_0$ solutions at the corresponding $r_0$ in the constant forcing case. The purple region labeled $PO$ indicates the location of periodic orbits and corresponds to the blue region in \cref{fig:vcmstable}. The light blue region immediately to the left indicates decay by one wavelength on each side of the localized structure per forcing cycle, the next region to the loss of two wavelengths per cycle, and so on. The solution grows by one wavelength on each side of the localized structure per forcing cycle in the region immediately to the right of the dark region, and so on. The white region to the left indicates solutions that collapse to the trivial state within one cycle.  }
    \label{fig:wholediagrho}
\end{figure}

\Cref{fig:wholediagrho} shows the $(r_0,\rho)$ parameter plane for $T=100$ and $T=200$, along with an extension of the predictions from the above asymptotic theory (\cref{eq:nPasymptotic,eq:nMasymptotic}).  The extension has been computed by replacing $\sqrt{p}$ in the denominator of the expressions by $\sqrt{\rho}$ as a means for correcting for the cases when $\rho \not\approx p$.  The modified theory is able to accurately predict the location of the transition between the zero and $\pm 1$ bands well outside of the limit in which it was derived, owing to the fact that the first nucleation/annihilation event necessarily occurs near the edge of the pinning region.  

\section{The low-frequency limit:  Adiabatic theory}

In this section we consider the remaining case, that of low-frequency forcing. In this regime we may neglect inertial effects that can cause delays in the onset of depinning or allow for the completion of depinning events within the pinning region. 
We note that, by applying the technique of matched asymptotics (see, for example,~\cite{hinch1991perturbation}), we can estimate the depinning delay to be $\sim |dr/dt|^{-1/3}$ where the derivative is evaluated at $r_{\pm}$. 

\subsection{Sweet spot structure}

Using the adiabatic approximation described above, the number $n_{\pm}$ of nucleation/annihilation events over the course of a forcing cycle can be estimated from the expression
\begin{equation}
n_{\pm}=\pm \int_{\mathcal{T}_{\pm}}\frac{dt}{T^{\mathrm{dpn}}_{\pm}(t)},
\label{eq:adiabatic}
\end{equation}
where $\mathcal{T}_{\pm}$ is the time spent outside of the pinning region and $T^{\mathrm{dpn}}_{\pm}(t)$ is the time between nucleation/annihilation events of the constant forcing problem with parameter $r(t)$. The super/subscript $+$ (resp. $-$) refers to regime $\mathcal{D}_+$ (resp. $\mathcal{D}_-$). We assume that the dynamics within the pinning region allow the system to either complete the nucleation process (corresponding to rounding $n_{\pm}$ up), or settle back down to the state already reached (corresponding to rounding $n_{\pm}$ down). We also suppose that the threshold between completing a nucleation event or settling back corresponds to $n_{\pm}+1/2$ and will use brackets, e.g. $[n_{\pm}]$, to denote the nearest integer.  We recall that leading order asymptotics near the edge of the pinning region predict that $(T^{\mathrm{dpn}}_{\pm})^{-1}=\Omega_{\pm} \delta_{\pm}^{1/2}/\pi$ \cite{burke2006} and use this prediction together with the assumption $r(t)=r_0+\rho \sin2\pi t/T$ to obtain 

\begin{equation}
n_{\pm}=\pm \frac{2\sqrt{2\rho}\Omega_{\pm} T} {\pi^2} \left[ K\left(\tfrac{1-\eta_{\pm}}{2}\right)- \tfrac{1+\eta_{\pm}}{2} E\left(\tfrac{1-\eta_{\pm}}{2}\right) \right],
\label{eq:adiabaticasymptotic}
\end{equation}
where  $\eta_{\pm}=|r_0-r_{\pm}|/\rho < 1$ and
\begin{equation}
K(\mathrm{m})= \int_0^{\pi/2} \frac{1}{\sqrt{1-\mathrm{m} \sin^2\theta}} \;d\theta\qquad
E(\mathrm{m})= \int_0^{\pi/2} \sqrt{1-\mathrm{m} \sin^2\theta}\;d\theta.
\end{equation}
are the complete elliptic integrals of the first and second kind~\cite{abramowitz1972handbook}.

\begin{figure}
\centering
    \mbox{
\subfloat[$\rho=p+10^{-3}$, theory and simulation]{\includegraphics[width=75mm]{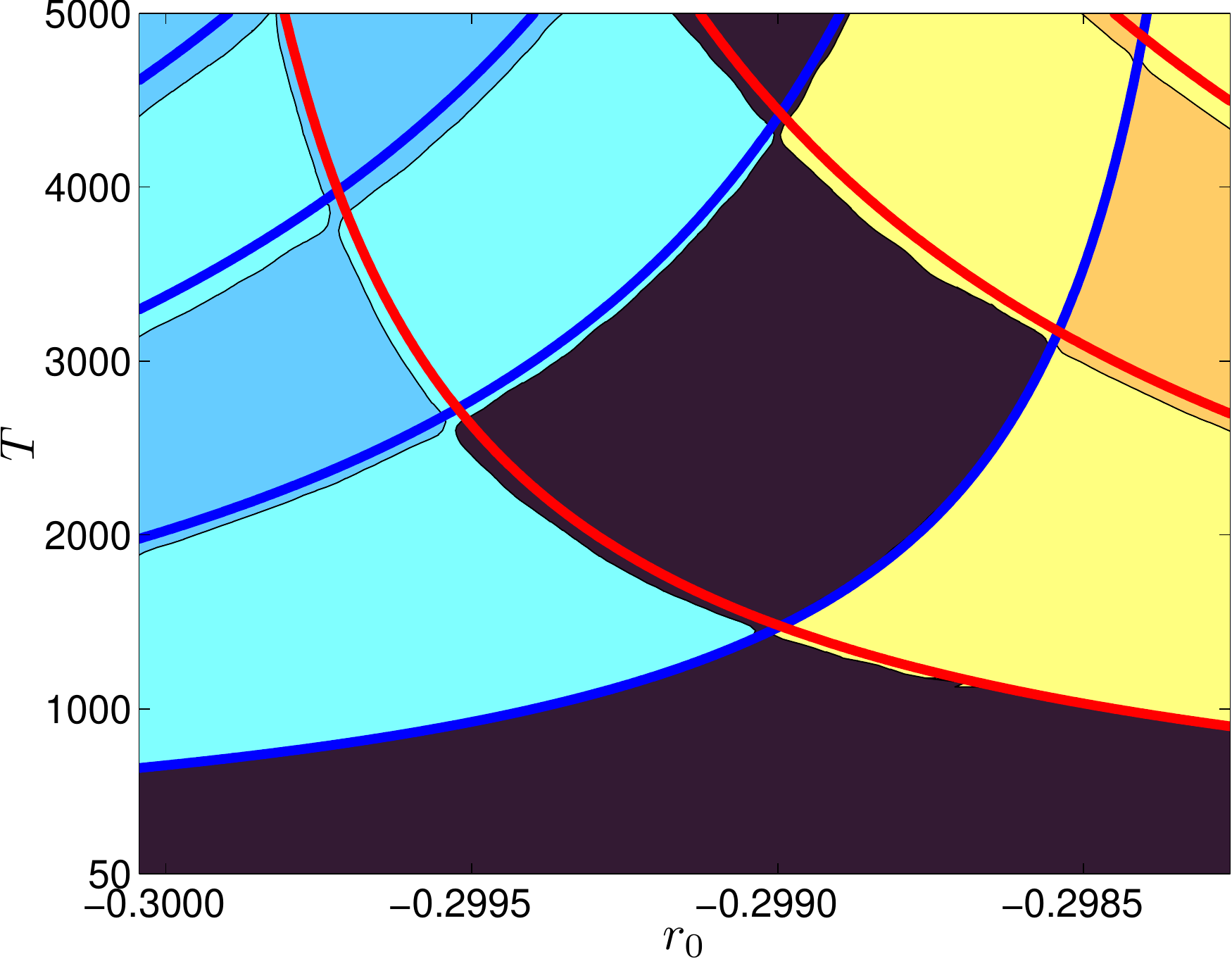}}\quad
\subfloat[$\rho=0.1$, theory only]{\includegraphics[width=75mm]{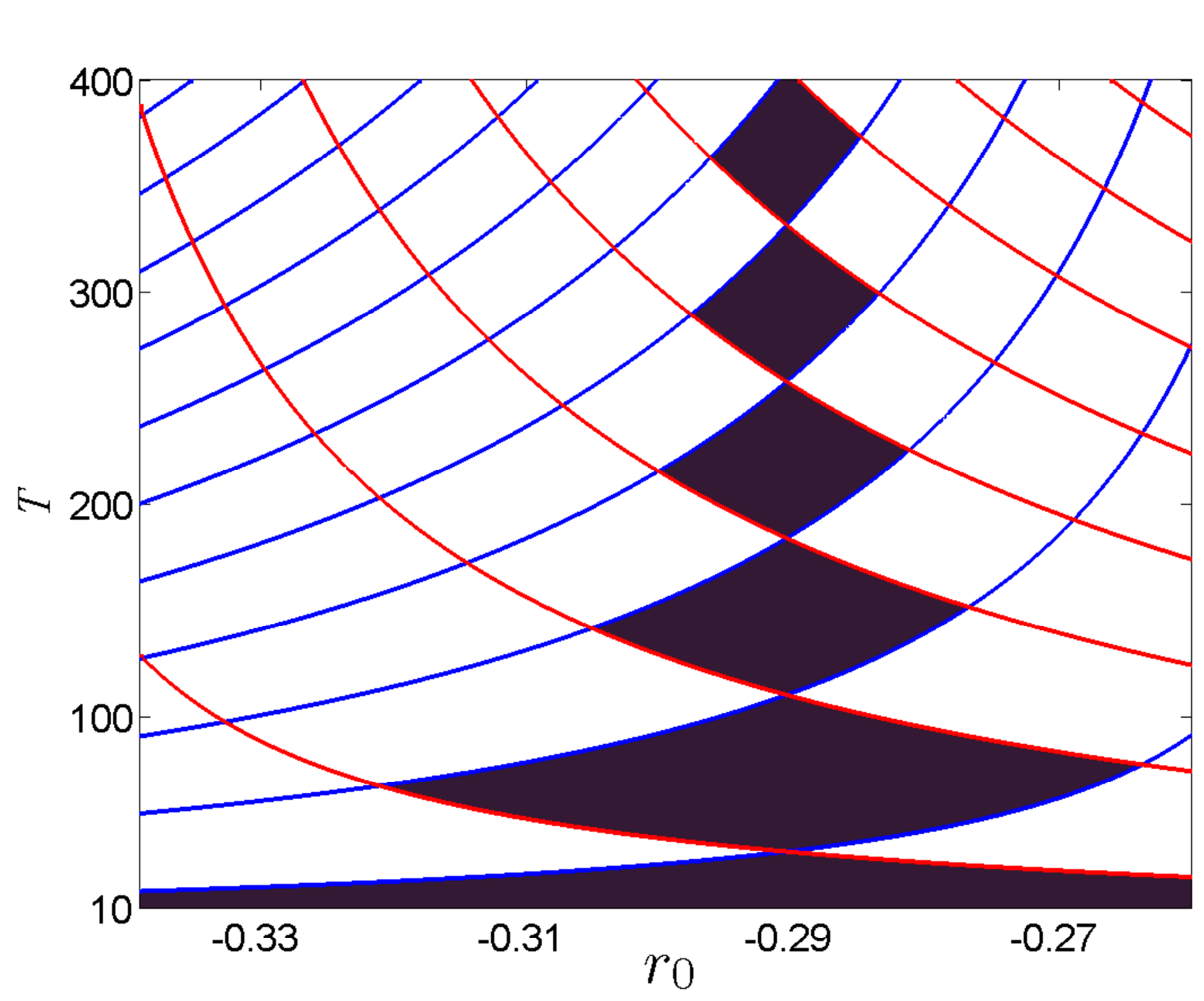}}
}
\caption{Predictions from adiabatic theory using the asymptotic approximation \cref{eq:adiabaticasymptotic} for the depinning time when (a) $\rho=p+10^{-3}\approx 0.04$ and (b) $\rho=0.1$. The colors in (a) refer to the numerical simulation results for the locations of $PO$ (dark), $O_{+n}$ (alternating yellow and orange), and $O_{-n}$ (alternating shades of blue).  The dark region in (b) is the adiabatic theory prediction of the $PO$ region.}
    \label{fig:adiabaticrho}
\end{figure}

The predictions of the adiabatic theory in \cref{eq:adiabaticasymptotic} are shown in \cref{fig:adiabaticrho} for $\rho=p+10^{-3} \approx 0.04$ and $\rho=0.1$. The red and blue lines indicate transitions between adjacent values of $[n_+]$ and $[n_-]$, respectively. The plot in \cref{fig:adiabaticrho}(a) is colored according to the simulation results to emphasize the quantitative accuracy of the adiabatic prediction for $\rho=p+10^{-3}$. The accuracy of these predictions diminishes with increasing $\rho$ as shown in \cref{fig:adiabaticrho}(b) for $\rho=0.1$ (cf. \cref{fig:wholediag}), although the predicted sweet spot and pinching structure continues to resemble the simulations. The dark region in this graph corresponds to the predicted location of $PO$ based on the theory (\cref{eq:adiabaticasymptotic}), i.e., $PO$ is the region where $[n_+]+[n_-]=0$.
The predictions for the $\rho=0.1$ case fail in three ways: (i) the pinched zones are spaced too far apart, (ii) there is no cliff demarcating the dominance of overall amplitude decay, and (iii) the region $PO$ does not slant as in the simulations (\cref{fig:wholediag}).  The qualitative disagreement occurs because of a breakdown of the asymptotic prediction for $T^{\mathrm{dpn}}_{\pm}$ when the system enters too far into regions $\mathcal{D}_{\pm}$. In addition to the quantitative disagreement of the depinning times, the theory omits the amplitude mode that destroys the localized states in $\mathcal{A}_-$.  We can account for (i) by making use of numerical fits in place of the asymptotic theory for $T^{\mathrm{dpn}}_{\pm}$ as described in \cref{fig:depinningtimecf} and can also extend the theory to include predictions about the cliff mentioned in (ii) as described in the next section.  The theory cannot, however, account for (iii) as the slanting is a result of the coupling between the amplitude and depinning modes which we have neglected.

\subsection{The cliff}
\begin{figure}
\centering
\includegraphics[width=120mm]{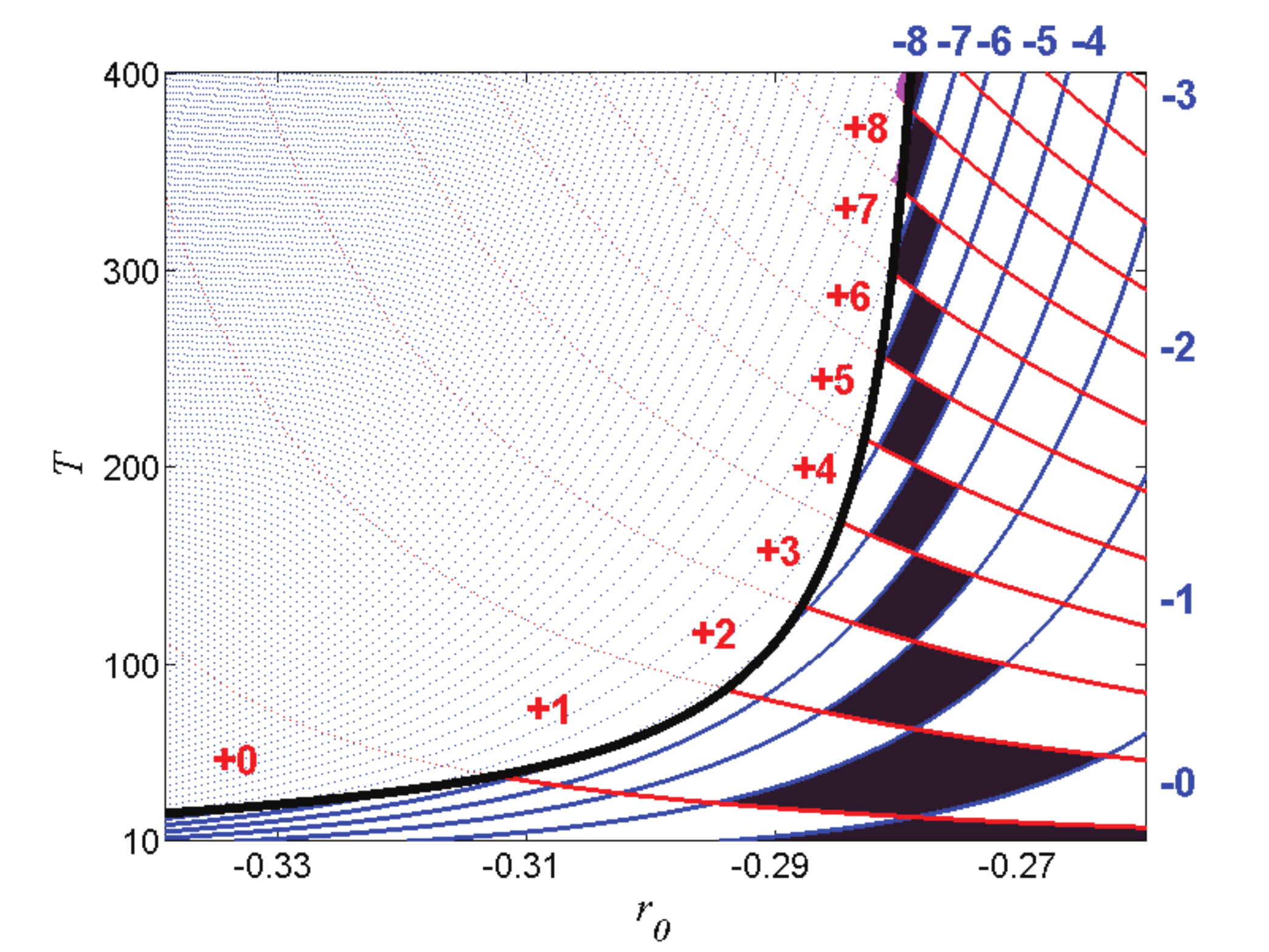}
\caption{Adiabatic prediction in the $(r_0,T)$ plane of the decay vs nucleation dynamics of a spatially localized initial condition of \cref{eq:SH} with time-periodic forcing. Positive (resp. negative) numbers represent $[n_+]$  (resp. $[n_-]$), the change in the number of wavelengths due to nucleation (resp. annihilation) events during one cycle. The results are obtained using the relation (\ref{eq:adiabaticasymptotic}) with $\rho = 0.1$ and $b=1.8$. The figure is plotted over the same $r_0$ interval as \cref{fig:wholediag}.}
\label{fig:adiabatic}
\end{figure}
We can approximate the dynamics of the overall amplitude decay of a localized state in $\mathcal{A}_-$ by computing the time $T^{\mathrm{col}}(r)$, $r<r_{sn}$, for a solution initialized with the periodic state at $r_{sn}$ to decay to the trivial state. This calculation mirrors the asymptotic calculation for $T^{\mathrm{dpn}}_{\pm}$. For $r=r_{sn}+\delta_{\mathrm{col}}$, $|\delta_{\mathrm{col}}|\ll1$, we find 
\begin{equation}
(T^{\mathrm{col}})^{-1} \approx \frac{\Omega_{\mathrm{col}}}{2\pi} |\delta_{\mathrm{col}}|^{1/2}.
\end{equation}
Substituting $T^{\mathrm{col}}$ into \cref{eq:adiabatic} in place of $T_{\pm}^{\mathrm{dpn}}$ leads to an equation analogous to \cref{eq:adiabaticasymptotic} for $n_{\mathrm{col}}$.  Since $n_{\mathrm{col}}$ is at most one we assume that the threshold for the solution to decay irrevocably is $n_{\mathrm{col}}=1/2$.  This procedure yields a prediction for the location of the cliff in parameter space.  For improved numerical accuracy for forcing cycles that penetrate far into $\mathcal{A}_-$, a numerical fit is useful (\cref{fig:depinningtimecf}). \Cref{fig:adiabatic} reveals the dramatic improvement in the $(r_0,T)$ phase diagram that results from this procedure applied to both $T^{\mathrm{dpn}}_{\pm}$ and $T^{\mathrm{col}}$ (bold black line). The hybrid adiabatic theory augmented with numerical fits from simulations of the system under constant forcing is in remarkably good agreement with the simulations even for $\rho=0.1$.  The predicted extent in $T$ of the sweet spots is $\triangle T \approx 45$, independently of $T$, which is within $5\%$ of the value computed from simulations, viz., $\triangle T \approx 43$.  In addition, the predicted  period at which the cliff occurs for a given value of of $r_0$ deviates from the value computed from simulations by $\triangle T \lesssim 10$, with the maximum deviation occurring for periods below $T=100$.  As expected, the agreement improves for larger periods and away from the cliff:  the top of the subregion $^{-3}O_{+7}^{+10}$ is predicted to be at $(r_0,T)\approx(-0.2633,373)$ but is located at $(r_0,T)\approx(-0.2637,377)$ in the simulations.  

\section{Conclusion}

We have considered the effects of parametric time-periodic forcing on the dynamics of localized structures in the Swift--Hohenberg equation with competing quadratic-cubic nonlinearities. In the high frequency limit averaging theory yields an averaged system that is also of Swift--Hohenberg-type. When oscillations are large enough, the time-varying forcing affects the averaged dynamics by modifying the coefficients of the nonlinear terms, thereby reducing the region of existence of spatially localized states (the pinning region in the case of constant forcing) and displacing it to larger values of $r_0$. For intermediate frequencies, the dynamics become more complex owing to depinning of the fronts bounding the localized structure over a significant fraction of the forcing cycle, resulting in breathing localized structures exhibiting behavior analogous to pinning, depinning, and amplitude collapse familiar from the constant forcing case. Of particular significance is the observation of a new resonance phenomenon between the forcing period and the time required to nucleate a new wavelength of the pattern. The presence of this resonance is responsible for the complex structure of the parameter space, which breaks up into regions labeled by a pair of integers $(m,q)$ denoting the number of wavelengths lost ($m$) per cycle and the number gained ($q$). When $n\equiv q-m=0$ the resulting state is periodic in time, and corresponds to a state that on average neither expands nor shrinks. We have described the resulting structure of the parameter space in terms of sweet-spots favoring the existence of such ``pinned'' states and pinched zones where the resonance was destructive and periodic localized structures absent. We found that these properties could be understood on the basis of appropriate asymptotics, valid either when the forcing cycle did not penetrate far into the depinning regions, or for low frequency forcing. In both cases we showed that the number of nucleation/annihilation events can be computed by adapting existing theory of the depinning process, and used these results to partition the parameter space. A similar approach was successful in obtaining the accumulation point of the decay regions beyond which all initial conditions collapse to the trivial state within one forcing cycle. Our calculations suggest that this accumulation is exponential and involves regions of frequency locking corresponding to all rational numbers. 

We found that asymptotic theory provided an excellent qualitative description of the resonance phenomenon, and moreover that quantitative agreement could often be obtained by augmenting the leading order nucleation theory with numerical fits to the nucleation times adopted from the time-independent case, thereby greatly extending the range of validity of the theory.

In view of the success of the Swift--Hohenberg equation in modeling localization in a great variety of systems with bistability between a homogeneous and a patterned state we expect that the model studied here, \cref{eq:SH}, captures faithfully the phenomenology arising from a resonance between the forcing period and the nucleation time in systems undergoing temporary depinning as a result of the forcing. As such we envisage numerous applications of the theory presented here to temporally forced systems such as models of vegetation growth in arid regions subject to seasonal forcing \cite{guttal2007self,siteur2014beyond}.

In future work it will be of interest to extend the present analysis to systems with time-periodic forcing that is not purely sinusoidal and in particular to periodic forcing with asymmetry between the rise and fall phases. In addition, many experimental systems exhibiting spatially localized states, including binary fluid convection \cite{Batiste2006spatially} and plane Couette flow \cite{Schneider2010snakes}, possess an additional midplane reflection symmetry whose effects are well modeled by the Swift--Hohenberg equation with a cubic-quintic nonlinearity \cite{HK11,Mercader13}. In future work we will investigate the properties of temporally-forced cubic-quintic Swift--Hohenberg equation with a view to elucidating the behavior expected when such systems are forced periodically.

\section*{Acknowledgments}
This work was supported in part by the National Science Foundation under Grants  DMS-1211953 and CMMI-1232902. 

\bibliography{TimeForcingBib}
\end{document}